\documentclass[aps,nofootinbib,notitlepage,superscriptaddress,twocolumn,10pt,prd,longbibliography]{revtex4-2}

\usepackage[T1]{fontenc}
\usepackage[utf8]{inputenc}
\usepackage{lmodern}
\usepackage[margin=2.25cm]{geometry}
\usepackage{amsmath,amssymb,bm,mathtools}
\usepackage{xcolor,enumitem}
\usepackage{pgfplots}
\pgfplotsset{compat=1.18}
\usepgfplotslibrary{groupplots}

\usepackage[colorlinks=true,linkcolor=blue!50!black,citecolor=blue!50!black,urlcolor=blue!50!black]{hyperref}
\newcommand{\pa}[1]{\left(#1\right)}
\definecolor{plotblue}{RGB}{42,112,196}
\definecolor{plotorange}{RGB}{222,112,46}
\definecolor{plotgreen}{RGB}{17,153,112}

\usepackage{graphicx}
\usepackage{amsmath,amssymb}
\usepackage{hyperref}
\usepackage{bm}
\usepackage{booktabs}
\usepackage[dvipsnames]{xcolor}
\usepackage{mathrsfs}
\usepackage{comment}

\newcommand{\dd}{\mathrm{d}}
\newcommand{\ii}{\mathrm{i}}
\newcommand{\eu}{\mathrm{e}}
\newcommand{\Ueff}{U_{\rm eff}}

\newcommand{\rr}{\bm{r}}
\newcommand{\kk}{\bm{k}}
\newcommand{\uu}{\bm{u}}
\newcommand{\xx}{\bm{x}}
\newcommand{\bn}{\bm{n}}
\newcommand{\bl}{\bm{\ell}}
\newcommand{\lp}{\left(}
\newcommand{\rp}{\right)}
\newcommand{\llp}{\left[}
\newcommand{\rrp}{\right]}
\newcommand{\be}{\begin{equation}}
\newcommand{\ee}{\end{equation}}

\newcommand{\Fzero}{F^{(0)}_{\rm newt}}

\newcommand{\Fasym}{F^{(0)}_{\rm newt,\infty}}

\begin{document}

\title{ Beyond the point lens: radial structure and the horizon in wave-optics gravitational lensing}
\author{Alice Garoffolo}
\affiliation{ Philippe Meyer Institute, Physics Department, Ecole Normale Sup\'erieure (ENS), Universit\'e PSL, 24 rue Lhomond, F-75231 Paris, France}
\author{Gianmassimo Tasinato}
\affiliation{Department of Physics, Swansea University, Swansea SA2 8PP, United Kingdom,} \affiliation{Dipartimento di Fisica e Astronomia, Universit\`a di Bologna, Via Irnerio 46, 40126 Bologna, Italy}

\medskip

\begin{abstract}
Gravitational-wave lensing in the wave-optics regime is sensitive to the structure of the lens beyond its total mass: to its exterior radial profile and, as we show, to the boundary condition at its centre. We develop an analytic framework to compute these effects for a scalar wave, as corrections to the wave lensed by a Newtonian point mass. The corrections are organized in inverse powers of the distance from the lens and computed with dressed radial kernels, which include the Newtonian interaction to all orders and resum the long-range phase that invalidates the Born expansion around the free wave. Retaining wavefront curvature, we find that the scattered wave is not the whole response: an inverse-square deformation changes the direct wave by an equal amount, even when the scattered wave is faint. Steeper terms of the exterior potential lead to short-distance divergences, which signal that the exterior geometry alone does not determine the lensed wave: the missing information is a set of local terms fixed by the inner boundary condition. For a Schwarzschild black hole we determine them by matching to the ingoing solution at the horizon. The horizon  enters the amplification factor as an absorptive contact term, which reproduces the horizon-area cross section; it  is absent for a non-absorbing lens of the same mass, and cannot be described by the thin-lens diffraction integral. It completes the correction quadratic in the mass, where it appears at the same order as finite-distance optical corrections, and largely cancels against them for an aligned source.  Together, these results connect Newtonian dressing, diffraction and inner boundary conditions, in a systematic description of corrections to point-mass gravitational-wave lensing.
\end{abstract}

\maketitle

%\tableofcontents

\section{Introduction}
\label{sec:introduction}

Gravitational-wave (GW) lensing gives access to the distribution of matter and curvature between a compact binary and the observer. When geometric optics applies, the signal is described by images with distinct magnifications and arrival times. In the wave-optics regime, diffraction and interference produce a complex amplification that varies across the chirp~\cite{Ohanian:1974ys,Deguchi:1986zz, Schneider:1992bmb,Nakamura:1997sw,Nakamura:1999uwi,Takahashi:2003ix}. See e.g. \cite{Leung:2023lmq,Grespan:2023cpa} for reviews. This frequency dependence provides a way to probe how the lens is structured: a point mass and a lens with an extended density profile can have the same total mass yet leave different modulations in the observed signal~\cite{Tanaka:2026bql,Choi:2026sjq,Cremonese:2021ahz}. This sensitivity has motivated studies of halo profiles \cite{Takahashi:2003ix,Tambalo:2022wlm} and dark matter around compact objects \cite{Gondolo:1999ef}, of compact and diffuse dark-matter structures \cite{Dai:2018enj,Lai:2018rto,Jung:2017flg,Diego:2019rzc,Oguri:2020ldf,Choi:2021bkx,Urrutia:2021qak,Guo:2022dre,Fairbairn:2022xln,Caliskan:2023zqm,Savastano:2023spl,Cheung:2024ugg,Ando:2026eam,Blinov:2026bxj}, and searches in the data of the LIGO--Virgo--KAGRA network~\cite{LIGOScientific:2023bwz,LIGOScientific:2025cwb}, including the analyses of GW231123 as a candidate lensed and diffracted event~\cite{Chan:2025kyu,Goyal:2025eqo,Chakraborty:2025pxt,Shan:2025dcd,Cheung:2026pky}. Wave optics has also been developed for systems of many lenses, with path-integral and multiplane formulations~\cite{Yamamoto:2003cd,Feldbrugge:2020tti} and statistical descriptions of the diffraction by populations of lenses, such as stellar fields and dark-matter halos~\cite{Diego:2019lcd,Zumalacarregui:2026uqs,Choi:2026lsa,Amoruso:2026txw}, and of the amplitude and phase fluctuations produced by the inhomogeneous distribution of matter~\cite{Takahashi:2005ug,Mizuno:2023pru,Kim:2025njb}. It also raises  questions about the point lens itself: how much of the waveform is fixed by the total mass, and which corrections carry information about the exterior geometry of the lens or about its inner boundary?

In this work, we set up the formalism for addressing these questions, and we apply it to the Schwarzschild black hole. Two tools are commonly used to compute the lensed waveform, and both have limitations when one asks these questions. The first is the thin-lens diffraction integral \cite{Takahashi:2003ix,Tambalo:2022plm,CarrilloGonzalez:2025gqm}; see also Ref.~\cite{Garoffolo:2026hyx} for wave-optics lensing with a modified wave equation. It sums over the paths that cross the plane of the lens, each weighted by a phase that combines the geometric travel time with the gravitational time delay integrated along the line of sight. It follows from the wave equation in the paraxial, or eikonal, approximation: small scattering angles, source and observer far from the lens, and paths that pass many wavelengths away from it. Corrections to the diffraction integral beyond the thin-lens, paraxial and eikonal approximations have been studied in~\cite{Suyama:2005mx,Braga:2024pik,Bruyere:2026gnt,Bruyere:2026xzb, Braga:2026bys}; in particular, Ref.~\cite{Bruyere:2026gnt} finds that for scalar waves in vacuum the first corrections beyond geometric optics appear at second order in the mass. For a black hole, for instance, the time delay integrated along the line of sight does not know that the horizon only admits ingoing waves, and  describing black hole lenses in this approach is a demanding task -- see e.g. \cite{Virbhadra:1999nm,Bozza:2001xd,Motohashi:2021zyv,Pijnenburg:2024btj,Chan:2025wgz,Saketh:2025cwf,Baraldo:1999ny,Nambu:2015aea,Nambu:2019sqn,Oancea:2022szu,Kubota:2024zkv,Li:2025gou}. Moreover, the two-dimensional oscillatory integral can be computed in closed form only in a few cases, such as the point lens, and is otherwise numerically challenging \cite{Feldbrugge:2019fjs,Villarrubia-Rojo:2024xcj}.

The second tool is the Born series around the free wave \cite{Peters:1974gj,Macquart:2004sh,Takahashi:2005sxa,Mizuno:2022xxp}, which solves the three-dimensional wave equation perturbatively in the lens potential (see Refs.~\cite{Yarimoto:2024uew,Tanaka:2026bql} for the Born expansion of the thin-lens diffraction integral). It retains the full geometry and every scattering angle, and gives a simple response to each term of the potential. However, it is not a useful expansion for corrections to a point lens. The Newtonian field acts over the whole path from the source to the observer, and, even when it is weak locally, its accumulated phase can be appreciable and spoils perturbation theory. In the Born series this phase appears as a logarithm of the distances of source and observer, which accompanies every Newtonian insertion, as for the Coulomb potential in quantum field theory~\cite{Dalitz:1951ah,Weinberg:1965nx}. As a consequence, higher Born terms are comparable in size  to the lower ones. As we show in Sec.~\ref{sec:born}, the second-order term with one Newtonian and one inverse-square insertion exceeds the first-order term of the inverse-cube correction --  of the same order in a mass expansion -- by many orders of magnitude. Related issues with the Born expansion in this setting were studied in Ref.~\cite{Mizuno:2022xxp}.

In this work we solve this problem by resumming the Newtonian potential contributions. We consider a static, spherically symmetric lens, whose effective potential $\Ueff$ we expand in inverse powers of the radius,
\be \label{eq_cp_expansion_intro}
	4\Ueff(r)=\sum_{p}\frac{c_p}{r^p},
\ee
and we call each term a radial channel. The Newtonian channel, $p=1$, is fixed by the mass. Faster-decaying channels describe relativistic corrections to the exterior geometry, as for a black hole in vacuum, or departures from the point-mass profile over a range of radii, as for black holes surrounded by an environment (see e.g. \cite{Barausse:2014tra,Ferrer:2017xwm,Cardoso:2021wlq,Ghosh:2026fbr}). Carrillo-Gonz\'alez, Heavens and Whiteway \cite{Carrillo-Gonzalez:2026} studied exterior relativistic corrections in a three-dimensional Green-function formulation, gave the exact finite-distance Newtonian solution \cite{Hostler:1963zz}, and suggested treating the remaining potential as a perturbation around it. We carry out this program explicitly: we build the Born series around the exact solution for a point lens, so that the Newtonian channel is treated non-perturbatively, and only the channels with $p>1$ are treated as perturbations, as in the distorted-wave Born approximation \cite{Gell-Mann:1953dcn}. The exact Coulomb wave~\cite{Gordon:1928} was used for the wave optics of a gravitational lens in \cite{Peters:1974gj,Herlt:1976,Deguchi:1986zz,Turyshev:2017pgm}, and for the propagation of gravitational waves from slow-motion sources in a Coulomb-type potential in~\cite{Asada:1997zu}.

This reorganization allows us to compute the response of the waveform to each radial channel, which we call the dressed radial kernels, while keeping the three-dimensional geometry. For distant source and observer, we represent a radial power as a superposition of screened potentials, so that the three-dimensional propagation integral is evaluated once, and the radial power enters only through the weight of a single integral over the inverse screening length, whose kernel follows from Nordsieck's integral~\cite{Nordsieck:1954zz}. For convergent powers we obtain closed forms in the Fraunhofer approximation, retaining the scattering angle and the Newtonian coupling without expanding either. The small-angle result shows directly how the long-range interaction dresses the free kernel: for the inverse-square channel $p=2$, it multiplies the leading small-angle kernel by a phase, whose expansion recovers the logarithm of the free-wave Born series. The Fraunhofer approximation, however, neglects the curvature of the wavefronts and fails near the forward direction. Keeping the curvature in a Fresnel calculation shows that the scattered wave is not the whole response: at small angles a second region contributes around the unlensed line of sight, and gives the response of the direct wave. For $p=2$, the direct and scattered corrections have equal magnitudes far from alignment. When the image picture applies, the secondary image passes closer to the lens and acquires a larger phase shift, which compensates for its smaller amplitude.

After the resummation, the large infrared logarithms of the distances are gone: they have become a phase. The channels that decay as $1/r^3$ or faster, however, still diverge at short distances. This divergence has a different, ultraviolet origin: it signals that the radial expansion of the potential has been extended to the centre of the lens, where it does not apply. We renormalize it by separating the contributions determined by the exterior potential from local terms, which must be fixed by the inner structure of the lens. This separation also underlies effective-field-theory treatments of black-hole scattering \cite{Goldberger:2004jt,Caron-Huot:2025tlq},  including absorption~\cite{Goldberger:2005cd,Ivanov:2024sds}.

Finally, we apply our formalism to a Schwarzschild black hole. There the inner structure is the horizon, and we fix the local terms by directly matching the low-frequency exterior solution to a purely ingoing solution at its position --  a boundary condition that the thin-lens diffraction integral cannot impose. At low frequency, the correction quadratic in the mass receives contributions from the inverse-square $1/r^2$ optical deformation, from a more steeply decaying term $1/r^4$ generated by the frequency dependence of the potential, and from the horizon. The matching reproduces the horizon-area limit of the scalar absorption cross section \cite{Das:1996we,Starobinskii:1973vzb,Page:1976df,Sanchez:1977si} -- see \cite{Unruh:1976fm} for the low-frequency matching formalism. The finite-distance comparison shows why the exterior and horizon terms must be evaluated at the same accuracy. For exact alignment and equal source and observer distances, the leading Fresnel inverse-square response gives the dominant correction, while at the next order in the large-distance expansion the finite-distance optical term cancels fifteen sixteenths of the horizon term in the field at the observer. Throughout, we use a minimally coupled massless scalar as a proxy for the GW amplitude, which isolates the propagation and matching problem; tensor waves can carry polarization information beyond a scalar amplification \cite{Cusin:2019rmt,Andersson:2020gsj,Dalang:2021qhu,Oancea:2022szu,Pijnenburg:2024btj,Kubota:2023dlz,Kubota:2024zkv,Li:2026dzr}.

\smallskip
Our work is organized as follows. Section~\ref{sec:waveBorn} introduces the wave equation, the Born series around a generic reference solution, and the propagation scales. Section~\ref{sec:born} analyzes the Born series around the free wave and motivates the Newtonian resummation. Section~\ref{sec:beyondborn} derives the dressed kernels and compares the Fraunhofer and Fresnel responses. Section~\ref{sec:v:f} matches the Schwarzschild solution and compares the complete quadratic correction with the leading optical benchmark. Section~\ref{sec:conclusions} discusses the implications and scope of these results. To help the reader, each section ends with a summary of its results.

\section{Wave equation and the Born series around a reference solution}
\label{sec:waveBorn}

A weak gravitational field need not produce a small change in a propagating wave: the accumulated phase depends on both the lens potential and the distances travelled. A controlled approximation therefore requires us to specify which part of the lensing is already included in the reference wave, and which additional effects are treated perturbatively. This section establishes that framework. We first recast the scalar wave equation as a scattering problem and distinguish the radial expansion of the potential from the Born expansion in successive interactions. We then formulate the latter around a general reference solution, allowing a direct comparison between free and Newtonian reference waves in the following sections. Finally, we identify the scales governing diffraction, gravitational focusing, and finite-distance effects, and introduce the eikonal description that will help interpret the subsequent calculations.

\subsection{Wave equation and radial channels}
\label{sec:waveeq}

As the GW propagates, it interacts with the lens potential.  We  formulate propagation in the lens background and distinguish two expansions: the expansion of the lens potential in radial channels, and the Born expansion in the number of interactions with that potential. Throughout, we set the light speed $c=1$ and use a massless scalar field as a proxy for the propagating GW amplitude. Such scalar description captures the lensing effects considered here --  while polarization transport, helicity conversion and the full spin-2 perturbation equations are outside the present setup.

We consider a static, spherically symmetric, asymptotically flat metric in isotropic coordinates,
\be \label{eq:iso}
	\dd s^2=-A(r)\,\dd t^2+D(r)\bigl(\dd r^2+r^2\dd\Omega^2\bigr) \,,
\ee
such that
\be
	A,D\longrightarrow1\quad \mbox{for} \qquad r\longrightarrow\infty \,.
\ee
Away from the source, the scalar obeys $\Box\phi=0$, where $\Box$ is the covariant operator on the curved background. Because of the invariance under time translation, each frequency mode of the scalar field evolves independently. For a mode of positive frequency $\omega$, write $\phi(t,\xx)=\eu^{-\ii\omega t}[A(r)D(r)]^{-1/4}\chi(\xx)$. This redefinition removes the metric-dependent first-derivative term~\cite{Carrillo-Gonzalez:2026} and gives
\be \label{eq:helmholtz}
	(\nabla^2+\omega^2)\chi=4\omega^2\Ueff(r)\chi,
\ee
where $\nabla^2$ is the flat three-dimensional Laplacian and
\begin{align}
	4\Ueff(r)&=1-\frac{D}{A} +\frac{1}{\omega^2}\left(\frac{h'}{2}+\frac{h^2}{4}+\frac{h}{r}\right), \nonumber \\
	\label{eq:Umaster} h(r)&\equiv\frac12(\ln AD)', \qquad {}'\equiv\frac{\dd}{\dd r}.
\end{align}
This reduction is exact for the  scalar equation \eqref{eq:helmholtz} wherever the metric is regular in these coordinates. A weak-field expansion of $\Ueff$ is a further approximation, to be  used only in the exterior region of the lens,  where it is justified. At leading weak-field order, $\Ueff$ reduces to the Newtonian potential. The term $D/A$ describes the change in propagation speed in these coordinates; the derivative terms also account for the spatial variation of the field normalization. 

When the effective potential admits a radial expansion, we write
\be \label{eq:cp_expansion}
	4\Ueff(r)=\sum_{p}\frac{c_p}{r^p}.
\ee
The index $p$ specifies a radial power, and we call each term a \emph{radial channel}. For instance, $p=1$ is the Newtonian channel, $4U_{\rm newt}=c_1/r$, which scales as $1/r$. Notice that $p$ need not be an integer. For example,  the Schwarzschild coefficients through $r^{-5}$ are
\begin{align}
	c_1^{\rm Sch}&=-4GM, &c_2^{\rm Sch}&=-\frac{15}{2}(GM)^2, \nonumber \\
	\label{eq:cpSchw} c_3^{\rm Sch}&=-9(GM)^3, &c_4^{\rm Sch}&=-\frac{129}{16}(GM)^4-\frac{(GM)^2}{4\omega^2}, \\
	c_5^{\rm Sch}&=-6(GM)^5. \nonumber
\end{align}
The first two coefficients agree with the potential obtained from the one-loop amplitude for the scattering of a massless particle off a massive one~\cite{Bjerrum-Bohr:2014zsa,Correia:2024jgr}. In this specific black-hole example the frequency-independent terms scale as $(GM)^p$. The derivative terms in Eq.~\eqref{eq:Umaster} have a different counting with $G$: the $\omega^{-2}$ contribution to $c_4$, for instance, is of order $G^2$. A calculation complete at a specified post-Minkowskian order must retain all contributions at that order, including repeated insertions of lower-order potentials in a Born expansion.

The asymptotic Newtonian term with $c_1=-4GM$ defines the reference lens mass. For an isolated, finite-mass environment, it is natural to absorb its contribution to the total mass into this coefficient. The remaining potential terms then decay faster than $1/r$ at large radius. For black holes surrounded by an environment, such as a dark matter distribution \cite{Sadeghian:2013laa,Eda:2013gg,Kavanagh:2020cfn, Cole:2022yzw}, the coefficients $c_p$ can depend on additional masses and length scales of the environment, and the radial powers need not be integers over an intermediate range of radii~\cite{Ghosh:2026fbr}.

\subsection{Lippmann--Schwinger equation around a reference solution}
\label{sec:LS}

\begin{figure}[t]
\centering
\includegraphics[width=8.23cm]{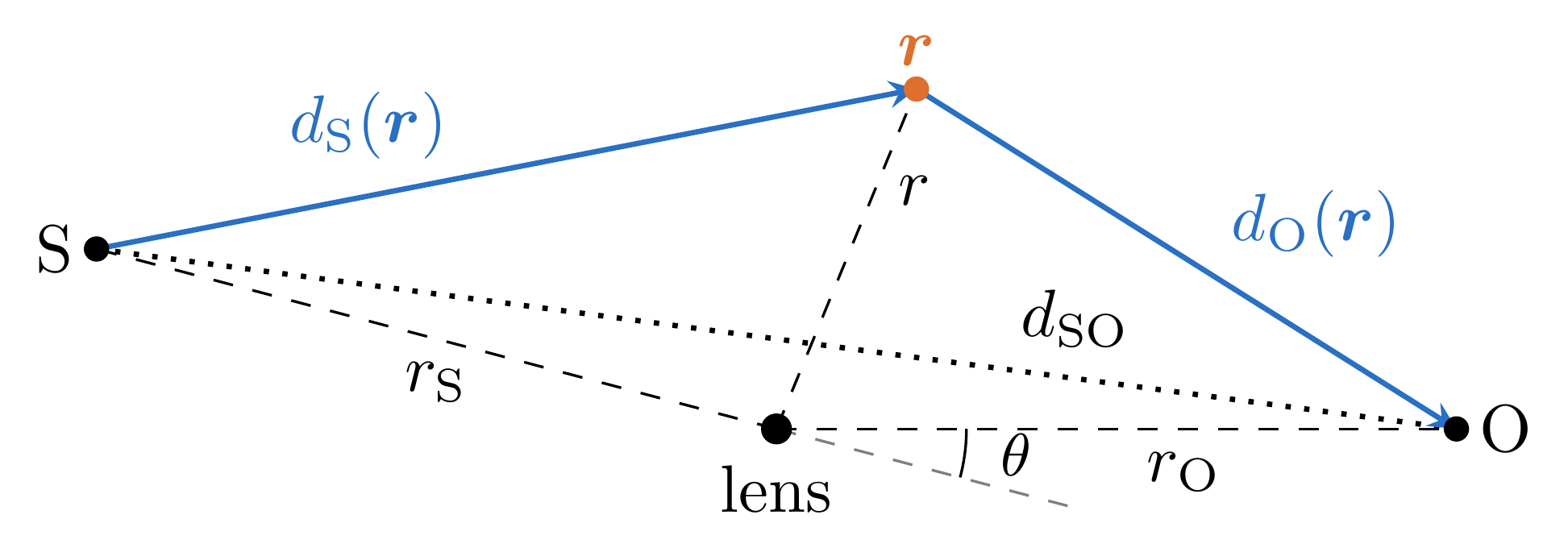}
\caption{Geometry of the problem. The lens is at the origin, the observer O is aligned with it, and the source S has a transverse offset. The point $\rr$ is an interaction point, at distance $r$ from the lens. The solid lines are the free propagation from the source to $\rr$ and from $\rr$ to the observer, and the dotted line is the unlensed propagation from source to observer. The scattering angle $\theta$ is the angle between the incident direction, from the source to the lens, and the outgoing direction, from the lens to the observer. The figure is not to scale: in lensing configurations $r_{\rm S},r_{\rm O}\gg r$.}
\label{fig:geometry}
\end{figure}

We use a reference system centred on the lens, as shown in Fig.~\ref{fig:geometry}. The source and the observer are at positions $\rr_{\rm S}$ and $\rr_{\rm O}$, at distances $r_{\rm S}\equiv|\rr_{\rm S}|$ and $r_{\rm O}\equiv|\rr_{\rm O}|$ from the lens, and $\bn_{\rm S}\equiv\rr_{\rm S}/r_{\rm S}$ and $\bn_{\rm O}\equiv\rr_{\rm O}/r_{\rm O}$ are the corresponding unit vectors. The wave interacts with the lens potential at a point $\rr$, at distance $r$ from the lens. The distances that enter the propagation of the wave are the distance between source and observer, and the distances of the interaction point from the source and from the observer,
\begin{align}
	d_{\rm SO}&\equiv|\rr_{\rm O}-\rr_{\rm S}|, \nonumber \\
	\label{eq:distances} d_{\rm S}(\rr)&\equiv|\rr-\rr_{\rm S}|,\qquad d_{\rm O}(\rr)\equiv|\rr_{\rm O}-\rr|.
\end{align}

We define the amplification factor $F$ of the lensed wave at the observer as
\be \label{eq:Fdef}
	F(\omega)=\frac{\chi(\rr_{\rm O})}{\chi_{\rm free}(\rr_{\rm O})}\,,
\ee
which depends on the GW frequency. In this expression $\chi_{\rm free}$ is a free wave which, for a unit point source, is given by
\be \label{eq:chi0}
	\chi_{\rm free}(\rr)=G_{\rm free}(\rr,\rr_{\rm S}).
\ee
with $G_{\rm free}$ the free retarded Green's function
\be \label{eq:G0}
	G_{\rm free}(\rr,\rr')=-\frac{\eu^{\ii\omega|\rr-\rr'|}}{4\pi|\rr-\rr'|},
\ee
satisfying
\be
	(\nabla^2+\omega^2)G_{\rm free}(\rr,\rr')=\delta^{(3)}(\rr-\rr').
\ee
As mentioned above, it is convenient to study lensing as a modification of a reference wave. However, we always define the amplification factor with respect to the free wave, so that it contains all the lensing information: it measures how much the wave is modified with respect to the case without the lens.

\bigskip
To solve the full wave equation~\eqref{eq:helmholtz}, we build a Born series around the solution of a simpler problem, the reference solution. To do so, we split the effective potential into a part $U_{\rm ref}$, which is retained exactly, and a remainder $\delta U_{\rm ref}$, which is treated perturbatively around $U_{\rm ref}$
\be \label{eq:Urefsplit}
	\Ueff(r)=U_{\rm ref}(r)+\delta U_{\rm ref}(r).
\ee
The reference Green function $G_{\rm ref}$ and the reference wave $\chi_{\rm ref}$ solve the wave equation,
\begin{align}
	\left(\nabla^2+\omega^2-4\omega^2U_{\rm ref}\right)G_{\rm ref}(\rr,\rr') &=\delta^{(3)}(\rr-\rr'), \nonumber \\
	\label{eq:Gref} \chi_{\rm ref}(\rr)&=G_{\rm ref}(\rr,\rr_{\rm S}),
\end{align}
by definition, with outgoing boundary conditions. Subtracting the reference equation from the full wave equation~\eqref{eq:helmholtz} and applying $G_{\rm ref}$ gives the Lippmann--Schwinger equation~\cite{Lippmann:1950zz}
\begin{align}
	\chi(\rr)={}&\chi_{\rm ref}(\rr) \nonumber \\
	\label{eq:LS} &+4\omega^2\int\dd^3\rr'\, G_{\rm ref}(\rr,\rr')\,\delta U_{\rm ref}(r')\,\chi(\rr').
\end{align}
Indeed, it can be checked that $\chi(\rr)$ defined this way solves exactly Eq.~\eqref{eq:helmholtz}. While this equation is exact, it is not in closed form. Iterating it generates the Born series around the reference solution,
\be \label{eq:Fseries}
	F=F^{(0)}_{\rm ref}+F^{(1)}_{\rm ref}+F^{(2)}_{\rm ref}+\cdots,
\ee
where the $n$th term contains $n$ insertions of $\delta U_{\rm ref}$. Inserting Eq.~\eqref{eq:LS} into the definition~\eqref{eq:Fdef}, the first three terms are
\begin{align}
	\label{eq:Fref0} F^{(0)}_{\rm ref}={}&\frac{\chi_{\rm ref}(\rr_{\rm O})}{\chi_{\rm free}(\rr_{\rm O})}, \\
	F^{(1)}_{\rm ref}={}&\frac{4\omega^2}{\chi_{\rm free}(\rr_{\rm O})} \int\dd^3\rr\,G_{\rm ref}(\rr_{\rm O},\rr) \nonumber \\
	\label{eq:Fref1} &\times\delta U_{\rm ref}(r)\,G_{\rm ref}(\rr,\rr_{\rm S}), \\
	F^{(2)}_{\rm ref}={}&\frac{16\omega^4}{\chi_{\rm free}(\rr_{\rm O})} \int\dd^3\rr_1\,\dd^3\rr_2\,G_{\rm ref}(\rr_{\rm O},\rr_1)\, \delta U_{\rm ref}(r_1) \nonumber \\
	\label{eq:Fref2} &\times G_{\rm ref}(\rr_1,\rr_2)\,\delta U_{\rm ref}(r_2)\, G_{\rm ref}(\rr_2,\rr_{\rm S}).
\end{align}
In $F^{(1)}_{\rm ref}$ the wave propagates with the reference Green function from the source to a single interaction point $\rr$, and from there to the observer. In $F^{(2)}_{\rm ref}$ it interacts at two points, first at $\rr_2$ and then at $\rr_1$, which are connected by a further reference Green function~\cite{Feynman:100771}. In this way we study lensing as a small perturbation around a known solution, instead of a small perturbation around the vacuum solution, and report the result in terms of $F$, i.e.\ of how much the wave differs from the unlensed one.

When the remainder $\delta U_{\rm ref}$ in Eq.~\eqref{eq:Urefsplit} admits a radial expansion as in Eq.~\eqref{eq:cp_expansion}, we write $4\,\delta U_{\rm ref}=\sum_p c_p/r^p$, where the sum runs over the channels that are not already contained in $U_{\rm ref}$. Each Born term then splits into channel contributions,
\be \label{eq:Frefsum}
	F^{(1)}_{\rm ref}=\sum_pF^{(1)}_{{\rm ref},p},\qquad F^{(2)}_{\rm ref}=\sum_{p,q}F^{(2)}_{{\rm ref},pq},
\ee
with
\begin{align}
	\label{eq:Fref1p} F^{(1)}_{{\rm ref},p}={}&\frac{\omega^2c_p}{\chi_{\rm free}(\rr_{\rm O})} \int\dd^3\rr\,G_{\rm ref}(\rr_{\rm O},\rr)\frac{1}{r^p}\,G_{\rm ref}(\rr,\rr_{\rm S}) \\
	F^{(2)}_{{\rm ref},pq}={}&\frac{\omega^4c_pc_q}{\chi_{\rm free}(\rr_{\rm O})} \int\dd^3\rr_1\,\dd^3\rr_2\,G_{\rm ref}(\rr_{\rm O},\rr_1)\, \frac{1}{r_1^p} \times \nonumber \\
	\label{eq:Fref2pq} &\times G_{\rm ref}(\rr_1,\rr_2)\,\frac{1}{r_2^q}\, G_{\rm ref}(\rr_2,\rr_{\rm S}).
\end{align}
The factors $4$ and $16$ of Eqs.~\eqref{eq:Fref1} and~\eqref{eq:Fref2} are absorbed by the factor $1/4$ that relates $\delta U_{\rm ref}$ to the coefficients $c_p$. In $F^{(2)}_{{\rm ref},pq}$ the wave interacts first with channel $q$ and then with channel $p$. For $p\ne q$, the mixed second-order term therefore contains both $F^{(2)}_{{\rm ref},pq}$ and $F^{(2)}_{{\rm ref},qp}$; for $p=q$, the ordered pair occurs only once.

In this paper, we use two choices of reference solution. In Section~\ref{sec:born} the reference is the free wave, $U_{\rm ref}=0$, so that every interaction with the lens, including the Newtonian one, is treated perturbatively. We will learn that this simple, intuitive prescription runs into problems. In Section~\ref{sec:beyondborn} the reference is the point lens, $U_{\rm ref}=U_{\rm newt}$, and only the remaining channels, $\delta U_{\rm ref}=\Ueff-U_{\rm newt}$, are treated perturbatively.

Finally, we point out that the individual coefficients of $\Ueff$ are not observables: a different choice of radial coordinate, or of the normalization of the field, changes the terms beyond the Newtonian one and moves contributions between the frequency-independent and derivative parts of the potential, while the scattering amplitude is unchanged~\cite{Correia:2024jgr}. The amplification factor is likewise independent of these choices, once the positions of source and observer are specified by the same physical distances, since $(AD)^{1/4}\to1$ far from the lens.

\subsection{Geometry and dimensionless quantities}
\label{sec:DimensionlessVariables}

The validity of a propagation approximation depends not only on the strength of the lens potential, but also on the source–lens–observer geometry and the wavelength. We therefore introduce the geometric quantities and dimensionless ratios that distinguish gravitational focusing from diffraction and identify when wavefront curvature and finite-distance effects must be retained.

The two endpoint distances combine into the effective distance
\be \label{eq:reff}
	r_{\rm eff}\equiv\frac{r_{\rm O}r_{\rm S}}{r_{\rm O}+r_{\rm S}},
\ee
and the position of the lens between source and observer is described by the location parameter
\be \label{eq:adef}
	a\equiv\frac{r_{\rm O}}{r_{\rm S}+r_{\rm O}}\,,
\ee
with $0<a<1$, so that $1/r_{\rm S}=a/r_{\rm eff}$ and $1/r_{\rm O}=(1-a)/r_{\rm eff}$.

Far from the lens, the wave arriving from the source and the wave leaving towards the observer are locally plane waves, with incident and outgoing wavevectors
\be \label{eq:kinkout}
	\kk_i=-\omega\bn_{\rm S} \,, \qquad \kk_f=\omega\bn_{\rm O} \,.
\ee
The momentum transferred to the wave by the lens is their difference, and its modulus is
\begin{align}
	\label{eq:kaskinkout} \kk&=\kk_f-\kk_i=\omega(\bn_{\rm O}+\bn_{\rm S}), \\
	\label{eq:kdef} k&=2\omega\sin(\theta/2)=\sqrt{2\omega^2(1-\mu)},
\end{align}
where $\theta$ is the scattering angle between $\kk_i$ and $\kk_f$, and $\mu\equiv\cos\theta=-\bn_{\rm O}\cdot\bn_{\rm S}$. Forward propagation has $\theta=0$, with source and observer on opposite sides of the lens, while $\theta=\pi$ is exact backscattering. A small-angle expansion requires $\theta\ll1$ (equivalently $k/\omega\ll1$), and replaces $k$ by $\omega\theta$. In this limit we call $\bm e$ the unit vector along the line from the source to the observer, so that $\bn_{\rm S}\simeq-\bm e$ and $\bn_{\rm O}\simeq\bm e$. This angle describes the scattering geometry fixed by the source and observer positions.

Several quantities play an important role in lensing. We first define the transverse vector
\be \label{eq:bdef}
	\bm b\equiv\frac{r_{\rm eff}}{\omega}\,\kk,\qquad b\equiv|\bm b|=\frac{r_{\rm eff}}{\omega} k,
\ee
with $0\leq b\leq2r_{\rm eff}$. In the small-angle limit, $\theta\ll1$, $\kk$ is perpendicular to the line of sight, and $\bm b$ goes from the lens to the unlensed source--observer line: $b$ is then the impact parameter of the unlensed path.

Two characteristic length scales play a central role in wave-optics lensing: the Fresnel scale, which governs the coherence of wave propagation, and the Einstein radius, which characterizes gravitational focusing; see e.g. \cite{Macquart:2004sh,Jow:2022pux,Leung:2023lmq}.

The Fresnel scale comes from the curvature of the free wavefronts, i.e.\ from the distance $|\rr-\rr'|$ in the free Green function~\eqref{eq:G0}. A free path from the source to the observer that crosses the plane of the lens at a transverse distance $\rho$ from the line of sight is longer than the direct path by $\rho^2/(2r_{\rm eff})$, and therefore acquires the additional phase $\omega\rho^2/(2r_{\rm eff})$. The Fresnel scale is the transverse distance at which this phase becomes of order one
\be \label{eq:rF}
	r_{\rm F}\equiv\sqrt{\frac{r_{\rm eff}}{\omega}}\,
\ee
and it represents the scale below which diffraction is important.

\begin{figure}[t]
\centering
\includegraphics[width=8.23cm]{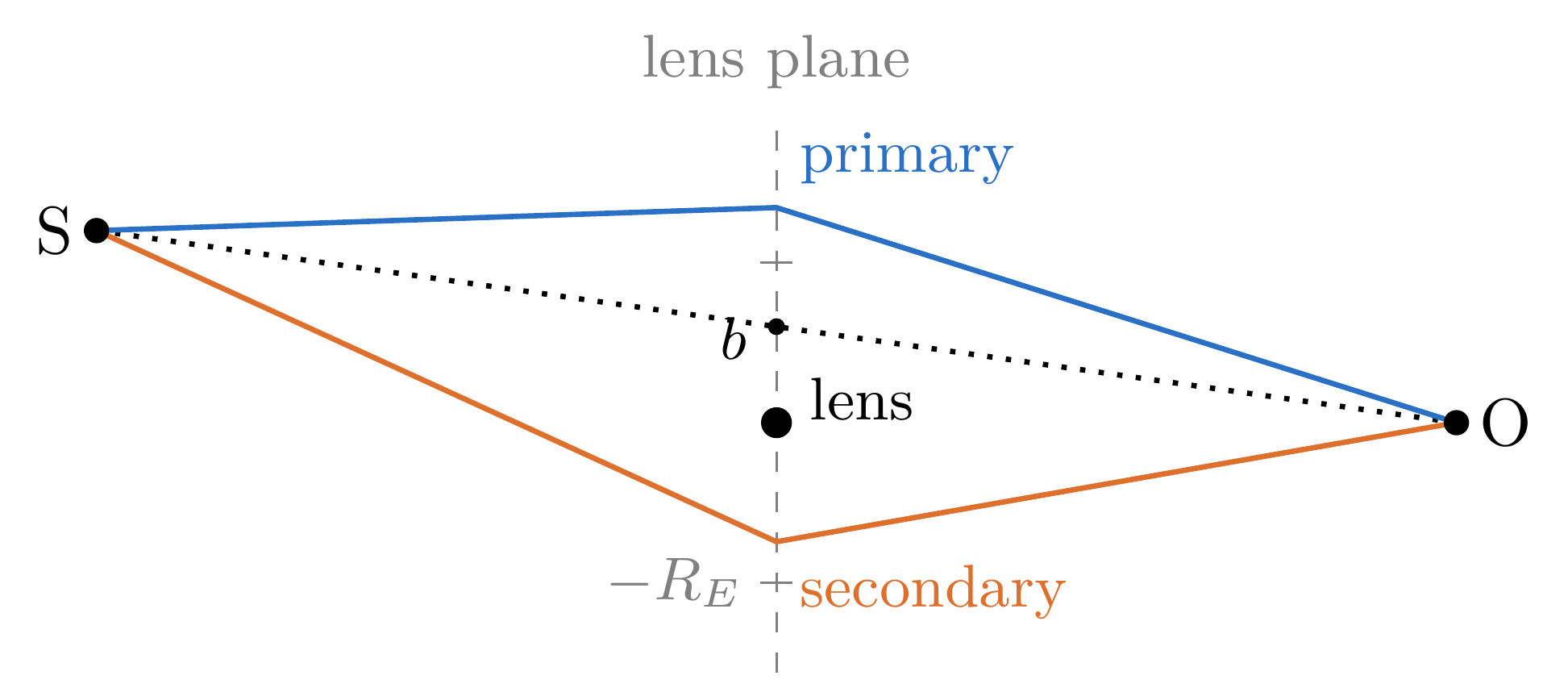}
\caption{Point lens in the geometric-optics regime. The observer O is aligned with the lens, and the source S has a transverse offset: the unlensed path (dotted) crosses the plane of the lens at the transverse offset $b=0.6\,R_E$. The wave reaches the observer along two paths, which cross the plane of the lens at the image positions of Eq.~\eqref{eq:v:d:image-positions}, $\zeta_+\simeq1.34\,R_E$ on the side of the source (primary image) and $\zeta_-\simeq-0.74\,R_E$ on the opposite side (secondary image), closer to the lens. The figure is not to scale.}
\label{fig:images}
\end{figure}

The gravitational lens introduces a second characteristic length scale, the Einstein radius. For a point-mass lens of reference mass $M$, it is given by
\be
	R_E\equiv\sqrt{4GM r_{\rm eff}}\,.
\ee
This scale characterizes the typical transverse positions of strongly lensed images, see Fig.~\ref{fig:images}.  The relative importance of gravitational focusing and diffraction is therefore controlled by the ratio of these two scales. This motivates the introduction of the dimensionless frequency parameter
\be
	w\equiv\frac{R_E^2}{r_{\rm F}^2}=4GM\omega\,.
\ee
When $w\ll1$, the Einstein radius lies within a broad diffraction region, and wave effects are important. Conversely, for $w\gg1$, gravitational focusing occurs on scales much larger than the Fresnel scale, allowing a geometric-optics description when the relevant image phases are sufficiently separated.

We also define the dimensionless transverse offset $y$ in units of Einstein radius, and a characteristic angular scale $\alpha_E$ as
\begin{align}
	\label{eq:dimlessInputs} y&\equiv\frac{b}{R_E}\,, \qquad \alpha_E \equiv\sqrt{\frac{4GM}{r_{\rm eff}}}\,.
\end{align}
For the analysis that follows, it is useful to introduce three additional combinations:
\begin{align}
	s&\equiv wy^2=\frac{\omega b^2}{r_{\rm eff}}=\frac{b^2}{r_{\rm F}^2}\,, \nonumber \\
	\label{eq:sepsx} \varepsilon&\equiv\frac{1}{\omega r_{\rm eff}}=\frac{\alpha_E^2}{w}\,, \\
	x&\equiv GMk=\frac{wy\alpha_E}{4}\,. \nonumber
\end{align}
Each of them characterizes a different physical effect: $s$ measures the transverse offset relative to the Fresnel scale and is therefore associated with diffraction effects; $\varepsilon$ measures the wavelength relative to the effective propagation distance and characterizes finite-distance corrections; finally, $x$ compares the gravitational radius with the characteristic Fourier scale $1/k$, controlling the weak-field expansion of the gravitational potential by testing whether the nonlocal Fourier contribution probes radii much larger than $GM$.

The radial expansion~\eqref{eq:cp_expansion} is an expansion in small $x$. This does not imply $w\ll1$: the weak-field radial expansion can remain valid even when gravitational lensing exhibits substantial wave-optics effects.

Useful identities that these parameters satisfy are
\be \label{eq:dimlessIdentities}
	\frac{k}{\omega}=y\alpha_E\,,\qquad (kr_{\rm eff})^2=\frac{s}{\varepsilon}\,,\qquad 1-\mu=\frac{s\varepsilon}{2}\,,
\ee
where we  use the definitions in Eqs.~\eqref{eq:reff} and~\eqref{eq:bdef}. In particular, the first identity shows that a small scattering angle does not require small $y$, since $\alpha_E$ can be very small.

One approximation which we will assume frequently is the large-distance one, $\varepsilon\ll1$, in which the source and the observer are located far away from the lens.

\subsection{Solution in the eikonal limit}
\label{sec:eikonal}

The eikonal limit provides a physical interpretation of the Born series by expressing successive interactions with the lens as an accumulated phase along each propagation path. We will use this description to estimate the relative sizes of the Born terms, motivate the resummation of the Newtonian interaction, and interpret the dressed response in terms of the images of the reference lens when geometric optics applies.

We work at small angle, $\theta\ll1$, with the unit vector $\bm e$ along the line of sight introduced in Sec.~\ref{sec:DimensionlessVariables}. We decompose positions as $\rr=\ell\,\bm e+\boldsymbol\zeta$, with $\ell=\bm e\cdot\rr$ the coordinate along the line of sight and $\boldsymbol\zeta$ perpendicular to $\bm e$, so that $r^2=\ell^2+\zeta^2$. The momentum transfer $\kk$ is then perpendicular to $\bm e$, $\kk\cdot\rr=\kk\cdot\boldsymbol\zeta$, and $\bm b=r_{\rm eff}\kk/\omega$ is the vector of modulus $b$ of Section~\ref{sec:DimensionlessVariables}, which goes from the lens to the unlensed line of sight.

When the wave passes many wavelengths away from the lens, $\omega b\gg1$, so that $\omega\zeta\gg1$ for the paths that contribute, and source and observer are far from it, the Born series can be resummed: the interactions with the lens are connected by straight-line propagation, and their sum gives the exponential of the phase accumulated along each path~\cite{Nakamura:1999uwi,CarrilloGonzalez:2025gqm,Takahashi:2003ix,Schneider:1992bmb,Bruyere:2026xzb}. The wave then reaches the observer along all the paths that cross the plane of the lens, $\ell=0$, at a transverse position $\boldsymbol\zeta$, each with a phase equal to $\omega$ times its time delay, and the amplification factor is given by the diffraction integral
\be \label{eq:overview-eikonal}
	F\simeq\frac{\omega}{2\pi\ii r_{\rm eff}}\int\dd^2\zeta\,\exp\left\{\ii\left[\frac{\omega|\boldsymbol\zeta-\bm b|^2}{2r_{\rm eff}}+\psi(\zeta)\right]\right\},
\ee
where the first term is $\omega$ times the geometric delay of the path, and
\be \label{eq:eikonalPsi}
	\psi(\zeta)=-2\omega\int\dd\ell\,\Ueff\!\left(\sqrt{\zeta^2+\ell^2}\right)
\ee
is $\omega$ times the gravitational delay along a straight line at transverse distance $\zeta$ from the lens. The prefactor normalizes $F$ to one in the absence of the lens, $\psi=0$. With the radial channels of Eq.~\eqref{eq:cp_expansion}, $\psi=\sum_p\psi_p$. Selecting $p>1$ for the radial channel powers,  the integral over $\ell$ converges at large distances and, with $c_p=\gamma_p(GM)^p$ and $\omega GM=w/4$, gives
\be \label{eq:v:d:delta-phi-p}
	\psi_p(\zeta)=-\frac{\sqrt\pi}{8}\,\frac{\Gamma\left(\frac{p-1}{2}\right)}{\Gamma\left(\frac p2\right)}\,\gamma_p\,w\left(\frac{GM}{\zeta}\right)^{p-1}.
\ee
This is the product of two factors: $w$, which measures how strongly the phase of the wave responds to a time delay, and $(GM/\zeta)^{p-1}$, which measures how strongly a path passing at distance $\zeta$ from the lens feels the channel. For the Newtonian channel, $p=1$, the integral diverges logarithmically at large distances. Integrating between the source and the observer, we obtain the Shapiro phase of a point mass,
\be \label{eq:eikonalShapiro}
	\psi_1(\zeta)\simeq\frac w2\ln\frac{4r_{\rm S}r_{\rm O}}{\zeta^2},
\ee
which depends on the path only through $-w\ln\zeta$,  growing with the distances of source and observer.

Expanding the exponential of Eq.~\eqref{eq:overview-eikonal} in powers of $\psi$ gives back the Born series in the eikonal limit, in which each interaction with channel $p$ contributes a factor $\ii\,\psi_p(\zeta)$. Since it resums the Born series, the diffraction integral does not depend on the choice of the reference wave of Section~\ref{sec:LS}. Expanding the whole exponential gives the Born series around the free wave, while keeping some channels in the exponential and expanding the others gives the Born series around the corresponding reference wave, as we do for the point lens in Section~\ref{sec:beyondborn}.

\subsection{Summary}
\label{sec:waveBorn:summary}

In this section we have set up the perturbation theory used in the rest of the paper. We wrote the wave equation of a massless scalar on a static, spherically symmetric background as a time-independent Schr\"odinger equation, Eq.~\eqref{eq:helmholtz}, with the effective potential of Eq.~\eqref{eq:Umaster}. Part of this potential describes the change of the propagation speed in the lens metric, and part comes from the derivatives of the metric through the field redefinition, as for the $\omega^{-2}$ term of the Schwarzschild coefficient $c_4$, Eq.~\eqref{eq:cpSchw}. We expanded the potential in radial channels, Eq.~\eqref{eq:cp_expansion}: the Newtonian channel is fixed by the mass of the lens, while the faster-decaying channels describe its structure.

To compute the lensed wave, we split   the potential into a reference part, which is solved exactly, and a remainder, Eq.~\eqref{eq:Urefsplit}. The Lippmann--Schwinger equation~\eqref{eq:LS} then generates a Born series, in which each term counts the insertions of the remainder and the wave propagates with the reference Green function between successive insertions, Eqs.~\eqref{eq:Fref0}--\eqref{eq:Fref2}. Each Born term splits into the contributions of the radial channels, Eqs.~\eqref{eq:Fref1p} and~\eqref{eq:Fref2pq}. Whatever the reference, the amplification factor is defined with respect to the free wave, Eq.~\eqref{eq:Fdef}, so that it contains the whole effect of the lens.

The geometry, Fig.~\ref{fig:geometry}, enters through the momentum transfer $k$, which sets the transverse scale probed by the scattered wave, the Fresnel scale $r_{\rm F}$, which measures the curvature of the free wavefronts, and the Einstein radius $R_E$, which measures the focusing of the point lens, Fig.~\ref{fig:images}. Their ratio $w=R_E^2/r_{\rm F}^2$ distinguishes wave optics from geometric optics, while the combinations of Eq.~\eqref{eq:sepsx} characterize diffraction ($s$), finite-distance effects ($\varepsilon$) and the weak-field radial expansion ($x$), which does not require $w\ll1$.

Finally, in the eikonal limit the Born series is resummed into the diffraction integral~\eqref{eq:overview-eikonal}, as shown in Ref.~\cite{CarrilloGonzalez:2025gqm} (see Refs.~\cite{Wallace:1973iu,Swift:1974pm} for the analogous resummation of the Born series in potential scattering), in which each path is weighted by the phase accumulated along it, Eq.~\eqref{eq:eikonalPsi}. The phase of a channel with $p>1$ does not depend on the distances of source and observer, Eq.~\eqref{eq:v:d:delta-phi-p}, while the Newtonian phase grows logarithmically with them, Eq.~\eqref{eq:eikonalShapiro}. Since it resums the Born series, the diffraction integral does not depend on the choice of the reference wave, and we use it to interpret the Born terms in the following sections.

\section{Born series around the free wave}
\label{sec:born}

The simplest perturbative treatment starts from the free wave and regards the entire lens potential as a correction. A weak potential, however, does not by itself guarantee that successive corrections to the wave are small: propagation over long distances can accumulate an appreciable phase.  How many interactions must be kept to predict the complex amplification to a given accuracy~\cite{Mizuno:2022xxp,Yarimoto:2024uew}?  We  examine how the radial structure of the potential and repeated interactions together determine the accuracy of this expansion. We first compute the response to a single insertion of each radial channel in the Fraunhofer limit, separating contributions fixed by the exterior potential from local terms requiring information about the inner lens. We then use the eikonal picture to estimate repeated interactions and show why neither radial falloff nor the number of Born insertions alone determines their relative importance. In particular, the long-range Newtonian field produces logarithmically enhanced corrections, whose origin and coefficient we establish through a direct second-order calculation. These results motivate incorporating the Newtonian interaction into the reference wave and clarify how the remaining perturbative terms should be ordered.

\subsection{The free wave as reference solution}
\label{sec:freeref}

We begin by choosing the free wave as the reference solution, so that every interaction with the lens is treated perturbatively. Writing the first two Born terms explicitly provides the starting point for comparing a single interaction with repeated scattering and assessing whether this expansion is controlled.

For the free wave, the reference potential vanishes and the whole effective potential is treated perturbatively:
\be
	U_{\rm ref}=0,\qquad \delta U_{\rm ref}=\Ueff,
\ee
where $\Ueff$ is given in Eq.~\eqref{eq:cp_expansion}. Then the reference Green function and wave are the free ones, $G_{\rm ref}=G_{\rm free}$ and $\chi_{\rm ref}=\chi_{\rm free}$, given in Eqs.~\eqref{eq:G0} and~\eqref{eq:chi0}, and Eq.~\eqref{eq:Fref0} gives $F^{(0)}_{\rm ref}=1$. Equation~\eqref{eq:Fseries} becomes the free-wave Born series
\be
	F(\omega) = 1 + F^{(1)}_{\rm free}(\omega) + F^{(2)}_{\rm free}(\omega) + \cdots \,,
\ee
where, for this choice of reference, we replace the label ``ref'' by ``free'' in the terms of Eqs.~\eqref{eq:Fref1}, \eqref{eq:Fref2}, \eqref{eq:Fref1p} and~\eqref{eq:Fref2pq}, and write $F^{(1)}_{{\rm free},p}$ and $F^{(2)}_{{\rm free},pq}$ for their channel contributions. The first and second Born terms are
\begin{align}
	F^{(1)}_{\rm free}={}&\frac{4\omega^2}{G_{\rm free}(\rr_{\rm O},\rr_{\rm S})}\int\dd^3\rr\,G_{\rm free}(\rr_{\rm O},\rr) \nonumber \\
	\label{eq:Fborn} &\times\Ueff(r)\,G_{\rm free}(\rr,\rr_{\rm S}), \\
	F^{(2)}_{\rm free}={}&\frac{16\omega^4}{G_{\rm free}(\rr_{\rm O},\rr_{\rm S})}\int\dd^3\rr_1\,\dd^3\rr_2\,G_{\rm free}(\rr_{\rm O},\rr_1) \nonumber \\
	\label{eq:F2born} &\times\Ueff(r_1)\,G_{\rm free}(\rr_1,\rr_2)\,\Ueff(r_2)\,G_{\rm free}(\rr_2,\rr_{\rm S}),
\end{align}
with the free Green function of Eq.~\eqref{eq:G0}, which depends on the distances of Eq.~\eqref{eq:distances}. Eqs.~\eqref{eq:Fborn} and~\eqref{eq:F2born} retain the full source--integration-point--observer geometry. The Born series is a good approximation to the full wave if each term is smaller than the previous one.

The integration point in Eq.~\eqref{eq:Fborn} is the position of a single interaction, so that the wave propagates from the source to that point and then to the observer, and contributions from all such points interfere. The next Born term contains two interaction points connected by another free Green function~\cite{Feynman:100771}. As for the Born expansion of the Coulomb scattering amplitude, which is not uniform in the forward direction, this second term can be as large as the first one~\cite{Lippstreu:2023vvg,Lippstreu:2026xfw}, as we show in Sec.~\ref{sec:breakdown}.

\subsection{First-order Born approximation in the Fraunhofer limit}
\label{sec:leading}

We first compute the response to a single insertion of each radial channel, establishing the baseline against which repeated interactions will be assessed. For distant source and observer, the Fraunhofer approximation reduces the propagation integral to a Fourier transform at the momentum transferred by the lens, as in the first-order Born approximation of wave-optics lensing~\cite{Takahashi:2005sxa,Choi:2021bkx,Mizuno:2022xxp,Yarimoto:2024uew,Tanaka:2026bql,CarrilloGonzalez:2025gqm}, which relates the response of extended lenses to their projected mass profile~\cite{Choi:2021bkx,Tanaka:2026bql,Wu:2025rvy}, through its form factor in the thin-lens limit~\cite{Choi:2026sjq}. Its evaluation reveals three distinct cases: convergent channels, the long-range Newtonian channel, and steeper channels whose short-distance divergences require matching to the inner lens. We explain how this matching separates the response fixed by the exterior potential from undetermined local contributions, then express the results in dimensionless form to compare the sizes of the different channels.

The effects of the finite distances of source and observer were studied in Ref.~\cite{Carrillo-Gonzalez:2026}, and we refer to that paper for those results. Here we are interested in setting up the result in the simplest way, in order to demonstrate an  issue with the $1/r$ potential. Namely, we consider situations where source, observer and lens are far from  the other. In a region satisfying $r\ll r_{\rm S},r_{\rm O}$, the distances defined in Eq.~\eqref{eq:distances} expand as
\begin{align}
	d_{\rm S}(\rr)&=r_{\rm S}-\bn_{\rm S}\cdot\rr +\frac{r^2-(\bn_{\rm S}\cdot\rr)^2}{2r_{\rm S}} +\mathcal O\left(\frac{r^3}{r_{\rm S}^2}\right), \nonumber \\
	\label{eq:distanceExpansion} d_{\rm O}(\rr)&=r_{\rm O}-\bn_{\rm O}\cdot\rr +\frac{r^2-(\bn_{\rm O}\cdot\rr)^2}{2r_{\rm O}} +\mathcal O\left(\frac{r^3}{r_{\rm O}^2}\right)\,,
\end{align}
which include both the linear terms, responsible for Fraunhofer diffraction, and the quadratic terms, responsible for Fresnel diffraction. The integral of Eq.~\eqref{eq:Fborn} has support where $\Ueff$ is not negligible, and there the distances enter the Green functions both in the phase and in the denominator. If this region has size $\mathcal L$, replacing $d_{\rm S}d_{\rm O}$ by $r_{\rm S}r_{\rm O}$ in the denominator requires $\mathcal L/r_{\rm S},\mathcal L/r_{\rm O}\ll1$, while dropping the quadratic terms of the phase requires, in addition, $\omega\mathcal L^2/r_{\rm eff}\ll1$. For a scale-free contribution the size of the region is set by the momentum transfer, $\mathcal L\sim1/k$, and these conditions become
\be \label{eq:farfieldConditions}
	\frac{1}{kr_{\rm S}}=a\sqrt{\frac{\varepsilon}{s}}\ll1, \quad \frac{1}{kr_{\rm O}}=(1-a)\sqrt{\frac{\varepsilon}{s}}\ll1, \quad \frac{1}{s}\ll1.
\ee
Thus $s\gg1$ describes the Fraunhofer limit\footnote{The denomination Fraunhofer has its usual wave-optics meaning here: across the contributing region the incident and outgoing spherical wavefronts can be replaced by their tangent planes, so their phase difference is linear in position. Retaining the quadratic terms instead describes Fresnel propagation. Neither approximation removes interference from the scattered wave. } of this scattering contribution: the transverse offset must be much larger than the Fresnel scale, and the approximation fails close to alignment, where wavefront curvature must be retained. In this limit we keep only the constant and linear terms in Eq.~\eqref{eq:distanceExpansion}, and approximate the external free Green functions in the integrals of Eqs.~\eqref{eq:Fref1} and~\eqref{eq:Fref2} as
\begin{align}
	G_{\rm free}(\rr_{\rm O},\rr) &\simeq-\frac{\eu^{\ii\omega r_{\rm O}}}{4\pi r_{\rm O}} \eu^{-\ii\kk_f\cdot\rr}, \nonumber \\
	\label{eq:planewaves} G_{\rm free}(\rr,\rr_{\rm S}) &\simeq-\frac{\eu^{\ii\omega r_{\rm S}}}{4\pi r_{\rm S}} \eu^{\ii\kk_i\cdot\rr}.
\end{align}
Under this approximation, the first-order amplification factor of Eq.~\eqref{eq:Fborn} becomes
\begin{align}
	\label{eq:Fleading} F^{(1)}_{\rm free}&\simeq-\frac{\omega^2}{4\pi}\,\mathcal P \sum_pc_p\mathcal J_p(k)\,, \\
	\label{eq:endpointFactor} \mathcal P&\equiv\frac{d_{\rm SO}}{r_{\rm S}r_{\rm O}}\eu^{\ii\omega(r_{\rm S}+r_{\rm O}-d_{\rm SO})}\,,
\end{align}
where the endpoint factor $\mathcal P$ collects the dependence on the source and observer positions, and
\be \label{eq:Jdef}
	\mathcal J_p(k)\equiv\int\dd^3\rr\, \frac{\eu^{-\ii\kk\cdot\rr}}{r^p} =\frac{4\pi}{k}\int_0^\infty\dd r\,r^{1-p}\sin(kr).
\ee
Finite-distance and Fresnel corrections to this Fraunhofer reduction are discussed in Ref.~\cite{Carrillo-Gonzalez:2026}.

We now evaluate the integrals of Eq.~\eqref{eq:Jdef} for a single radial channel. Their evaluation depends on the radial power $p$, and we distinguish three cases, depending on whether the integral converges or requires a prescription at large or small radii.

\subsubsection{Convergent integrals: $1<p<3$.} In this interval for the radial power $p$ the ordinary radial integral converges both at small and large radii, absolutely for $2<p<3$ and conditionally for $1<p\leq2$. It can be evaluated analytically, giving
\be \label{eq:jp}
	\mathcal J_p(k)=4\pi\Gamma(2-p)\sin\left[\frac{\pi(2-p)}{2}\right]k^{p-3},
\ee
with the removable limit understood at $p=2$. Since no other length scale is available, the dependence on $k$ is the power $k^{p-3}$. For $1<p<3$ this power is not analytic in $k^2$: as we discuss below for the singular channels, such contributions come from the whole exterior region, up to radii $r\sim1/k$, and are determined by the radial channel alone. Correspondingly, a power-law tail spoils the effective-range expansion of the phase shifts at low energy~\cite{Frank:1971xx}, and the terms that are not analytic in the momentum transfer are the long-range, classical part of the interaction~\cite{Holstein:2004dn}.

\subsubsection{The Newtonian contribution: $p=1$.}
For $p\leq1$, the ordinary oscillatory integral does not converge at infinity. The Newtonian case $p=1$ is well defined for nonzero $k$ by Abel regulation: inserting $\eu^{-\eta r}$ in Eq.~\eqref{eq:Jdef} and taking $\eta\to0^+$ gives
\be \label{eq:JNewton}
	\mathcal J_1(k)=\frac{4\pi}{k^2}\,.
\ee

Although the Newtonian potential is long ranged, Eq.~\eqref{eq:JNewton} contains no logarithm of the distances, and no infrared divergence for $k\neq0$. At fixed momentum transfer the oscillating factor $\eu^{-\ii\kk\cdot\rr}$ cuts off the $1/r$ tail at $r\sim1/k$, and the result is the power $k^{-2}$ of Rutherford scattering. The large-distance logarithm anticipated in the Introduction appears only at second order in Born expansion, Sec.~\ref{sec:breakdown}, where a Newtonian insertion carries an internal momentum transfer that is integrated down to zero. The divergence of Eq.~\eqref{eq:JNewton} for $k\to0$ is instead a forward divergence: it occurs where the Fraunhofer approximation fails, and it is regulated by the curvature of the wavefronts, see Sec.~\ref{sec:beyondborn}.

\subsubsection{Short-distance divergences: $p\geq3$.}
For $p\geq3$ the integral of Eq.~\eqref{eq:Jdef} diverges at the origin, since its integrand behaves as $r^{2-p}$ at small radii. Physically, this ultraviolet divergence means that the radial expansion~\eqref{eq:cp_expansion}, which holds only at radii much larger than $GM$, has been extended to $r\to0$. We therefore renormalize it, adapting the method of Ref.~\cite{Goldberger:2004jt} to the present context (for singular potentials in quantum mechanics, where a power-law singularity defined as a distribution is fixed only up to local terms $(\nabla^2)^n\delta^{(3)}(\rr)$, see Refs.~\cite{Frank:1971xx,Beane:2000wh}), see also Ref.~\cite{Carrillo-Gonzalez:2026} for a dimensional-regularization treatment of the same divergences, and Ref.~\cite{Tanaka:2026bql} for a finite-part treatment of the divergences from power-law tails at large radius. We split the integral at a radius $r_0$ with
\be \label{eq:r0range}
	GM\ll r_0\ll\frac1k,
\ee
which is possible when $x\ll1$. For $r>r_0$ we use the radial channel, while the region $r<r_0$ must be described by the inner, or ultraviolet, completion of the lens. The Fourier transform of the channel is then the sum of two contributions,
\be \label{eq:Jtot}
	\mathcal J_p^{\rm tot}(k)=\mathcal J_p^{(r_0)}(k)+\mathcal J_p^{\rm UV}(k),
\ee
where
\be \label{eq:Jr0def}
	\mathcal J_p^{(r_0)}(k)\equiv\frac{4\pi}{k}\int_{r_0}^\infty\dd r\,r^{1-p}\sin(kr)
\ee
is the integral of Eq.~\eqref{eq:Jdef} restricted to $r\geq r_0$, which converges for every $p>1$, and $\mathcal J_p^{\rm UV}$ is the contribution of the region $r<r_0$, determined by the inner completion. The contribution of the channel to Eq.~\eqref{eq:Fleading} is
\be \label{eq:F1-split}
	F^{(1)}_{{\rm free},p}=-\frac{\omega^2}{4\pi}\,\mathcal P\,c_p\,\mathcal J_p^{\rm tot}(k).
\ee
The splitting radius $r_0$ is arbitrary, since it only marks where one description is replaced by the other, and the sum~\eqref{eq:Jtot} cannot depend on it: the dependence of $\mathcal J_p^{\rm UV}$ on $r_0$ must cancel that of $\mathcal J_p^{(r_0)}$. This is equivalent to the usual renormalization procedure, in which the divergences are removed by counterterms, i.e.\ local terms in the potential whose coefficients depend on the cutoff. We now compute the two contributions.

In the exterior region, for $p\neq2n+3$, with $n=0,1,\ldots$, the integral of Eq.~\eqref{eq:Jr0def} is
\begin{align}
	\mathcal J_p^{(r_0)}(k)={}&4\pi\Gamma(2-p)\sin\left[\frac{\pi(2-p)}{2}\right]k^{p-3} \nonumber \\
	\label{eq:Jcut-exact} &-4\pi\sum_{n\geq0}\frac{(-k^2)^n}{(2n+1)!}\,\frac{r_0^{\,2n+3-p}}{2n+3-p}.
\end{align}
The first term has the same form as Eq.~\eqref{eq:jp}. It does not depend on $r_0$, it is not analytic in $k^2$, and it comes from radii $r\sim1/k$: it is determined by the exterior potential alone~\cite{Holstein:2004dn}. All the dependence on $r_0$ is in the second term, which is a power series in $k^2$. Its terms with $2n+3>p$ vanish for $r_0\to0$, while those with $2n+3<p$ are the power divergences.

For $p=2n+3$, the exponent of the $n$-th term of the sum vanishes, and both this term and the first term of Eq.~\eqref{eq:Jcut-exact} have a pole. The poles cancel, since $\mathcal J_p^{(r_0)}$ is finite, and the limit $p\to2n+3$ gives a logarithm. In this case $k^{p-3}=k^{2n}$ is itself analytic in $k^2$, and the nonanalytic dependence on $k$ takes the form $k^{2n}\ln k$; for $n=0$ this is the correspondence between $\ln k$ and a $1/r^3$ term of the potential~\cite{Holstein:2004dn}. The limit gives
\begin{align}
	\mathcal J_{2n+3}^{(r_0)}(k) ={}& - 4 \pi\frac{ (-k^2)^n \left[\ln(kr_0)+\gamma_{\rm E}-H_{2n+1}\right]}{(2 n + 1)!} \nonumber \\
	\label{eq:Jcut} &-4\pi\sum_{m\neq n}\frac{(-k^2)^m}{(2m+1)!}\,\frac{r_0^{\,2(m-n)}}{2(m-n)},
\end{align}
where $H_n$ are the harmonic numbers
\be \label{eq:CnHn}
	H_n\equiv\sum_{m=1}^{n}\frac1m,\quad H_0\equiv0.
\ee
In Eq.~\eqref{eq:Jcut} the terms of the sum with $m<n$ are finite at fixed $r_0$ but diverge for $r_0\to0$, while those with $m>n$ vanish for $r_0\to0$, thus the sum is effectively limited.

The coefficient of $\ln k$ is fixed by the exterior potential, and it is equal to that of $\ln r_0$. For the first three singular channels, up to terms that vanish for $r_0\to0$,
\begin{align}
	\mathcal J_3^{(r_0)}(k)&=-4\pi\left[\ln(kr_0)+\gamma_{\rm E}-1\right], \nonumber \\
	\label{eq:J345cut} \mathcal J_4^{(r_0)}(k)&=\frac{4\pi}{r_0}-\pi^2k, \\
	\mathcal J_5^{(r_0)}(k)&=\frac{2\pi}{r_0^2}+\frac{2\pi}{3}k^2\left[\ln(kr_0)+\gamma_{\rm E}-\frac{11}{6}\right]. \nonumber
\end{align}

In the inner region, $r<r_0$, we have $kr\ll1$, and we can expand the sine in Eq.~\eqref{eq:Jdef} in powers of $kr$. Whatever the potential is in this region, its contribution is a power series in $k^2$,
\be \label{eq:inner-series}
	\sum_{n\geq0}\frac{4\pi\,(-k^2)^n}{(2n+1)!}\int_0^{r_0}\dd r\,r^{2+2n}\,4\Ueff(r),
\ee
whose coefficients are moments of the inner potential. Since $(-k^2)^n$ is the Fourier transform of $(\nabla^2)^n\delta^{(3)}(\rr)$, the inner region acts on the wave as a set of local terms at the centre of the lens,
\be \label{eq:local-terms}
	4\omega^2\,\delta U_{\rm loc}=\sum_{n\geq0}g_{p,n}(r_0)\,(\nabla^2)^n\delta^{(3)}(\rr),
\ee
whose coefficients $g_{p,n}$ cannot be computed from the exterior potential. For a black hole, which has a horizon instead of an inner potential, they are obtained by matching wave solutions, as we do in Sec.~\ref{sec:v:f:matching2}. The contribution of the inner completion is therefore
\be \label{eq:JUV}
	\mathcal J_p^{\rm UV}(k)=\frac{1}{\omega^2c_p}\sum_{n\geq0}g_{p,n}(r_0)\,(-k^2)^n.
\ee

Since Eq.~\eqref{eq:Jtot} cannot depend on $r_0$, the dependence of the local coefficients on $r_0$ must cancel that of the exterior integral, order by order in $k^2$. This is possible because all the dependence on $r_0$ of the exterior integral is analytic in $k^2$, as is the contribution of the inner region. The local terms of Eq.~\eqref{eq:local-terms} play the role of counterterms: their dependence on $r_0$ cancels the divergences, while their finite part is fixed by matching. For $2n+3\neq p$, Eq.~\eqref{eq:Jcut-exact} shows that the combinations
\be \label{eq:ghat}
	\hat g_{p,n}\equiv g_{p,n}(r_0)-\frac{4\pi\,\omega^2c_p\,r_0^{\,2n+3-p}}{(2n+1)!\,(2n+3-p)}
\ee
do not depend on $r_0$, and
\be \label{eq:F1-renormalized}
	F^{(1)}_{{\rm free},p}=-\frac{\mathcal P}{4\pi}\Big[\omega^2c_p\,\mathcal J_p(k)+\sum_{n}\hat g_{p,n}\,(-k^2)^n\Big],
\ee
where $\mathcal J_p(k)$ is the first term of Eq.~\eqref{eq:Jcut-exact}, so that the dependence on the cutoff has disappeared. The exterior potential determines $\mathcal J_p$, while the inner completion determines the constants $\hat g_{p,n}$, which must be obtained by matching. Their natural size is $\hat g_{p,n}\sim\omega^2c_p(GM)^{2n+3-p}$, so that, relative to the exterior term, the local term $(-k^2)^n$ scales as $x^{2n+3-p}$. For $x\ll1$ only the terms with $2n\leq p-3$, which are those needed to cancel a divergence, must be retained.

For $p=2n+3$, the same procedure removes $\ln r_0$ from Eq.~\eqref{eq:Jcut}, but the remaining $\ln k$ needs a scale to be dimensionless, which can only come from the inner completion. We write its remaining, $r_0$-independent part in terms of a scale $\bar\mu$, defined by
\be \label{eq:MSfree}
	g_{p,n}(r_0)=\frac{4\pi\,\omega^2c_p}{(2n+1)!}\left[\ln(\bar\mu r_0)+\gamma_{\rm E}-H_{2n+1}+\frac{H_n}{2}\right],
\ee
where the constants are chosen for convenience. With this local coefficient, the terms proportional to $k^{2n}$ become $\omega^2c_p\mathcal J_{2n+3}(k)$, with
\be \label{eq:Jodd}
	\mathcal J_{2n+3}(k)= -\frac{2\pi\,(-k^2)^n}{(2n+1)!} \left[\ln\frac{k^2}{\bar\mu^2}-H_n\right].
\ee

For odd $p$ the inner completion therefore enters through the scale $\bar\mu$, which must be fixed by matching. Its natural value is $\bar\mu\sim1/(GM)$, and a different choice changes $\mathcal J_{2n+3}$ by a term proportional to $k^{2n}$, i.e.\ by a local term. For $p=2n+3$, Eq.~\eqref{eq:F1-renormalized} holds with $\mathcal J_p$ given by Eq.~\eqref{eq:Jodd} and without the local term with $(-k^2)^n$.

The independence of the cutoff can also be written as an equation for the local coefficients. Since $\mathcal J_p^{\rm tot}$ of Eq.~\eqref{eq:Jtot} does not depend on $r_0$, the derivatives of its two parts with respect to $\ln r_0$ must cancel,
\be
	\frac{\dd\mathcal J_p^{(r_0)}}{\dd\ln r_0}=-\frac{\dd\mathcal J_p^{\rm UV}}{\dd\ln r_0}.
\ee
On the left, the derivative of the integral~\eqref{eq:Jr0def} with respect to its lower limit gives $-r_0$ times its integrand at $r_0$, i.e.\ $-(4\pi/k)\,r_0^{2-p}\sin(kr_0)=-4\pi\sum_n(-k^2)^n r_0^{\,2n+3-p}/(2n+1)!$. On the right, Eq.~\eqref{eq:JUV} gives $-\sum_n(-k^2)^n\,(\dd g_{p,n}/\dd\ln r_0)/(\omega^2c_p)$. Equating the coefficients of $(-k^2)^n$ gives
\be \label{eq:rg-free}
	\frac{\dd g_{p,n}}{\dd\ln r_0}=\frac{4\pi\,\omega^2c_p}{(2n+1)!}\,r_0^{\,2n+3-p},
\ee
the renormalization-group equation of the local coefficients, valid for every $p$. Integrating it gives a power of $r_0$ plus an integration constant, Eq.~\eqref{eq:ghat}, and, when the exponent vanishes, $p=2n+3$, a logarithm of $r_0$, Eq.~\eqref{eq:MSfree}. The integration constant, $\hat g_{p,n}$ or $\bar\mu$, is what the exterior potential cannot determine, and it is fixed by matching, as in the renormalization of singular potentials in quantum mechanics~\cite{Beane:2000wh} and in the Born series for wave scattering off a black hole~\cite{Caron-Huot:2025tlq}.

The logarithms obtained in this way are single logarithms, $\ln r_0$, for two reasons: the right-hand side of Eq.~\eqref{eq:rg-free} does not contain $g_{p,n}$, and the coefficient $c_p$ is a constant. The first condition holds at first order in the potential. At higher Born orders, a local term can combine with a further insertion of the potential, for instance the term $g_{p,0}\,\delta^{(3)}(\rr)$ with the $p=2$ channel, whose product diverges logarithmically at the origin. The right-hand side then depends on the local coefficients, which run with $r_0$, and powers of logarithms can appear, as in the effective-field-theory treatments of Refs.~\cite{Goldberger:2005cd,Ivanov:2024sds,Caron-Huot:2025tlq,Galley:2015kus,Ivanov:2025ozg}. The second condition is not automatic. If the coefficient varied with the radius, $c_p\to c_p(r)$, the same derivation would give Eq.~\eqref{eq:rg-free} with $c_p$ replaced by its value at the cutoff, $c_p(r_0)$. For example, for a slowly varying coefficient $c_p(r)\propto1/\ln(r/\ell)$, with $\ell$ a fixed length, in a logarithmic channel, $p=2n+3$, the right-hand side becomes $\propto1/\ln(r_0/\ell)$, and its integral is $g_{p,n}\propto\ln\ln(r_0/\ell)$, a logarithm of a logarithm. In our case this does not happen. The wave is a test field on a fixed background: it does not backreact on the metric, so that the mass of the lens, and the coefficients $c_p$ determined by it, are not renormalized by the scattering. Unlike the couplings of a quantum field theory, which run because of loop corrections, the $c_p$ are fixed external parameters, and they do not depend on $r_0$.

For $p\neq2n+3$, the renormalized amplitude~\eqref{eq:F1-renormalized} can also be obtained in position space, if the radial channel $c_p/r^p$ is replaced in the potential by $c_p\,{\rm FP}\,r^{-p}$, where
\begin{align}
	{\rm FP}\frac{1}{r^p}\equiv\lim_{r_0\to0}\Bigg[&\frac{\Theta(r-r_0)}{r^p} \nonumber \\
	\label{eq:FPgeneral} &+\sum_{2n+3<p}\frac{4\pi\,r_0^{\,2n+3-p}\,(\nabla^2)^n\delta^{(3)}(\rr)}{(2n+1)!\,(2n+3-p)}\Bigg]
\end{align}
is the finite part of $r^{-p}$, whose Fourier transform is $\mathcal J_p(k)$, and the local terms of Eq.~\eqref{eq:local-terms} are added with $g_{p,n}$ replaced by $\hat g_{p,n}$. We use this form for the Schwarzschild lens in Sec.~\ref{sec:v:f:complete}.

To summarize, the first five functions $\mathcal J_p$ are
\begin{align}
	\mathcal J_1(k)&=\frac{4\pi}{k^2}\,, \nonumber \\
	\mathcal J_2(k)&=\frac{2\pi^2}{k}\,, \nonumber \\
	\label{eq:J15} \mathcal J_3(k)&=-2\pi\ln\frac{k^2}{\bar\mu^2}\,, \\
	\mathcal J_4(k)&=-\pi^2k\,, \nonumber \\
	\mathcal J_5(k)&=\frac{\pi}{3}k^2\left(\ln\frac{k^2}{\bar\mu^2}-1\right). \nonumber
\end{align}
$\mathcal J_1$ and $\mathcal J_2$ are the convergent integrals, while $\mathcal J_3$, $\mathcal J_4$ and $\mathcal J_5$ are the renormalized exterior contributions, to which the local terms of Eq.~\eqref{eq:F1-renormalized} are added: $\hat g_{4,0}$ for $p=4$ and $\hat g_{5,0}$ for $p=5$, while for $p=3$, and for the logarithm of $p=5$, the inner completion enters through $\bar\mu$. When comparing channels for a black hole, we set $\bar\mu=1/(GM)$, which gives $\ln(k^2/\bar\mu^2)=\ln x^2$. These short-distance logarithms must be distinguished from the large-distance logarithms associated with the Newtonian channel, which we encounter in Sec.~\ref{sec:breakdown}.

Eq.~\eqref{eq:J15} gives a simple physical interpretation of the radial channels. The Newtonian case $p=1$ is most strongly enhanced toward forward scattering, as $k^{-2}$; the $1/r^2$ correction scales as $k^{-1}$. More rapidly decaying exterior terms $p>1$ are less forward enhanced and probe shorter scales, while eventually requiring information about the inner lens. The logarithms at odd $p\geq3$ are the marginal cases of this short-distance expansion.

\subsubsection{Radial channels in dimensionless quantities.}
We now express the first Born result in the dimensionless variables of Section~\ref{sec:DimensionlessVariables}, to compare the size of the different radial channels. We write $c_p=\gamma_p(GM)^p$, where the dimensionless coefficients $\gamma_p$ may depend on frequency and on the parameters of the lens environment, and normalize each channel to the Newtonian one,
\be \label{eq:Tpdimless}
	T_p\equiv(GM)^{p-1}\frac{\mathcal J_p(k)}{\mathcal J_1(k)}\,.
\ee
From Eq.~\eqref{eq:J15}, with $\bar\mu=1/(GM)$, the first five are
\begin{align}
	T_1&=1\,, \nonumber \\
	T_2&=\frac{\pi}{2}x\,, \nonumber \\
	\label{eq:T0schw} T_3&=-\frac{x^2}{2}\ln x^2\,, \\
	T_4&=-\frac{\pi}{4}x^3\,, \nonumber \\
	T_5&=\frac{x^4}{12}\left(\ln x^2-1\right). \nonumber
\end{align}
For coefficients $\gamma_p$ of order unity, the higher channels are therefore suppressed by powers of $x=GMk$, with additional logarithms for odd $p\geq3$. This hierarchy compares different radial powers within a fixed Born order, and it is distinct from the Born expansion, which counts interactions: a small $x$ suppresses the higher channels, but it does not guarantee that repeated Newtonian interactions are negligible.

In the small-angle limit
\be \label{eq:dSOsmallangle}
	d_{\rm SO}\simeq r_{\rm S}+r_{\rm O}-\frac{b^2}{2r_{\rm eff}},
\ee
so that the endpoint factor of Eq.~\eqref{eq:endpointFactor} is
\be \label{eq:Psmallangle}
	\mathcal P\simeq \frac{\eu^{\ii s/2}}{r_{\rm eff}}\,.
\ee
Using $c_p\mathcal J_p=4\pi GM\gamma_pT_p/k^2$, Eq.~\eqref{eq:bdef} and $R_E^2=4GMr_{\rm eff}$, Eq.~\eqref{eq:Fleading} becomes
\be \label{eq:FleadingDimless}
	F^{(1)}_{\rm free}\simeq-\frac{\eu^{\ii s/2}}{4y^2}\sum_{p\geq1}\gamma_pT_p\,,
\ee
where the Newtonian channel gives $\gamma_1T_1=-4$. The first Born term is therefore a small correction to the free wave only for $y\gg1$, i.e.\ when the transverse offset is much larger than the Einstein radius. In this regime the point lens produces a faint second image, with amplitude $1/y^2$ relative to the first one, and Eq.~\eqref{eq:FleadingDimless} describes its interference with the direct wave. For $y\lesssim1$ the two images have comparable amplitudes, and the first Born term cannot reproduce the multiple-image structure of the point lens.

\subsection{Issues of the free-wave Born expansion}
\label{sec_compissues}
The response to a single interaction does not establish whether a perturbative prediction is reliable: repeated interactions with lower radial channels can compete with, or exceed, the higher-channel corrections we wish to calculate. We use the eikonal picture  of Sec.~\ref{sec:eikonal} to identify the parameters controlling this competition, distinguishing the accumulated Newtonian phase, which motivates resummation, from the hierarchy among the remaining interactions, which determines a consistent truncation. In this case, the Born series is the expansion of the exponential of the lensing phase $\psi=\sum_q\psi_q$ in Eq.~\eqref{eq:overview-eikonal}, so that the $n$th Born term is
\be \label{eq:eikonalBornTerm}
	F^{(n)}_{\rm free}\simeq\frac{\omega}{2\pi\ii r_{\rm eff}}\int\dd^2\zeta\,\eu^{\frac{\ii\omega|\boldsymbol\zeta-\bm b|^2}{2r_{\rm eff}}}\,\frac{[\ii\,\psi(\zeta)]^n}{n!}.
\ee
A term with insertions of the channels $q_1,\dots,q_n$ contains the product $\psi_{q_1}\cdots\psi_{q_n}\propto\zeta^{-D}$, with $D=\sum_i(q_i-1)$, as it can be understood from Eq.~\eqref{eq:v:d:delta-phi-p}. In the Fraunhofer limit, where we neglect the quadratic term at the exponential, the transverse integral is $\int\dd^2\zeta\,\eu^{-\ii\kk\cdot\boldsymbol\zeta}\zeta^{-D}$. The integral over $\zeta$ is dominated by $\zeta\sim1/k$: at larger $\zeta$ the factor $\eu^{-\ii\kk\cdot\boldsymbol\zeta}$ oscillates and the contributions cancel, while at smaller $\zeta$ the area is too small to contribute. This contribution is proportional to $k^{D-2}$, which is nonanalytic in $k$, and it is fixed by the exterior of the lens. Its size can be estimated by evaluating each phase at $\zeta\sim1/k$, where $GM/\zeta\sim x$: the size of the term is the product of the phases evaluated there. Thus, every insertion of a $p$ radial channel carries a power of $x^{p-1} \ll 1$ for $p>1$.   For $D\geq2$, however, the integrand grows so fast at small $\zeta$ that the region close to the lens also contributes. As for the singular channels of Sec.~\ref{sec:leading}, this contribution consists of local terms, analytic in $k^2$, which must be fixed by the inner structure of the lens, and it is not included in this estimate.

The first issue concerns the Newtonian insertions, which have $q=1$ and therefore do not change $D$: they are not suppressed by powers of $x$, and the phase of the Newtonian channel is not small. By Eq.~\eqref{eq:eikonalShapiro}, at $\zeta\sim1/k$,
\be \label{eq:breakdownPsi1}
	\psi_1\simeq\frac w2\ln\left(4k^2r_{\rm S}r_{\rm O}\right),
\ee
which we recover with the full second-order calculation in Sec.~\ref{sec:breakdown}. Because of the logarithm of the distances, $\psi_1$ is of order one or larger in astrophysical configurations even when $w\ll1$, as shown by the values of $\ln(k^2r_{\rm S}r_{\rm O})$ in the table below. The terms of the expansion of $\eu^{\ii\psi_1}$, $(\ii\psi_1)^n/n!$, then do not decrease until $n$ exceeds $\psi_1$: a truncation after $n$ Newtonian insertions leaves out a term of the same size as those that are kept, and approximates a phase of modulus one by a polynomial in $\psi_1$. The number of Newtonian insertions that must be retained therefore grows with $\psi_1$, i.e.\ with the logarithm of the distances, and the Newtonian channel must be resummed.

The second issue concerns the other channels, $q>1$: the Born series in these channels is not ordered by the number of insertions. Their phases, Eq.~\eqref{eq:v:d:delta-phi-p}, do not depend on the distances of source and observer, and at $\zeta\sim1/k$ they are small, $\psi_q\sim\gamma_q\,w\,x^{q-1}\ll1$, so that the series converges. However, a term with more insertions is not necessarily smaller. Compare the first Born term of channel $p$, the second Born term of channels $p$ and $q$, and the first Born term of channel $p+q$:\footnote{These estimates hold in the eikonal limit, $\omega\zeta\gg1$, and refer to the contribution from $\zeta\sim1/k$. In Sec.~\ref{sec:breakdown} we compute explicitly, without the eikonal approximation, the second Born term with one Newtonian and one $p=2$ insertion, and compare it with the first Born term of the $p=3$ channel: their ratio, Eq.~\eqref{eq:finalratio}, contains the factor $\omega/k=1/(y\alpha_E)$ found here, together with the logarithm of the distances of the first issue.}
\begin{align}
	\psi_p\,\psi_q&\sim w^2x^{p+q-2}, \nonumber \\
	\label{eq:breakdownCounting} \psi_{p+q}&\sim w\,x^{p+q-1}.
\end{align}

The last two are of the same order in $G$, but $\psi_p\psi_q/\psi_{p+q}\sim w/x=4\omega/k\gg1$:  each interaction comes with the frequency, through $w=4GM\omega$, while each additional power of $1/r$ comes with $GM/\zeta\sim GMk$. The second Born terms of the lower channels are therefore larger than the first Born terms of the higher channels, and a truncation at first Born order that retains the higher channels is inconsistent.

The series, which is convergent, has to be truncated according to the size of its terms. A term with $n$ insertions is of order $w^nx^D$, and relative to the first Born term of channel $p$ it is of order $w^{n-1}x^{D-(p-1)}$. Hence, if $p$ is the last channel that is retained at first Born order, all the Born terms with
\be \label{eq:breakdownPrescription}
	w^{n-1}x^{\sum_i(q_i-1)-(p-1)}\gtrsim1
\ee
must be retained as well. In wave optics, $w\lesssim1$, this requires all the terms with $\sum_i(q_i-1)\leq p-1$. For example, retaining the first Born term of the $p=4$ channel requires the second Born terms with channels $(2,2)$ and $(2,3)$, and the third Born term with channels $(2,2,2)$. In geometric optics, $w\gg1$, also terms with $\sum_i(q_i-1)>p-1$ must be retained, as long as $w^{n-1}$ compensates the additional powers of $x$, since each insertion still costs $\psi_q\sim wx^{q-1}\ll1$. A Newtonian insertion does not change the power of $x$ in Eq.~\eqref{eq:breakdownPrescription}, and costs $\psi_1$, which is not small, so that all of them must be retained: this is the first issue.

The eikonal limit, however, assumes small angles and paths that pass many wavelengths away from the lens. In the next subsection we show, with the full second-order Born calculation, that the first issue is not an artifact of this limit, and we compare the resulting terms for a Schwarzschild lens.

\subsection{Second-order Born approximation}
\label{sec:breakdown}

The eikonal estimates of Sec.~\ref{sec_compissues} suggest that the long-range Newtonian field can substantially enhance the response to another radial channel, but we must establish whether this conclusion survives beyond the straight-path approximation. We therefore calculate the mixed second-order Born contribution directly from the three-dimensional propagation integral, identifying the origin and coefficient of the large-distance logarithm and showing when Newtonian resummation becomes necessary.

The second-order Born contribution of two radial channels, $c_p/r_1^p$ and $c_q/r_2^q$, is $F^{(2)}_{{\rm ref},pq}$ of Eq.~\eqref{eq:Fref2pq} with the free wave as reference,
\begin{align}
	F^{(2)}_{{\rm free},pq}={}&\frac{\omega^4c_pc_q}{G_{\rm free}(\rr_{\rm O},\rr_{\rm S})}\int\dd^3\rr_1\,\dd^3\rr_2\,G_{\rm free}(\rr_{\rm O},\rr_1)\frac{1}{r_1^p} \nonumber \\
	\label{eq:F2start} &\hspace{1.8em}\times G_{\rm free}(\rr_1,\rr_2)\frac{1}{r_2^q}G_{\rm free}(\rr_2,\rr_{\rm S}).
\end{align}
As for the first Born term, we work in the Fraunhofer limit. We approximate the two external Green functions by their plane-wave forms of Eq.~\eqref{eq:planewaves}, with $\rr\to\rr_1$ on the observer side and $\rr\to\rr_2$ on the source side, and the wavevectors of Eqs.~\eqref{eq:kinkout} and~\eqref{eq:kaskinkout}. For the internal Green function $G_{\rm free}(\rr_1,\rr_2)$ we use instead its Fourier representation
\be \label{eq:G0Fourier}
	\frac{\eu^{\ii\omega|\rr_1-\rr_2|}}{|\rr_1-\rr_2|}=\int\frac{\dd^3u}{(2\pi)^3}\,\eu^{\ii\uu\cdot(\rr_1-\rr_2)}\frac{4\pi}{u^2-(\omega+\ii0)^2},
\ee
where $\ii0$ denotes the outgoing prescription, $\ii0=\lim_{\epsilon\to0^+}\ii\epsilon$. The two position integrals then reduce to the radial channel functions $\mathcal J_p(|\uu-\kk_f|)$ and $\mathcal J_q(|\uu-\kk_i|)$ introduced in Eq.~\eqref{eq:Jdef}. With the endpoint factor $\mathcal P$ of Eq.~\eqref{eq:endpointFactor}, the plane-wave reduction of Eq.~\eqref{eq:F2start} becomes
\begin{align}
	F^{(2)}_{{\rm free},pq}&\simeq\frac{\omega^4c_pc_q}{16\pi^2}\,\mathcal P\,\mathcal I_{pq}\,, \nonumber \\
	\label{eq:F2u} \mathcal I_{pq}&\equiv\int\frac{\dd^3u}{(2\pi)^3}\,\mathcal J_p(|\uu-\kk_f|) \\
	&\quad\times\frac{4\pi}{u^2-(\omega+\ii0)^2}\mathcal J_q(|\uu-\kk_i|). \nonumber
\end{align}

We now focus on a Newtonian insertion on the source side, $q=1$, and initially restrict the other channel to $1<p<3$, so that its first Born contribution is free of ultraviolet divergences. We write the internal momentum as $\uu=\kk_i+\bl$, choose the $z$ axis along $\kk_i\simeq\omega\bm e$, decompose $\bl=(\bl_\perp,\ell_z)$ and define $t\equiv|\bl_\perp|$. Using the Newtonian Fourier transform of Eq.~\eqref{eq:JNewton}, the two potentially singular factors in Eq.~\eqref{eq:F2u} become
\be \label{eq:breakdownSoftDenominators}
	\mathcal J_1(|\uu-\kk_i|)=\frac{4\pi}{\ell_z^2+t^2},\qquad u^2-\omega^2=\ell_z^2+2\omega\ell_z+t^2.
\ee
The Newtonian contribution is singular at zero momentum transfer, $\bl=0$, whereas the internal propagator is singular when $|\uu|=\omega$, corresponding to the free-wave dispersion relation, or on-shell condition. Since $|\kk_i|=\omega$, these singular regions overlap. Physically, a long-wavelength component of the Newtonian field can transfer very little momentum while leaving the intermediate wave close to free propagation.

In the region $|\bl|\ll k$, the remaining channel function varies slowly and can be approximated as $\mathcal J_p(|\bl-\kk|)\simeq\mathcal J_p(k)$, so that Eq.~\eqref{eq:F2u} with $q=1$ becomes
\be
	\mathcal I_{p1}\simeq\mathcal J_p(k)\int\frac{\dd^3\ell}{(2\pi)^3}\,\frac{4\pi}{\ell_z^2+t^2}\,\frac{4\pi}{\ell_z^2+2\omega\ell_z+t^2-\ii0}.
\ee
Writing $\dd^3\ell=2\pi t\,\dd t\,\dd\ell_z$, this is
\begin{align}
	\mathcal I_{p1}\simeq{}&4\mathcal J_p(k)\int t\,\dd t \nonumber \\
	&\times\int_{-\infty}^{\infty}\frac{\dd\ell_z}{(\ell_z^2+t^2)(\ell_z^2+2\omega\ell_z+t^2-\ii0)}.
\end{align}
The integral over $\ell_z$ is done by closing the contour in the upper half plane. For $0<t\ll\omega$, it encloses the poles $\ell_z=\ii t$ and $\ell_z\simeq-t^2/(2\omega)+\ii0$, with residues $-1/(4\omega t^2)$ and $1/(2\omega t^2)$, respectively, so that
\begin{align}
	\int_{-\infty}^{\infty}&\frac{\dd\ell_z}{(\ell_z^2+t^2)(\ell_z^2+2\omega\ell_z+t^2-\ii0)} \simeq \frac{\ii\pi}{2\omega t^2}.
\end{align}
The integral over $t$ is therefore logarithmic, and its limits are set by the approximations we made. At large $t$ it is cut off at $t\sim k$, where $\mathcal J_p(|\bl-\kk|)$ starts to vary. At small $t$, a momentum transfer $t$ probes the Newtonian field at distances $r\sim1/t$ from the lens, while the plane-wave form of the source-side Green function, Eq.~\eqref{eq:planewaves}, holds only for $r\ll r_{\rm S}$: the integral is cut off at $t\sim1/r_{\rm S}$, below which the wave resolves the position of the source. To logarithmic accuracy,
\be \label{eq:breakdownSoftIntegral}
	\mathcal I_{p1}\simeq\mathcal J_p(k)\frac{2\pi\ii}{\omega}\int_{1/r_{\rm S}}^{k}\frac{\dd t}{t}=\mathcal J_p(k)\frac{2\pi\ii}{\omega}\ln(kr_{\rm S}),
\ee
while the terms that are not enhanced by the logarithm are not fixed by this calculation. Using Eq.~\eqref{eq:F2u}, the term of channel $p$ in Eq.~\eqref{eq:Fleading} and $c_1=-4GM$, and adding the insertion on the observer side, which gives $\ln(kr_{\rm O})$, we obtain
\begin{align}
	\frac{F^{(2)}_{{\rm free},p1}+F^{(2)}_{{\rm free},1p}}{F^{(1)}_{{\rm free},p}} ={}&\frac{\ii w}{2}\left[\ln(k^2r_{\rm S}r_{\rm O})+C_p(\theta)\right] \nonumber \\
	\label{eq:F21} &+\hbox{finite-distance corrections}.
\end{align}
Here $C_p(\theta)$ denotes the endpoint-independent finite contribution, which need not be real at finite angle, and Eq.~\eqref{eq:F21} holds for fixed nonzero momentum transfer and $1<p<3$\footnote{The dimensionless variables of Section~\ref{sec:DimensionlessVariables} give the identity $k^2r_{\rm S}r_{\rm O}=s/(\varepsilon\,a(1-a))$.}. The correction is controlled by $w\ln(k^2r_{\rm S}r_{\rm O})$, rather than by an additional power of $x$, so that repeated Newtonian interactions need not be negligible even when the higher radial channels are strongly suppressed. The logarithm is the Newtonian phase accumulated by the wave between the region $r\sim1/k$ around the lens and the source,
\be \label{eq:breakdownOuterPhase}
	-2\omega\int_{1/k}^{r_{\rm S}}\dd\ell\,U_{\rm newt}\simeq\frac w2\ln(kr_{\rm S}),
\ee
and similarly on the observer side, in agreement with the Shapiro phase of Eq.~\eqref{eq:eikonalShapiro} at $\zeta\sim1/k$. When $|w\ln(k^2r_{\rm S}r_{\rm O})|/2$ is not small, successive terms of the Born series are comparable, and a few of them are a poor approximation to this phase, as anticipated in Sec.~\ref{sec_compissues}. This phase is accumulated between the region $r\sim1/k$ and the endpoints: moving the source from $r_{\rm S}$ to $r_{\rm S}+\dd r_{\rm S}$ multiplies the response by $1+\frac{\ii w}{2}\,\dd\ln r_{\rm S}$, so that the logarithms $(\ii wL/2)^m/m!$, with $L=\ln(k^2r_{\rm S}r_{\rm O})$, of the Born series around the free wave, Eq.~\eqref{eq:F21}, sum to a power, $(k^2r_{\rm S}r_{\rm O})^{\ii w/2}$,  in analogy with the exponentiation of the Coulomb phase~\cite{Dalitz:1951ah,Weinberg:1965nx,Lippstreu:2023vvg}, and with the infrared logarithms of the tail effect in gravitational-wave emission, which exponentiate into a phase that can be removed by a redefinition of time~\cite{Galley:2015kus,Ivanov:2025ozg}.

Both issues appear when the mixed term with one Newtonian and one $p=2$ insertion is compared with the first Born term of the $p=3$ channel, which is of the same order $(GM)^3$. With Eqs.~\eqref{eq:F21} and~\eqref{eq:FleadingDimless}, and the Schwarzschild coefficients of Eq.~\eqref{eq:cpSchw}, we find
\be \label{eq:finalratio}
	\left|\frac{F^{(2)}_{{\rm free},21}+F^{(2)}_{{\rm free},12}}{F^{(1)}_{{\rm free},3}}\right| \simeq\frac{5\pi}{3}\,\frac{1}{y\alpha_E}\,\frac{\ln(k^2r_{\rm S}r_{\rm O})}{|\ln x^2|},
\ee
where the logarithm of the distances comes from the first issue, and the factor $1/(y\alpha_E)$ from the second. The table gives this ratio for $w=0.1$ and $y=30$, with the lens midway between source and observer, in three configurations with $kr_{\rm S,O}\gg1$, $k/\omega\ll1$ and $x\ll1$: two cosmological lenses with $r_{\rm eff}=0.5\,$Gpc, and a compact geometry with $r_{\rm eff}=10^6GM$, which holds for any mass, since at fixed $w$ and $y$ the entries depend on the mass and on the distances only through $r_{\rm eff}/(GM)$.
\begin{table}[t]
\caption{\label{tab:finalratio}Ratio of Eq.~\eqref{eq:finalratio} for $w=0.1$, $y=30$ and the lens midway between source and observer, $a=1/2$. In this case $\ln(k^2r_{\rm S}r_{\rm O})=2\ln(2wy/\alpha_E)$ and $x=wy\alpha_E/4$, so that the ratio depends on the configuration only through $\alpha_E$, i.e.\ through $r_{\rm eff}/(GM)$.}
\small
\resizebox{\columnwidth}{!}{
\begin{tabular}{@{}lccc@{}}
\toprule
Configuration & $\alpha_E$ & $\ln(k^2r_{\rm S}r_{\rm O})$ & Ratio in Eq.~\eqref{eq:finalratio}\\
\midrule
$M=10^2M_\odot$, $r_{\rm eff}=0.5\,$Gpc
  & $1.96\times10^{-10}$ & $48.3$ & $9.5\times10^8$\\
$M=10^6M_\odot$, $r_{\rm eff}=0.5\,$Gpc
  & $1.96\times10^{-8}$ & $39.1$ & $9.7\times10^6$\\
any $M$, $r_{\rm eff}=10^6GM$
  & $2.0\times10^{-3}$ & $16.0$ & $1.1\times10^2$\\
\bottomrule
\end{tabular}}
\end{table}

The first issue is specific to the Newtonian channel, and it is resolved by resumming it. As we are going to discuss, the natural reorganization of the problem is to include $U_{\rm newt}=-GM/r$ in the reference wave operator and treat only the remaining (non-Newtonian) potential $\delta U_{\rm newt}$ perturbatively. The reference Green function $G_{\rm newt}$ sums the Newtonian interactions. The leading response to $\delta U_{\rm newt}$ has the same single-insertion structure as before, but with the Newtonian Green function $G_{\rm newt}$ on both sides of the insertion. On expanding those Green functions in $U_{\rm newt}$, one recovers $F^{(1)}_{{\rm free},p}$, $F^{(2)}_{{\rm free},p1}+F^{(2)}_{{\rm free},1p}$, and all terms with further Newtonian insertions, so that the logarithm of the distances is resummed into a phase. This use of the exact Newtonian reference problem in three-dimensional lensing was advocated in \cite{Carrillo-Gonzalez:2026}; the following sections make this suggestion concrete in our setup, and develop a quantitative method to evaluate the resulting distorted channel kernels explicitly. The second issue, instead, is not related to the long range of the potential, and it remains after the resummation: also around the point lens, the terms must be retained according to the size of the products of the phases $\psi_q$, as in Eq.~\eqref{eq:breakdownPrescription}.

\subsection{Summary}
\label{sec:born:summary}

In this section we have computed the Born series around the free wave, in which every interaction with the lens, including the Newtonian one, is treated as a perturbation, Eqs.~\eqref{eq:Fborn} and~\eqref{eq:F2born}.

At first order, in the Fraunhofer limit, which requires the transverse offset to be much larger than the Fresnel scale, Eq.~\eqref{eq:farfieldConditions}, the response of each radial channel is the Fourier transform of $r^{-p}$ at the momentum transfer $k$, Eqs.~\eqref{eq:Fleading} and~\eqref{eq:Jdef}. For the convergent channels, $1<p<3$, it is a power $k^{p-3}$, Eq.~\eqref{eq:jp}, which is not analytic in $k^2$ and is fixed by the radial channel alone. The Newtonian channel gives the Rutherford form $k^{-2}$, Eq.~\eqref{eq:JNewton}, which contains no logarithm of the distances. The channels with $p\geq3$ diverge at short distances, because the radial expansion has been extended to the centre of the lens. When the integral is split at a radius $r_0$, the dependence on $r_0$ is always analytic in $k^2$, Eqs.~\eqref{eq:Jcut-exact} and~\eqref{eq:Jcut}, as is the contribution of the inner region, which acts as a set of local terms. The two cancel, and the renormalized response, Eq.~\eqref{eq:F1-renormalized}, is the sum of a nonanalytic part, determined by the exterior potential, and of local terms, which are fixed by the inner structure of the lens through matching. For odd $p$ the inner structure enters through the scale of a logarithm. The first five channel functions are given in Eq.~\eqref{eq:J15}. In dimensionless form, Eq.~\eqref{eq:FleadingDimless}, the higher channels are suppressed by powers of $x=GMk$, and the first Born term is small only for $y\gg1$.

Beyond first order, the eikonal limit shows two issues, Sec.~\ref{sec_compissues}. The first is that the Newtonian phase is not small, Eq.~\eqref{eq:breakdownPsi1}: it contains the logarithm of the distances of source and observer, which is large in astrophysical configurations even when $w\ll1$, so that all the Newtonian insertions must be retained. The second is that the Born terms are not ordered by their number of insertions, Eq.~\eqref{eq:breakdownCounting}: repeated interactions with lower non-Newtonian channels can be larger than a single interaction with a higher one, and the series must be truncated according to the size of its terms, Eq.~\eqref{eq:breakdownPrescription}.

The full second-order calculation of Sec.~\ref{sec:breakdown} shows that the first issue is not an artifact of the eikonal limit. A Newtonian insertion at small momentum transfer gives a logarithmic integral, which is cut off by the distances of source and observer, so that the mixed term with one Newtonian insertion is enhanced by $\ln(k^2r_{\rm S}r_{\rm O})$ with respect to the first-order term, Eq.~\eqref{eq:F21}. This logarithm is the Newtonian phase accumulated between the lens and the endpoints, Eq.~\eqref{eq:breakdownOuterPhase}. For Schwarzschild, the mixed term with one Newtonian and one inverse-square insertion exceeds the first Born term of the inverse-cube channel, of the same order in the mass, by many orders of magnitude, Eq.~\eqref{eq:finalratio}. Section~\ref{sec:beyondborn} removes the first issue by putting the Newtonian field into the reference Green function, while the second sets the accuracy of any subsequent truncation.

\section{Born series around the Newtonian solution}
\label{sec:beyondborn}

The preceding analysis shows why the Newtonian field should be included in the reference propagation: its accumulated effect need not be small even when the additional lens structure produces only a weak correction. We therefore expand around the exact Newtonian point-lens solution, retaining its interactions to all orders while treating the remaining potential perturbatively. After establishing the reference Green function and a physical picture of the response, we derive the dressed radial kernels in the Fraunhofer limit, making explicit how Newtonian propagation modifies the free-wave results and how short-distance sensitivity enters. We then retain wavefront curvature to determine when the scattering contribution alone is insufficient and recover the response of both wave branches. This connects the analytic kernels to diffraction near alignment and, where geometric optics applies, to the images of the reference lens.

\subsection{The Born series with the Newtonian solution as reference}
\label{sec:newtref}

To separate the lensing produced by the total mass from the effect of additional radial structure, we now include the Newtonian potential in the reference wave operator. The resulting Born series counts only interactions with the remaining potential, while its Green functions already contain arbitrarily many Newtonian interactions, incorporating the long-range phase without expanding it.

Hence, we choose the point lens as reference solution, $U_{\rm ref}=U_{\rm newt}$, where
\be
	U_{\rm newt}= \frac{c_1}{4r}= -\frac{GM}{r} \,.
\ee
The reference Green function $G_{\rm newt}$ and the reference wave $\chi_{\rm newt}$ then solve Eq.~\eqref{eq:Gref} with the Newtonian potential,
\begin{align}
	\left(\nabla^2+\omega^2-4\omega^2U_{\rm newt}\right)G_{\rm newt}(\rr,\rr') &=\delta^{(3)}(\rr-\rr'), \nonumber \\
	\label{eq:GN} \chi_{\rm newt}(\rr)&=G_{\rm newt}(\rr,\rr_{\rm S}).
\end{align}
The remaining potential, which is treated perturbatively, contains all the channels of Eq.~\eqref{eq:cp_expansion} except the Newtonian one,
\be \label{eq:UNsplit}
	\delta U_{\rm newt}(r)\equiv\Ueff(r)-U_{\rm newt}(r)=\frac14\sum_{p\geq2}\frac{c_p}{r^p}.
\ee
It describes the radial channels of the vacuum black hole example of Eq.~\eqref{eq:cpSchw}, but it also applies to a specified environmental profile, including fractional powers over finite radial intervals.

The Born series around the point lens is then Eq.~\eqref{eq:Fseries} with this reference,
\be \label{eq:FNseries}
	F=F^{(0)}_{\rm newt}+F^{(1)}_{\rm newt}+F^{(2)}_{\rm newt}+\cdots,
\ee
where, from Eqs.~\eqref{eq:Fref0}--\eqref{eq:Fref2} with $G_{\rm ref}=G_{\rm newt}$ and $\delta U_{\rm ref}=\delta U_{\rm newt}$,
\begin{align}
	\label{eq:F0newt} F^{(0)}_{\rm newt}={}&\frac{G_{\rm newt}(\rr_{\rm O},\rr_{\rm S})}{G_{\rm free}(\rr_{\rm O},\rr_{\rm S})}, \\
	F^{(1)}_{\rm newt}={}&\frac{4\omega^2}{G_{\rm free}(\rr_{\rm O},\rr_{\rm S})}\int\dd^3\rr\,G_{\rm newt}(\rr_{\rm O},\rr) \nonumber \\
	\label{eq:F1newt} &\times\delta U_{\rm newt}(r)\,G_{\rm newt}(\rr,\rr_{\rm S}), \\
	F^{(2)}_{\rm newt}={}&\frac{16\omega^4}{G_{\rm free}(\rr_{\rm O},\rr_{\rm S})}\int\dd^3\rr_1\,\dd^3\rr_2 \nonumber \\
	\label{eq:F2newt} &\times G_{\rm newt}(\rr_{\rm O},\rr_1)\,\delta U_{\rm newt}(r_1)\,G_{\rm newt}(\rr_1,\rr_2) \\
	&\times\delta U_{\rm newt}(r_2)\,G_{\rm newt}(\rr_2,\rr_{\rm S}). \nonumber
\end{align}
Each term of Eq.~\eqref{eq:FNseries} counts insertions of $\delta U_{\rm newt}$. Between any two such insertions the wave propagates with the Newtonian Green function $G_{\rm newt}$, so Newtonian interactions have already been included to all orders. This is the distorted-wave Born expansion: the reference wave is already lensed by the Newtonian potential before the additional interactions in $\delta U_{\rm newt}$ act. We nevertheless express the results in terms of the amplification factor $F$, defined with respect to the free wave, which contains the full effect of the lens.

\subsection{An exact reference solution, and the distorted wave approach}
\label{sec:exactref}

The distorted-wave approach, in which the Coulomb interaction is solved exactly and the remaining interactions are treated relative to the Coulomb waves (see e.g.\ Refs.~\cite{Gell-Mann:1953dcn,Kong:1999sf}), is convenient for two reasons. First, it resums the long-range $1/r$ potential to all orders, since $U_{\rm newt}$ is included in the reference and $\delta U_{\rm newt}$ starts from $p=2$, Eq.~\eqref{eq:UNsplit}. Second, the Green function of the Newtonian point lens, i.e. $G_{\rm newt}$ of Eq.~\eqref{eq:GN}, is known in closed form: it is the Hostler--Pratt solution. The problem was indeed addressed and solved by Hostler and Pratt~\cite{Hostler:1963zz,Hostler:1964zz} (see also \cite{Schwinger:1964zzb}) in the context of the Coulomb potential in quantum mechanics, where the scattering solution goes back to Gordon~\cite{Gordon:1928}. It has been recently applied to GW lensing in Ref.~\cite{Carrillo-Gonzalez:2026}.  It is the starting point for our generalizations. We write it between the source and the observer, and let
\be \label{eq:pointlensXY}
	X\equiv r_{\rm S}+r_{\rm O}+d_{\rm SO},\qquad Y\equiv r_{\rm S}+r_{\rm O}-d_{\rm SO},
\ee
with $d_{\rm SO}$ defined in Eq.~\eqref{eq:distances}. For the outgoing Coulomb Green function, with the regular condition at the origin, the Hostler--Pratt representation of the Green function is
\begin{align}
	G_{\rm newt}(\rr_{\rm O},\rr_{\rm S})={}& -\frac{\Gamma(1-\frac{\ii w}{2})}{4\pi\ii\omega d_{\rm SO}} (\partial_X-\partial_Y) \nonumber \\
	\label{eq:HP} &\times\left[ W_{\frac{\ii w}{2},\frac12}(-\ii\omega X) M_{\frac{\ii w}{2},\frac12}(-\ii\omega Y)\right].
\end{align}
The variables $X$ and $Y$ are differentiated as independent arguments before their geometric values are inserted. $W$ and $M$ are Whittaker functions; the outgoing prescription is understood through $\omega\to\omega+\ii0$ with the coupling continued consistently. This closed form sums the partial-wave representation of the same Coulomb reference problem \cite{Hostler:1963zz,Hostler:1964zz} (see also~\cite{Braga:2026bys} for an alternative derivation). The Green function between the source and a generic point $\rr$ follows by replacing $\rr_{\rm O}\to\rr$ in Eqs.~\eqref{eq:pointlensXY} and~\eqref{eq:HP}, so that $d_{\rm SO}\to d_{\rm S}(\rr)$, and similarly for the observer.

The free limit of this expression checks both normalization and phase. At $w=0$, $W_{0,1/2}(z)=\eu^{-z/2}$ and $M_{0,1/2}(z)=\eu^{z/2}-\eu^{-z/2}$, so that
\begin{align}
	(\partial_X-\partial_Y) [W_{0,1/2}(-\ii\omega X)&M_{0,1/2}(-\ii\omega Y)] \nonumber \\
	\label{eq:pointlensFreeCheck} &=\ii\omega\eu^{\ii\omega(X-Y)/2}.
\end{align}
Because $X-Y=2d_{\rm SO}$, Eq.~\eqref{eq:HP} reduces to the outgoing $G_{\rm free}$. This is a useful check: at $w=0$ the Newtonian potential term $4\omega^2U_{\rm newt} $ in Eq.~\eqref{eq:GN} vanishes, and the reference equation becomes the free one. As mentioned above, the use of this exact reference Green function in finite-distance GW lensing was discussed in Ref.~\cite{Carrillo-Gonzalez:2026}. Our next step is to extract a form convenient for the integrals containing $\delta U_{\rm newt}$.

As before, we focus on the limit in which source and observer are far from the lens. Since $\bn_{\rm S}\cdot\bn_{\rm O}=-\cos\theta$, Eq.~\eqref{eq:distances} gives $d_{\rm SO}^2=(r_{\rm S}+r_{\rm O})^2-4r_{\rm S}r_{\rm O}\sin^2(\theta/2)$. Moreover, $d_{\rm SO}\geq|r_{\rm S}-r_{\rm O}|$ by the triangle inequality. The two arguments of Eq.~\eqref{eq:pointlensXY} therefore satisfy
\begin{align}
	X&\geq2\max(r_{\rm S},r_{\rm O}), \nonumber \\
	\label{eq:pointlensXYsmall} Y&=\frac{4r_{\rm S}r_{\rm O}\sin^2(\theta/2)}{X}.
\end{align}
Hence $\omega X\gg1$ as soon as one of the two endpoints, either the source or the observer, is many wavelengths away from the lens, $\omega\max(r_{\rm S},r_{\rm O})\gg1$, whatever the position of the other. Instead, $\omega Y$ is not necessarily large: $Y$ is the difference between the length of the path through the lens and that of the straight path, and it vanishes at alignment, $\theta=0$. Therefore only the Whittaker function with argument $-\ii\omega X$ can be expanded for large argument, while the one with argument $-\ii\omega Y$ is kept exactly. For large $|z|$, the Whittaker function has the asymptotic expansion~\cite{DLMF}
\be \label{eq:pointlensWhittakerAsymptotic}
	W_{\frac{\ii w}{2},\frac12}(z)\simeq z^{\ii w/2}\,\eu^{-z/2}\left[1+\frac{\ii w(2-\ii w)}{4z}+\mathcal O\left(\frac{1}{z^2}\right)\right].
\ee
With $z=-\ii\omega X$, the first correction has modulus $w\sqrt{4+w^2}/(4\omega X)$, so that the leading term can be used when $(1+|w|)^2/(\omega X)\ll1$.

We now compute the derivatives in Eq.~\eqref{eq:HP}. From Eq.~\eqref{eq:pointlensWhittakerAsymptotic}, $\partial_XW=(\ii\omega/2)\,W\,[1+\mathcal O(w/\omega X)]$, where the correction comes from the derivative of the power $X^{\ii w/2}$. At leading order in $1/(\omega X)$, and writing $z=-\ii\omega Y$, so that $\partial_Y=-\ii\omega\,\dd/\dd z$, we then have
\begin{align}
	(\partial_X-\partial_Y)[WM]&\simeq W\left(\frac{\ii\omega}{2}M-\partial_YM\right) \nonumber \\
	\label{eq:pointlensDerivativeSplit} &=\ii\omega\,W\left(\frac{\dd M}{\dd z}+\frac M2\right).
\end{align}
To compute the derivative of $M$, we use its expression in terms of the confluent hypergeometric function~\cite{DLMF},
\be \label{eq:pointlensWhittakerIdentity}
	M_{\frac{\ii w}{2},\frac12}(z)=z\,\eu^{-z/2}\,{}_1F_1\Bigl(1-\frac{\ii w}{2};2;z\Bigr).
\ee
Differentiating, and using $\dd[z\,{}_1F_1(a;2;z)]/\dd z={}_1F_1(a;1;z)$, which follows from the series of ${}_1F_1$, we find
\begin{align}
	\frac{\dd M_{\frac{\ii w}{2},\frac12}}{\dd z}={}&\eu^{-z/2}\,{}_1F_1\Bigl(1-\frac{\ii w}{2};1;z\Bigr) \nonumber \\
	&-\frac z2\,\eu^{-z/2}\,{}_1F_1\Bigl(1-\frac{\ii w}{2};2;z\Bigr).
\end{align}
The second term cancels against $M/2$ in Eq.~\eqref{eq:pointlensDerivativeSplit}, so that $\dd M/\dd z+M/2=\eu^{-z/2}\,{}_1F_1(1-\ii w/2;1;z)=\eu^{z/2}\,{}_1F_1(\ii w/2;1;-z)$, where in the last step we used Kummer's transformation ${}_1F_1(a;b;z)=\eu^{z}\,{}_1F_1(b-a;b;-z)$. Inserting $z=-\ii\omega Y$, and the leading term of Eq.~\eqref{eq:pointlensWhittakerAsymptotic} for $W$, we obtain
\begin{align}
	(\partial_X-\partial_Y)[WM] \simeq{}&\ii\omega(-\ii\omega X)^{\ii w/2} \eu^{\ii\omega(X-Y)/2} \nonumber \\
	\label{eq:pointlensCombinedDerivative} &\times{}_1F_1\Bigl(\frac{\ii w}{2};1;\ii\omega Y\Bigr).
\end{align}
This step must be performed before dropping the position dependence in the fast phase of $W$. The combination $X-Y=2d_{\rm SO}$ gives the full free propagation phase, including the wavefront curvature.

Collecting the results so far, we obtain the useful asymptotic formula
\begin{align}
	G_{\rm newt}&(\rr_{\rm O},\rr_{\rm S})\simeq G_{\rm free}(\rr_{\rm O},\rr_{\rm S})\, \times \nonumber \\
	\label{eq:pointlensHypergeometricExactGeometry} &\times \Gamma\Bigl(1-\frac{\ii w}{2}\Bigr)(-\ii\omega X)^{\ii w/2} {}_1F_1\Bigl(\frac{\ii w}{2};1;\ii\omega Y\Bigr),
\end{align}
with the \emph{exact geometric} $X$, $Y$ and $d_{\rm SO}$ still retained.

We now compute the amplification factor of the point lens, $F^{(0)}_{\rm newt}$ of Eq.~\eqref{eq:F0newt}. Dividing Eq.~\eqref{eq:pointlensHypergeometricExactGeometry} by $G_{\rm free}(\rr_{\rm O},\rr_{\rm S})$ gives
\be \label{eq:pointlensRawAmplification}
	F^{(0)}_{\rm newt}\simeq \Gamma\left(1-\frac{\ii w}{2}\right) (-\ii\omega X)^{\ii w/2} \,{}_1F_1\left(\frac{\ii w}{2};1;\ii\omega Y\right),
\ee
where $X$ and $Y$ are evaluated at the positions of the source and of the observer, as in Eq.~\eqref{eq:pointlensXY}. No large-argument expansion of the hypergeometric function has been used. In particular, this expression keeps diffraction when $\omega Y$ is small.

Equations~\eqref{eq:pointlensHypergeometricExactGeometry} and~\eqref{eq:pointlensRawAmplification} are the leading term of an expansion in $1/(\omega X)$. The first correction is needed in Sec.~\ref{sec:v:f} and is obtained by keeping the second term of Eq.~\eqref{eq:pointlensWhittakerAsymptotic}, both in $W$ and in $\partial_X W$. To this order $\partial_X W=(\ii\omega/2)\,W\,[1-\ii w/z]$ with $z=-\ii\omega X$, and Eq.~\eqref{eq:pointlensDerivativeSplit} acquires the additional term $-(\ii w/2z)\,M$. Using Eq.~\eqref{eq:pointlensWhittakerIdentity} and Kummer's transformation we find
\begin{align}
	\Fzero \simeq{}& \Gamma\!\Big(1-\frac{\ii w}{2}\Big)\, (-\ii\omega X)^{\ii w/2} \nonumber \\
	\label{eq:F0-finite-distance} & \bigg\{\Big[1-\frac{w(2-\ii w)}{4\,\omega X}\Big]\, {}_1F_1\!\Big(\frac{\ii w}{2};1;\ii\omega Y\Big) \\
	& -\frac{\ii w}{2}\,\frac{Y}{X}\, {}_1F_1\!\Big(1+\frac{\ii w}{2};2;\ii\omega Y\Big)\bigg\} +\mathcal{O}\!\big((\omega X)^{-2}\big). \nonumber
\end{align}
At small angle $1/(\omega X)\simeq \varepsilon\,a(1-a)/2$ and $Y/X\simeq s\,\varepsilon\,a(1-a)/4$, so both corrections are of relative order $w\,\varepsilon$: they are finite-distance corrections to the point-lens wave, of first order in $\varepsilon$, and they start at \emph{linear} order in the mass. We denote by $\Fasym$ the leading form, Eq.~\eqref{eq:pointlensRawAmplification}.

To compare with the standard result, we separate the dependence on the ratio $X/Y$, which does not depend on frequency. For $Y>0$, we write $-\ii\omega X$ as the product of $X/Y$ and $-\ii\omega Y$. With the principal branch, the phase of $-\ii$ is $-\pi/2$, so that it contributes a factor $\eu^{\pi w/4}$ to the power $\ii w/2$, and we obtain
\be \label{eq:pointlensSeparatedPhase}
	(-\ii\omega X)^{\ii w/2}=\eu^{\ii(w/2)\ln(X/Y)}\,\eu^{\pi w/4}\,(\omega Y)^{\ii w/2}.
\ee
The first factor is a phase linear in $\omega$, i.e.\ a common arrival time of the whole signal, and it can be removed by a change of normalization of the amplification factor. In the small-angle limit, $X\simeq2(r_{\rm S}+r_{\rm O})$, and Eq.~\eqref{eq:pointlensXYsmall}, with Eqs.~\eqref{eq:reff}, \eqref{eq:kdef} and~\eqref{eq:bdef}, gives $Y\simeq k^2r_{\rm eff}/(2\omega^2)=b^2/(2r_{\rm eff})$, in agreement with Eq.~\eqref{eq:dSOsmallangle}, so that
\be \label{eq:XYsmallangle}
	\omega Y\simeq\frac{s}{2},\qquad \frac{X}{Y}\simeq\frac{4r_{\rm S}r_{\rm O}}{b^2}.
\ee
Since $s=kb$, by Eqs.~\eqref{eq:bdef} and~\eqref{eq:sepsx}, the second relation also gives
\be \label{eq:XYphase}
	(k^2r_{\rm S}r_{\rm O})^{\ii w/2}\,\eu^{-\ii(w/2)\ln(X/Y)}\simeq\left(\frac s2\right)^{\ii w},
\ee
which we use for the Born terms around the point lens. With this normalization, Eqs.~\eqref{eq:pointlensRawAmplification} and~\eqref{eq:pointlensSeparatedPhase} give
\begin{align}
	&\eu^{-\ii(w/2)\ln(X/Y)}F^{(0)}_{\rm newt} \nonumber \\
	\label{eq:FhatNewt} &\qquad\simeq\eu^{\pi w/4}\Gamma\left(1-\frac{\ii w}{2}\right) \left(\frac{s}{2}\right)^{\ii w/2} \,{}_1F_1\left(\frac{\ii w}{2};1;\frac{\ii s}{2}\right),
\end{align}
which is the standard point-lens amplification factor of wave-optics lensing~\cite{Takahashi:2003ix,Peters:1974gj,Deguchi:1986zz,Nakamura:1997sw}.

For $s\gg1+w^2$, the confluent hypergeometric function in Eq.~\eqref{eq:FhatNewt} has the large-argument expansion~\cite{DLMF}
\begin{align}
	{}_1F_1 &\lp\frac{\ii w}{2};1;\frac{\ii s}{2}\rp \simeq \nonumber \\
	\label{eq:pointlensKummerLarge} \simeq& \frac{(-\ii s/2)^{-\ii w/2}}{\Gamma(1-\ii w/2)}\left[1-\frac{\ii w^2}{2s}+\mathcal O\left(\frac{1}{s^2}\right)\right] + \\
	&+\frac{\eu^{\ii s/2}(\ii s/2)^{\ii w/2-1}}{\Gamma(\ii w/2)} \left[1+\mathcal O\left(\frac{1+w^2}{s}\right)\right], \nonumber
\end{align}
so that Eq.~\eqref{eq:FhatNewt} is split into two contributions,
\be \label{eq:pointlensLargeS}
	\eu^{-\ii(w/2)\ln(X/Y)}F^{(0)}_{\rm newt} \simeq1-\frac{\ii w^2}{2s}+B+\mathcal O(s^{-2}),
\ee
with
\be \label{eq:v:c:scattered-branch}
	B=\frac{\eu^{\ii s/2}}{y^2} \left(\frac s2\right)^{\ii w} \frac{\Gamma(1-\ii w/2)}{\Gamma(1+\ii w/2)},
\ee
where we used $\Gamma(\ii w/2)=\Gamma(1+\ii w/2)/(\ii w/2)$ and $w/s=1/y^2$.

The two contributions can be identified from their phases. Since $F$ is the wave divided by the free wave, which is proportional to $\eu^{\ii\omega d_{\rm SO}}$, a term of $F$ with phase $\omega\Delta$ arrives with a delay $\Delta$ with respect to a wave propagating along the straight line from the source to the observer. The first contribution in Eq.~\eqref{eq:pointlensLargeS} has no oscillating phase: up to the common delay removed by the normalization, it arrives together with the wave travelling along the straight line, and we call it the direct wave. The contribution $B$ carries instead the phase $\eu^{\ii s/2}$, and by Eq.~\eqref{eq:XYsmallangle} $s/2\simeq\omega Y=\omega(r_{\rm S}+r_{\rm O}-d_{\rm SO})$. The total phase of this wave is therefore $\omega(r_{\rm S}+r_{\rm O})$, which is the travel time from the source to the lens and from the lens to the observer: $B$ is the wave deflected by the point mass, and we call it the scattered wave. Its amplitude, $|B|=1/y^2$ for real $w$, is proportional to $1/k^2$, as the Rutherford amplitude for the scattering by a $1/r$ potential.

In the geometric-optics regime, $w\gg1$, the two contributions are the two images of the point lens. For $y\gg1$, using Stirling's formula for the gamma functions, the phase of $B$ becomes $\omega\Delta t-\pi/2$, where $\Delta t\simeq4GM(y^2/2+\ln y^2+1)$ is the time delay between the two images and $-\pi/2$ is the phase of the secondary image, while $|B|=1/y^2$ is the square root of the magnification of the secondary image, $|\mu_-|\simeq1/y^4$. At lower frequencies the two contributions remain wave contributions, without separately identifiable rays. Both the correction $-\ii w^2/(2s)$ and $B$ are of order $1/s$, since $1/y^2=w/s$, so that $1+B$ alone would omit a term of the same order. The expansion holds at fixed $w$; at large $w$ it also requires $s\gg w^2$.

\subsection{First-order Born around point lens: overall picture}
\label{sec:overview}

Before deriving the dressed kernels, we use the eikonal description of Sec.~\ref{sec:eikonal} to build an intuitive picture of how an additional radial potential changes a wave already lensed by the point mass. The picture has three messages. In geometric optics, the correction multiplies each image of the point lens by the extra phase $\ii\,\delta\Phi$ accumulated along its path, so that images passing closer to the lens are affected more. In wave optics, the images are not resolved, and the correction is $\ii\,\delta\Phi$ averaged over a region of the size of the Fresnel scale. Finally, the Fraunhofer approximation keeps only one of the two images, which explains why it fails near alignment.

We keep the Newtonian phase $\psi_1$ in the exponential of Eq.~\eqref{eq:overview-eikonal}, and expand in the other channels. Dropping the constant phase $(w/2)\ln(4r_{\rm S}r_{\rm O})$ of Eq.~\eqref{eq:eikonalShapiro}, the exponent becomes $\Phi(\boldsymbol\zeta)+\delta\Phi(\zeta)$, with\footnote{The scale $\zeta_{\rm ref}$ in $\Phi$ is kept for dimensional reasons. It only adds a constant phase, which can be pulled out of the integral and is fixed by the normalization.}
\begin{align}
	\label{eq:v:d:transverse-phase} \Phi(\boldsymbol\zeta)={}&\frac{\omega|\boldsymbol\zeta-\bm b|^2}{2r_{\rm eff}}-w\ln (\zeta / \zeta_{\rm ref}), \\
	\label{eq:v:d:additional-phase} \delta\Phi(\zeta)={}&\sum_{p\geq2}\psi_p(\zeta).
\end{align}
$\Phi$ is $\omega$ times the time delay of a path in the field of the point mass, and $\delta\Phi$ is $\omega$ times the additional delay produced by $\delta U_{\rm newt}$, with $\psi_p$ given in Eq.~\eqref{eq:v:d:delta-phi-p}. The first two terms of the Born series around the point lens are
\begin{align}
	\label{eq:pointlensDiffraction} F^{(0)}_{\rm newt}\simeq{}&\frac{\omega}{2\pi\ii r_{\rm eff}}\int\dd^2\zeta\,\eu^{\ii\Phi(\boldsymbol\zeta)}, \\
	\label{eq:v:d:F1-diffraction} F^{(1)}_{\rm newt}\simeq{}&\frac{\omega}{2\pi\ii r_{\rm eff}}\int\dd^2\zeta\,\eu^{\ii\Phi(\boldsymbol\zeta)}\,\ii\,\delta\Phi(\zeta).
\end{align}
The first is the diffraction integral of the point lens, which gives Eq.~\eqref{eq:FhatNewt} up to a constant phase. The second is the same integral, with each path weighted by $\ii\,\delta\Phi(\zeta)$.

In geometric optics, $w\gg1$, the integrals are dominated by the stationary points of $\Phi$, which are the two images of the point lens. Writing $\boldsymbol\zeta=R_E\boldsymbol\chi$, with $\chi$ the signed coordinate along $\bm b$, the condition $\nabla\Phi=0$ gives the lens equation and the image positions,
\be \label{eq:v:d:image-positions}
	\chi^2-y\chi-1=0,\qquad \chi_\pm=\frac{y\pm\sqrt{y^2+4}}2,
\ee
and the inverse of the determinant of the Hessian of $\Phi$, in units of $(\omega/r_{\rm eff})^2$, gives their magnifications,
\be \label{eq:v:d:magnifications}
	|\mu_\pm|=\frac{1}{|1-\chi_\pm^{-4}|}=\frac{y^2+2}{2y\sqrt{y^2+4}}\pm\frac12 .
\ee
Evaluating Eqs.~\eqref{eq:pointlensDiffraction} and~\eqref{eq:v:d:F1-diffraction} at the stationary points,
\begin{align}
	F^{(0)}_{\rm newt}\simeq{}&|\mu_+|^{1/2}\eu^{\ii\Phi(\zeta_+)}+|\mu_-|^{1/2}\eu^{\ii\Phi(\zeta_-)-\ii\pi/2}, \nonumber \\
	\label{eq:v:d:F1-saddle} F^{(1)}_{\rm newt}\simeq{}&\ii\,\delta\Phi(\zeta_+)\,|\mu_+|^{1/2}\eu^{\ii\Phi(\zeta_+)} \\
	&+\ii\,\delta\Phi(\zeta_-)\,|\mu_-|^{1/2}\eu^{\ii\Phi(\zeta_-)-\ii\pi/2}, \nonumber
\end{align}
where $-\pi/2$ is the Morse phase of the secondary image, which is a saddle point of $\Phi$. The first Born term is therefore the two images of the point lens, each multiplied by $\ii\,\delta\Phi$ evaluated at its position.

\begin{figure*}[t]
\centering
\includegraphics[width=\textwidth]{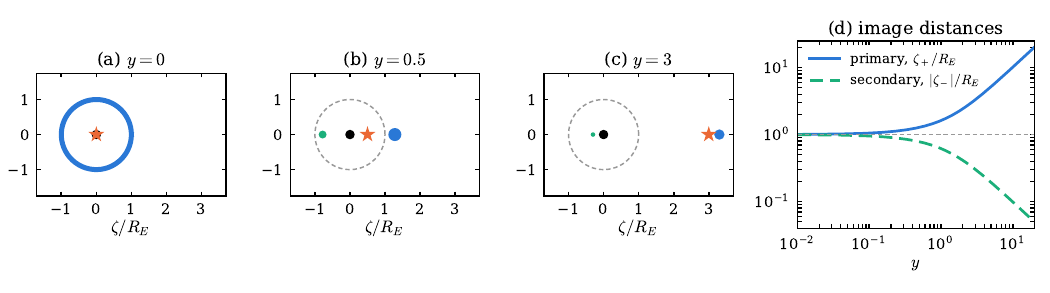}
\caption{Images of the point lens in the lens plane, in units of the Einstein radius, Eqs.~\eqref{eq:v:d:image-positions} and~\eqref{eq:v:d:magnifications}. The black dot is the lens, the star the position of the source, $\zeta=b=R_Ey$, the dashed circle the Einstein radius, and the area of each image is proportional to its magnification. (a) At alignment the two images merge into the Einstein ring. (b) Close to alignment the two images are at distances of order $R_E$ from the lens, with comparable magnifications. (c) Far from alignment the primary image (blue) is close to the position of the source, far from the lens, while the secondary image (green) is faint and close to the lens. (d) Distances of the two images from the lens as functions of $y$: $\zeta_+\simeq R_Ey$ and $|\zeta_-|\simeq R_E/y$ for $y\gg1$.}
\label{fig:v:images}
\end{figure*}

Since $\psi_p\propto\zeta^{1-p}$, the correction is larger for the image that passes closer to the lens, and this depends on the position of the source, Fig.~\ref{fig:v:images}. At alignment, $y=0$, both images are at the Einstein radius and form the Einstein ring. Far from alignment, $y\gg1$, the primary image passes far from the lens, at $\zeta_+\simeq R_Ey=b$, with $|\mu_+|\simeq1$, while the secondary image passes close to it, at $|\zeta_-|\simeq R_E/y$, and is faint, $|\mu_-|\simeq1/y^4$. For $p=2$, $\psi_2(\zeta)=-\pi\gamma_2wGM/(8\zeta)$, and with $GM/R_E=\alpha_E/4$,
\begin{align}
	y=0:&\quad\delta\Phi(R_E)=-\frac{\pi}{32}\gamma_2\,w\alpha_E, \nonumber \\
	\label{eq:v:d:image-phases} y\gg1:&\quad\delta\Phi(\zeta_+)\simeq-\frac{\pi}{32}\gamma_2\,\frac{w\alpha_E}{y}, \\
	&\quad\delta\Phi(\zeta_-)\simeq-\frac{\pi}{32}\gamma_2\,w\alpha_E\,y. \nonumber
\end{align}
Far from alignment, the phase of the secondary image grows with $y$, while that of the primary image decreases.

Whether the first Born term is enough depends on two different conditions. To describe the phase of each image correctly, its extra phase must be small, $|\delta\Phi(\zeta_\pm)|\ll1$. For the correction to the total wave to be small, it is enough that each image contributes little, i.e.
\be \label{eq:v:d:born-condition}
	|\mu_\pm|^{1/2}\,|\delta\Phi(\zeta_\pm)|\ll1.
\ee
Close to alignment the two conditions coincide, and require $|\delta\Phi(R_E)|\ll1$. Far from alignment, the primary image always satisfies them, since $\delta\Phi(\zeta_+)\propto y^{1-p}\to0$. For the secondary image, $\delta\Phi(\zeta_-)\propto y^{p-1}$ grows, but its amplitude $|\mu_-|^{1/2}\simeq1/y^2$ decreases, so that its contribution to Eq.~\eqref{eq:v:d:born-condition} scales as $y^{p-3}$. For $1<p<3$ the faintness wins: even when the phase of the secondary image is not small, as for $p=2$ when $y\gtrsim1/(w\alpha_E)$, the first Born term remains a good approximation for the total wave. For $p\geq3$ it does not: the potential grows close to the lens faster than the secondary image fades, and this image probes the region where the expansion in radial channels fails. This is the short-distance divergence of Sec.~\ref{sec:v:e}, which requires information on the inner structure of the lens. Note that the distances of source and observer enter only through the positions of the images, i.e.\ through $\alpha_E$ and $y$.

The same picture shows why the Born expansion around the free wave fails near alignment. Its first term in the Newtonian channel is of relative order $1/y^2$, Eq.~\eqref{eq:FleadingDimless}, i.e.\ of the order of the amplitude $|\mu_-|^{1/2}$ of the secondary image. It is a small correction only for $y\gg1$, when the secondary image is faint, and not when two images of comparable brightness form.

For $w\lesssim1$ the images are not resolved, and the correction is averaged over the Fresnel disk. The geometric term of $\Phi$ oscillates rapidly for $|\boldsymbol\zeta-\bm b|\gg r_{\rm F}=R_E/\sqrt w$, while the Shapiro term varies slowly, so that the integrals come from a disk of radius $r_{\rm F}$ around $\bm b$. The ratio $F^{(1)}_{\rm newt}/F^{(0)}_{\rm newt}$ is then approximately $\ii\,\delta\Phi$ averaged over this disk. Since $r_{\rm F}>R_E$, the disk extends to regions where $\delta\Phi$ is smaller than at the Einstein radius: at alignment, the relative correction of the $p=2$ channel is of order $\delta\Phi(r_{\rm F})\sim w^{3/2}\alpha_E$, instead of the geometric-optics value $w\alpha_E$ of Eq.~\eqref{eq:v:d:image-phases}. Far from alignment, $b\gg r_{\rm F}$, the disk around $\bm b$ does not contain the lens and gives the response of the direct wave, while the scattered wave $B$ of Eq.~\eqref{eq:v:c:scattered-branch} comes from a region of size $1/k$ around the lens.

Finally, the Fraunhofer approximation keeps only the secondary image. It drops the curvature term in $\omega|\boldsymbol\zeta-\bm b|^2/(2r_{\rm eff})=s/2-\kk\cdot\boldsymbol\zeta+\omega\zeta^2/(2r_{\rm eff})$, so that $\Phi\simeq s/2-\kk\cdot\boldsymbol\zeta-w\ln\zeta$. The lens equation~\eqref{eq:v:d:image-positions} then becomes $y\chi+1=0$, whose only solution, $\chi=-1/y$, is the secondary image: the primary image is lost. Physically, the Fraunhofer result is the amplitude for the wave to be deflected by the angle $\theta$, i.e.\ the Fourier transform of the potential at momentum $\kk$. The secondary image passes close to the lens and is deflected by $\theta$; the primary image passes far from it, is almost undeflected, and belongs to the forward wave, with zero momentum transfer. This is why the Fraunhofer result fails at $k=0$. Keeping the curvature of the wavefronts recovers the response of the primary image. In geometric optics, by Eq.~\eqref{eq:v:d:F1-saddle}, it is larger than that of the secondary image by $y^{2(2-p)}$: negligible for $p>2$, comparable for $p=2$, and dominant for $1<p<2$. In Sec.~\ref{sec:v:d} we find the same ratio in wave optics, in the form $s^{2-p}$.

We derive these results from Eq.~\eqref{eq:F1newt}, with the Green functions of Eq.~\eqref{eq:Cwave}, in the next subsections: in the Fraunhofer limit, for any scattering angle, in Sec.~\ref{sec:v:a}, including the singular channels in Sec.~\ref{sec:v:e}; in its small-angle limit in Sec.~\ref{sec:v:c}; and with the curvature of the wavefronts in Sec.~\ref{sec:v:d}, where we recover Eq.~\eqref{eq:v:d:F1-diffraction}.

\subsection{First-order Born approximation in the Fraunhofer limit}
\label{sec:v}
\label{sec:v:a}

We now turn this physical picture into a quantitative response to each radial channel, retaining the Newtonian interaction without expanding its strength or the scattering angle. The Fraunhofer approximation makes the calculation analytically tractable and reveals both how Newtonian propagation dresses the exterior response and where forward or short-distance sensitivity requires a more complete treatment. To do so, we compute the first Born term around the point lens, $F^{(1)}_{\rm newt}$ of Eq.~\eqref{eq:F1newt}, with $\delta U_{\rm newt}$ given in Eq.~\eqref{eq:UNsplit}. The two Green functions describe propagation from the source to an interaction point and then to the observer. There is one insertion of $\delta U_{\rm newt}$, with arbitrarily many Newtonian interactions on either side. Approximating these propagation legs is a separate decision from truncating the response at first order in $\delta U_{\rm newt}$. We use the geometry and the dimensionless variables of Section~\ref{sec:DimensionlessVariables}.

To begin with a simple case, we compute the first Born term around the point lens in the Fraunhofer limit, similarly to the computation of Section~\ref{sec:leading} for the Born expansion around the free wave. We assume that the lens is far from the source and from the observer, so that the wave that reaches the interaction region is a plane wave times a Coulomb modification, and we neglect finite-distance effects and the curvature of the wavefronts. The Coulomb factors are instead retained in full, so that neither the Newtonian strength nor the scattering angle is expanded.

In the first Born term, Eq.~\eqref{eq:F1newt}, we need the Green functions between the source and an interaction point $\rr$, and between $\rr$ and the observer. For the source leg, $G_{\rm newt}(\rr,\rr_{\rm S})$, Eq.~\eqref{eq:pointlensHypergeometricExactGeometry} holds with $\rr_{\rm O}$ replaced by $\rr$, so that $r_{\rm O}\to r$ and $d_{\rm SO}\to d_{\rm S}(\rr)$, and the two arguments $X$ and $Y$ of Eq.~\eqref{eq:pointlensXY} are replaced by $r+r_{\rm S}\pm d_{\rm S}(\rr)$. For $r\ll r_{\rm S}$, the expansion of $d_{\rm S}(\rr)$ in Eq.~\eqref{eq:distanceExpansion} gives
\begin{align}
	r+r_{\rm S}+d_{\rm S}(\rr)&=2r_{\rm S}+r-\bn_{\rm S}\cdot\rr +O(r^2/r_{\rm S}), \nonumber \\
	\label{eq:pointlensOneDistantEndpoint} r+r_{\rm S}-d_{\rm S}(\rr)&=r+\bn_{\rm S}\cdot\rr -\frac{r^2-(\bn_{\rm S}\cdot\rr)^2}{2r_{\rm S}} \\
	&\quad+O(r^3/r_{\rm S}^2). \nonumber
\end{align}
Approximating the first one by $2r_{\rm S}$ in the slowly varying Newtonian power, and keeping the leading term of the second one in the ${}_1F_1$, gives
\begin{align}
	G_{\rm newt}(\rr,\rr_{\rm S})\simeq{}& G_{\rm free}(\rr,\rr_{\rm S})\, \Gamma\left(1-\frac{\ii w}{2}\right) (-2\ii\omega r_{\rm S})^{\ii w/2} \nonumber \\
	\label{eq:Cwave} &\times{}_1F_1\left(\frac{\ii w}{2};1; \ii\omega(r+\bn_{\rm S}\cdot\rr)\right).
\end{align}
The observer leg, $G_{\rm newt}(\rr_{\rm O},\rr)$, is obtained in the same way, by replacing ${\rm S}\to{\rm O}$ in Eq.~\eqref{eq:Cwave}, and involves $r_{\rm O}$ and $\bn_{\rm O}$.

In the Green function of Eq.~\eqref{eq:Cwave}, and in its observer-side counterpart, we keep in the free Green function only the constant and linear terms of Eq.~\eqref{eq:distanceExpansion}, as in Eq.~\eqref{eq:planewaves}. This gives
\begin{align}
	G_{\rm newt}(\rr,\rr_{\rm S})\simeq{}& -\frac{\eu^{\ii\omega r_{\rm S}}}{4\pi r_{\rm S}} \Gamma\left(1-\frac{\ii w}{2}\right)(-2\ii\omega r_{\rm S})^{\ii w/2} \nonumber \\
	\label{eq:v:a:source-leg} &\times \eu^{\ii\kk_i\cdot\rr} {}_1F_1\left(\frac{\ii w}{2};1;\ii\omega(r+\bn_{\rm S}\cdot\rr)\right), \\
	G_{\rm newt}(\rr_{\rm O},\rr)\simeq{}&-\frac{\eu^{\ii\omega r_{\rm O}}}{4\pi r_{\rm O}}\, \Gamma\left(1-\frac{\ii w}{2}\right)(-2\ii\omega r_{\rm O})^{\ii w/2} \nonumber \\
	\label{eq:v:a:observer-leg} &\times \eu^{-\ii\kk_f\cdot\rr} {}_1F_1\left(\frac{\ii w}{2};1;\ii\omega(r+\bn_{\rm O}\cdot\rr)\right),
\end{align}
with the wavevectors of Eq.~\eqref{eq:kinkout}. The hypergeometric functions carry the focusing and scattering produced by the point lens, so that the reference wave is not merely a free wave multiplied by a phase. Let $\mathcal L$ be the size of the region around the lens that contributes to the integral in Eq.~\eqref{eq:F1newt}. The approximations made in Eqs.~\eqref{eq:Cwave}, \eqref{eq:v:a:source-leg} and~\eqref{eq:v:a:observer-leg} neglect phases of order $|w|\mathcal L/r_{\rm S,O}$, from replacing $r+r_{\rm S,O}+d_{\rm S,O}(\rr)$ by $2r_{\rm S,O}$ in the Newtonian power, and of order $\omega\mathcal L^2/r_{\rm S,O}$, from the quadratic terms of Eq.~\eqref{eq:distanceExpansion}; they also require the condition below Eq.~\eqref{eq:pointlensWhittakerAsymptotic} at both endpoints. For a contribution controlled by the momentum transfer, $\mathcal L\sim1/k$, these conditions become those of Eq.~\eqref{eq:farfieldConditions}: in particular, as for the free wave in Sec.~\ref{sec:leading}, the Fraunhofer approximation requires $s\gg1$. When the curvature phase is retained, as in Sec.~\ref{sec:v:d}, the region $\zeta\sim b$ around the unlensed line of sight also contributes, and there the first condition reads $|w|b/r_{\rm eff}=|w|k/\omega=4x\ll1$; moreover, the cubic terms of the expansion of $d_{\rm S,O}(\rr)$, which we always neglect, give phases of order $\omega\mathcal L^3/r_{\rm S,O}^2$, which must also be small. This does not prove that the integral over an unbounded power law is dominated by that region: another region can contribute at the same order, as we show in Sec.~\ref{sec:v:d}.

Inserting Eqs.~\eqref{eq:v:a:source-leg} and~\eqref{eq:v:a:observer-leg} in Eq.~\eqref{eq:F1newt}, we obtain, in analogy with Eq.~\eqref{eq:Fleading},
\be \label{eq:v:a:fraunhofer-response}
	F^{(1)}_{\rm newt}\simeq-\frac{\omega^2}{4\pi}\,\mathcal P\sum_pc_p\,\mathcal J^{\rm newt}_{p, {\rm Fraun}}(k),
\ee
where $\mathcal P$ is the endpoint factor of Eq.~\eqref{eq:endpointFactor}, the sum runs over the channels of $\delta U_{\rm newt}$ in Eq.~\eqref{eq:UNsplit}, and
\begin{align}
	\mathcal J^{\rm newt}_{p, {\rm Fraun}}(k)&=\Gamma\left(1-\frac{\ii w}{2}\right)^2 \nonumber \\
	&\times(-2\ii\omega r_{\rm S})^{\frac{\ii w}{2}}(-2\ii\omega r_{\rm O})^{\frac{\ii w}{2}} \nonumber \\
	\label{eq:v:a:channel-kernel} &\times\int\dd^3\rr\,\frac{\eu^{-\ii\kk\cdot\rr}}{r^p} \\
	&\times{}_1F_1\left(\frac{\ii w}{2};1;\ii\omega(r+\bn_{\rm S}\cdot\rr)\right) \nonumber \\
	&\times{}_1F_1\left(\frac{\ii w}{2};1;\ii\omega(r+\bn_{\rm O}\cdot\rr)\right). \nonumber
\end{align}
At $w=0$ the prefactor and the hypergeometric functions are equal to one, and $\mathcal J^{\rm newt}_{p, {\rm Fraun}}$ reduces to the free channel function $\mathcal J_p$ of Eq.~\eqref{eq:Jdef}. A profile that is a sum of channels is handled by summing their contributions.

To compute $\mathcal J^{\rm newt}_{p, {\rm Fraun}}$, we write the power law as a superposition of Yukawa potentials,
\be \label{eq:v:a:yukawa-representation}
	\frac1{r^p}=\frac1{\Gamma(p-1)} \int_0^\infty\mathrm{d}\lambda\,\lambda^{p-2} \frac{e^{-\lambda r}}r, \qquad p>1.
\ee
where $\lambda$ is an inverse range. Inserting it in Eq.~\eqref{eq:v:a:channel-kernel}, $\mathcal J^{\rm newt}_{p, {\rm Fraun}}$ becomes an integral over $\lambda$,
\begin{align}
	\mathcal J^{\rm newt}_{p, {\rm Fraun}}(k)={}& \Gamma\lp 1 - \frac{\ii w}{2}\rp^2 (-2\ii\omega r_{\rm S})^{\frac{\ii w}{2}}(-2\ii\omega r_{\rm O})^{\frac{\ii w}{2}} \nonumber \\
	\label{eq:v:a:weighted-yukawa} &\times \frac{1}{\Gamma(p-1)}\int_0^\infty\dd\lambda\,\lambda^{p-2}K(\lambda),
\end{align}
where we have defined
\begin{align}
	K(\lambda)\equiv{}&\int\dd^3\rr\,\frac{\eu^{-\lambda r}}{r}\,\eu^{-\ii\kk\cdot\rr} \times \nonumber \\
	\label{eq:v:a:yukawa-kernel} &\times{}_1F_1\left(\frac{\ii w}{2};1;\ii\omega(r+\bn_{\rm S}\cdot\rr)\right) \\
	&\times{}_1F_1\left(\frac{\ii w}{2};1;\ii\omega(r+\bn_{\rm O}\cdot\rr)\right). \nonumber
\end{align}
This changes the order of the calculation: the three-dimensional propagation problem is evaluated once in $K(\lambda)$, and the choice of radial channel enters only through its inverse-range weight. Small $\lambda$ samples the long-distance tail, and large $\lambda$ samples the inner region. Interestingly, the integral $K(\lambda)$ can be done analytically. Nordsieck's integral~\cite{Nordsieck:1954zz}, specialized to the elastic kinematics $|\kk_i|=|\kk_f|=\omega$ and $\kk=\kk_f-\kk_i$, gives
\be \label{eq:v:a:nordsieck}
	\begin{aligned}
		K(\lambda)={}&4\pi\lambda^{\mathrm{i} w-2} (\lambda-2\mathrm{i}\omega)^{-\mathrm{i} w} \\
		&\times{}_2F_1\!\left( 1-\frac{\mathrm{i} w}{2},1-\frac{\mathrm{i} w}{2}; 1;-\frac{k^2}{\lambda^2}\right).
	\end{aligned}
\ee
One derivation applies Euler integral representations to the two Kummer functions and performs the remaining Yukawa Fourier integral before the parameter integrations. An outgoing/Abel prescription is understood when oscillatory integrations are interchanged, and complex powers are continued consistently with that prescription. As a normalization check,
\be
	K(\lambda)\rightarrow\frac{4\pi}{\lambda^2+k^2} \qquad \mbox{for} \qquad w\rightarrow0\,,
\ee
which is the free Fourier transform of a Yukawa potential.

In Eq.~\eqref{eq:v:a:weighted-yukawa}, the factor $(\lambda-2\ii\omega)^{-\ii w}$ of Eq.~\eqref{eq:v:a:nordsieck} is the product of $(-2\ii\omega)^{-\ii w}$ and $[1-\lambda/(2\ii\omega)]^{-\ii w}$, and the first factor cancels the frequency dependence of the prefactor. Hence
\begin{align}
	\label{eq:v:a:one-dimensional-integral} \mathcal J^{\rm newt}_{p, {\rm Fraun}}(k)={}&\frac{4\pi\,\Gamma(1-\ii w/2)^2}{\Gamma(p-1)}\,(r_{\rm S}r_{\rm O})^{\ii w/2} I^{\rm newt}_p (k)
\end{align}
where we have defined
\begin{align}
	I^{\rm newt}_p(k)={}& \int_0^\infty\dd\lambda\,\lambda^{p+\ii w-4}\left(1-\frac{\lambda}{2\ii\omega}\right)^{-\ii w} \times \nonumber \\
	\label{eq:Ip} &\times{}_2F_1\!\left(1-\frac{\ii w}{2},1-\frac{\ii w}{2};1;-\frac{k^2}{\lambda^2}\right).
\end{align}
This is the central result of the reduction: a single inverse-range integral replaces the three-dimensional channel integral. It retains the full scattering-angle dependence and the unexpanded Newtonian strength within the Fraunhofer approximation, and at $w=0$ it gives back the free channel function $\mathcal J_p$.

The integral in Eq.~\eqref{eq:Ip} converges only for a range of $p$, and the ${}_2F_1$ plays a different role at the two ends. At large $\lambda$, which corresponds to small radii, its argument $-k^2/\lambda^2$ tends to zero and ${}_2F_1\to1$, so that the integrand behaves as $(-2\ii\omega)^{\ii w}\lambda^{p-4}$, and the integral converges for $p<3$. At small $\lambda$, which corresponds to large radii, the argument is large and negative, and the ${}_2F_1$ decreases as $(k^2/\lambda^2)^{-1+\ii w/2}$, up to a logarithm of $\lambda/k$ when $w\neq0$~\cite{DLMF}. This factor, proportional to $\lambda^{2-\ii w}$, turns the power $\lambda^{p+\ii w-4}$ into $\lambda^{p-2}$, so that the integral converges for $p>1$. The ${}_2F_1$ therefore regularizes the integral at large radii: it comes from the plane-wave factor $\eu^{-\ii\kk\cdot\rr}$ in Eq.~\eqref{eq:v:a:channel-kernel}, whose oscillations cancel the contribution of the tail of the potential at $r\gg1/k$. Hence $\mathcal J^{\rm newt}_{p, {\rm Fraun}}$ is finite for $1<p<3$ and $k>0$, as the free channel functions of Section~\ref{sec:leading}. The channels with $p\geq3$ are singular at the origin and require the subtraction discussed in Sec.~\ref{sec:v:e}.

At $k=0$, i.e.\ for $\theta=0$, instead, the ${}_2F_1$ is equal to one for all $\lambda$, and there is nothing to regularize the integral at small $\lambda$: the integrand behaves as $\lambda^{p-4+\ii w}$, and the integral diverges for $p<3$. Equivalently, for $k>0$ the ${}_2F_1$ cuts the integral at $\lambda\sim k$, so that the integral scales as $k^{p-3}$, which diverges for $k\to0$. This divergence has the same origin as for the free channel functions, which scale as $k^{p-3}$, Eq.~\eqref{eq:jp}: at $k=0$ the plane-wave factor in Eq.~\eqref{eq:v:a:channel-kernel} is equal to one, so that all points of the potential contribute with the same phase, and the volume integral of the tail $1/r^p$ diverges at large radii for $p<3$. Resumming the Newtonian potential does not remove this divergence, because it comes from the tail of $\delta U_{\rm newt}$ and not from the $1/r$ potential: as we show below, the Newtonian dressing only multiplies the free scaling $k^{p-3}$ by the phase $k^{\ii w}$. In the full problem the large radii are cut off by the quadratic phase of Eq.~\eqref{eq:distanceExpansion}, which we neglected, i.e.\ at the Fresnel scale. The divergence is therefore an artifact of the Fraunhofer approximation, which, as recalled above, requires $s\gg1$ and is not valid at $k=0$. As discussed in Sec.~\ref{sec:overview}, this failure has a simple interpretation: the Fraunhofer limit keeps only the response of the secondary image of the point lens, and misses the primary image, whose response is at zero momentum transfer. We retain the wavefront curvature in Sec.~\ref{sec:v:d}.

\subsubsection{Convergent channels: $1<p<3$}
\label{sec:v:b}
For $1<p<3$, we compute the integral $I^{\rm newt}_p$ of Eq.~\eqref{eq:Ip} exactly, at any scattering angle and without expanding in $w$. A Mellin--Barnes representation of the ${}_2F_1$ in Eq.~\eqref{eq:Ip} converts $I^{\rm newt}_p$ into a contour integral, whose gamma-function poles organize the result into three residue families: poles at even and odd nonnegative values of the contour variable generate two families, while poles at $t=3-p-\ii w+2n$ generate a third. Accordingly, we write
\be
	I^{\rm newt}_p= I_p^{\rm I}+I_p^{\rm II}+I_p^{\rm III}
\ee
These three terms are parts of the same spatial kernel. Their separation is useful because their dependence on $k$ differs; it is not a separation into independent scattering processes or disjoint spatial regions. The three contributions are
\begin{align}
	I_p^{\rm I}={}&P_p^{\rm I}\times{}_4F_3\!\left(\begin{matrix} \frac{p-1}2,\frac{p-1}2,\frac{1+\ii w}2,\frac{\ii w}2 \\ \frac12,\frac{p-1+\ii w}2,\frac{p-1+\ii w}2 \end{matrix};\frac{k^2}{4\omega^2}\right), \nonumber \\[0.6em]
	\label{eq:v:b:three-families} I_p^{\rm II}={}&P_p^{\rm II}\times{}_4F_3\!\left(\begin{matrix} \frac p2,\frac p2,\frac{1+\ii w}2,\frac{2+\ii w}2 \\ \frac32,\frac{p+\ii w}2,\frac{p+\ii w}2 \end{matrix};\frac{k^2}{4\omega^2}\right), \\[0.6em]
	I_p^{\rm III}={}&P_p^{\rm III}\times{}_4F_3\!\left(\begin{matrix} \frac{3-p}2,\frac{4-p}2,\frac{2-\ii w}2,\frac{2-\ii w}2 \\ 1,\frac{4-p-\ii w}2,\frac{5-p-\ii w}2 \end{matrix};\frac{k^2}{4\omega^2}\right), \nonumber
\end{align}
with
\begin{align}
	P_p^{\rm I} &= k^{p-3+\ii w} \times \frac{ \Gamma\!\left(\frac{p-1}{2}\right)^2 \Gamma\!\left(\frac{3-p-\ii w}{2}\right)}{ 2\Gamma\!\left(1-\frac{\ii w}{2}\right)^2 \Gamma\!\left(\frac{p-1+\ii w}{2}\right)}, \nonumber \\[0.7em]
	\label{eq:v:b:prefactors} P_p^{\rm II} &= k^{p-3+\ii w} \times \lp \frac{k}{4\omega} \rp \, \frac{ w \Gamma\!\left(\frac p2\right)^2 \Gamma\!\left(\frac{2-p-\ii w}{2}\right)}{ \Gamma\!\left(1-\frac{\ii w}{2}\right)^2 \Gamma\!\left(\frac{p+\ii w}{2}\right)}, \\[0.7em]
	P_p^{\rm III} &=(-2\ii\omega)^{p-3+\ii w} \frac{\Gamma(p-3+\ii w)\Gamma(3-p)}{\Gamma(\ii w)}. \nonumber
\end{align}
The two-row notation specifies the parameters of a generalized hypergeometric series~\cite{DLMF}; it is not a matrix operation:
\begin{align}
	{}_4F_3\!\left(\begin{matrix} a_1,a_2,a_3,a_4 \\ b_1,b_2,b_3 \end{matrix};x\right) ={}&\sum_{n=0}^{\infty} \frac{\prod_{j=1}^{4}(a_j)_n}{\prod_{j=1}^{3}(b_j)_n} \times\frac{x^n}{n!}
\end{align}
where $(a)_n$ are the Pochhammer symbols defined as $(a)_n \equiv \Gamma(a + n)/\Gamma(a)$. The argument of the hypergeometric functions in Eq.~\eqref{eq:v:b:three-families} is $k^2/(4\omega^2)=\sin^2(\theta/2)$. Since a hypergeometric function is equal to one when its argument vanishes, each of them is $1+\mathcal O(k^2/\omega^2)$, and contains all the finite-angle corrections of its family.

The dependence on $k$ at small angle is instead carried by the prefactors. $P_p^{\rm I}$ scales as $k^{p-3+\ii w}$, and $P_p^{\rm II}$ is $P_p^{\rm I}$ times a further factor $wk/\omega=4x$, with $x=GMk$ of Eq.~\eqref{eq:sepsx}, and a ratio of gamma functions, which is of order one except near $p=2$, where it is of order $1/w$. $P_p^{\rm III}$ does not depend on $k$, so that $I_p^{\rm III}$ is a power series in $k^2$. Since $P_p^{\rm I}$ grows as $k^{p-3}$ when $k\to0$, while $P_p^{\rm III}$ stays constant, their ratio is of order $(k/\omega)^{3-p}$. Hence, for $k/\omega\ll1$, $I_p^{\rm II}$ and $I_p^{\rm III}$ are suppressed with respect to $I_p^{\rm I}$ at least by $k/\omega$ and by $(k/\omega)^{3-p}$, respectively. These estimates hold at fixed $p$ and $w$, and the gamma functions can change them when $w$ is large or $p$ is close to $2$ or $3$.

Note that the small-angle regime is compatible with the Fraunhofer approximation. The small-angle limit requires $k/\omega\ll1$, i.e.\ a small argument $k^2/(4\omega^2)$ of the hypergeometric functions in Eq.~\eqref{eq:v:b:three-families}. The Fraunhofer approximation requires $s\gg1$, which in terms of the same ratio reads $(k/\omega)^2\gg\varepsilon$, since $s=(k/\omega)^2/\varepsilon$ by Eq.~\eqref{eq:dimlessIdentities}. Since $\varepsilon$ is extremely small for astrophysical distances, there is a wide window $\sqrt\varepsilon\ll k/\omega\ll1$ in which both hold. Only for $k/\omega\lesssim\sqrt\varepsilon$, which includes alignment, $k=0$, the Fraunhofer result is not valid, and there $P_p^{\rm I}$ diverges for $p<3$ and $P_p^{\rm II}$ for $p<2$, which is the divergence discussed above.

The family $I_p^{\rm III}$ comes from inverse ranges $\lambda\sim\omega$ in Eq.~\eqref{eq:Ip}, i.e.\ from radii of the order of the wavelength, $r\sim1/\omega$, as shown by its prefactor $P_p^{\rm III}\propto\omega^{p-3}$. If $\delta U_{\rm newt}$ is the weak-field expansion of a black-hole metric, which holds for $r\gg GM$, this contribution is therefore reliable only for $1/\omega\gg GM$, i.e.\ $w\ll1$. This limitation does not affect the small-angle results of Sec.~\ref{sec:v:c}, where $I_p^{\rm III}$ is subleading.

The free limit provides a check of Eqs.~\eqref{eq:v:b:three-families} and~\eqref{eq:v:b:prefactors}. At $w=0$, the hypergeometric function in $I_p^{\rm I}$ is equal to one, since one of its upper parameters vanishes, and $P_p^{\rm I}$ tends to $k^{p-3}\Gamma((p-1)/2)\Gamma((3-p)/2)/2$. For $p\neq2$, $P_p^{\rm II}$ and $P_p^{\rm III}$ vanish at $w=0$, because $P_p^{\rm II}$ is proportional to $w$ and $P_p^{\rm III}$ to $1/\Gamma(\ii w)$. Hence
\be
	\lim_{w\to0}I^{\rm newt}_p=\frac{k^{p-3}}{2}\,\Gamma\left(\frac{p-1}{2}\right)\Gamma\left(\frac{3-p}{2}\right),
\ee
and Eq.~\eqref{eq:v:a:one-dimensional-integral} gives the same $\mathcal J_p$ of the free channel function of Eq.~\eqref{eq:jp}.\footnote{Equation~\eqref{eq:jp} contains $\Gamma(2-p)\sin[\pi(2-p)/2]$, which, by the reflection and duplication formulas of the gamma function, is equal to $\Gamma((p-1)/2)\Gamma((3-p)/2)/[2\Gamma(p-1)]$.} At $p=2$, instead, the poles of $\Gamma((2-p-\ii w)/2)$ in $P_p^{\rm II}$ and of $\Gamma(p-3+\ii w)$ in $P_p^{\rm III}$ compensate their vanishing factors, so that $I_2^{\rm II}$ and $I_2^{\rm III}$ do not vanish separately at $w=0$. As an example, for $p=2$ coincident hypergeometric parameters reduce the three families to lower-order functions,
\begin{align}
	I^{\rm newt}_2={}&\frac{\pi k^{\ii w-1}\Gamma((1-\ii w)/2)} {2\Gamma(1-\ii w/2)^2\Gamma((1+\ii w)/2)} \nonumber \\
	&\times{}_2F_1\!\left(\frac{\ii w}{2},\frac12;\frac{1+\ii w}2;\frac{k^2}{4\omega^2}\right) \nonumber \\
	&+\frac{\ii k^{\ii w}}{2\omega}\frac{\sinh(\pi w/2)}{\pi w/2} \nonumber \\
	\label{eq:v:b:p-two} &\times{}_3F_2\!\left(1,1,\frac{1+\ii w}2;\frac32,1+\frac{\ii w}{2};\frac{k^2}{4\omega^2}\right) \\
	&+\frac{(-2\ii\omega)^{\ii w-1}}{\ii w-1} \nonumber \\
	&\times{}_2F_1\!\left(\frac12,1-\frac{\ii w}{2};\frac{3-\ii w}2;\frac{k^2}{4\omega^2}\right), \nonumber
\end{align}
where the three lines are $I_2^{\rm I}$, $I_2^{\rm II}$ and $I_2^{\rm III}$, and the ratio $\sinh(\pi w/2)/(\pi w/2)$ is understood by continuity at $w=0$. For $w\to0$ the first line tends to $\pi/(2k)$, in agreement with the general result and with $\mathcal J_2=2\pi^2/k$ of Eq.~\eqref{eq:J15}, while the second and third lines tend to $+\ii/(2\omega)$ and $-\ii/(2\omega)$ times the same function, ${\rm artanh}[k/(2\omega)]/[k/(2\omega)]$, and cancel exactly. Hence $I_2^{\rm III}$ cannot be omitted from the all-angle result.

\subsubsection{Forward and backward scattering}
At $\theta=0$, $k=0$, source, lens and observer are aligned, and the point lens forms an Einstein ring. There the Fraunhofer kernel diverges. For $k\neq0$, the plane-wave factor in Eq.~\eqref{eq:v:a:channel-kernel} oscillates at large radii, and the contributions of the tail of the potential cancel. At $k=0$ this factor is equal to one, all radii contribute with the same phase, and the integral of the tail diverges for $p<3$. This is a large-distance divergence, unrelated to the short-distance one of the singular channels, and it is removed by the curvature of the wavefronts, Sec.~\ref{sec:v:d}. The point-lens amplification $F^{(0)}_{\rm newt}$ of Eq.~\eqref{eq:pointlensRawAmplification}, instead, is finite at alignment, where diffraction regulates the divergent magnification of the Einstein ring.

At $\theta=\pi$, $k=2\omega$, the wave is scattered backwards, with source and observer on the same side of the lens. The integral $I^{\rm newt}_p$ of Eq.~\eqref{eq:Ip} is finite there, but each of the three families of Eq.~\eqref{eq:v:b:three-families} diverges separately, as $-\ln[1-\sin^2(\theta/2)]$, because their hypergeometric series converge too slowly at argument one. The three divergences cancel in the sum, so that at $\theta=\pi$ one should use the integral of Eq.~\eqref{eq:Ip}, or the limit of the complete sum, rather than the separate families.

The Mellin--Barnes representation is derived for complex values of $\ii w$ with $0<{\rm Re}(\ii w)<3-p$, where the contour is well defined, and the result is then continued analytically to imaginary $\ii w$, i.e.\ real $w$. This continuation is exact and is not an expansion in $w$. When gamma-function poles in Eq.~\eqref{eq:v:b:prefactors} coincide, as for $p=2$, the families must be combined before taking the limit.

\subsubsection{Singular channels: $p\geq3$}
\label{sec:v:e}
As for the Born expansion around the free wave, Sec.~\ref{sec:leading}, the channel functions diverge for $p\geq3$, and the origin of the divergence is the same: near the origin the Coulomb waves are regular, so that the integrand of Eq.~\eqref{eq:v:a:channel-kernel} behaves as $1/r^p$, and the radial expansion~\eqref{eq:cp_expansion} has been extended to the centre of the lens, where it does not apply. This divergence comes from small radii, and is therefore different from the one at $k=0$ discussed above, which comes from large radii. We renormalize it as in Sec.~\ref{sec:leading}. In the representation of Eq.~\eqref{eq:v:a:yukawa-representation}, large inverse ranges $\lambda$ correspond to small radii, so that we cut the integral at an inverse range $\Lambda$, which plays the role of $1/r_0$, with
\be
	\omega,\,k\ll\Lambda\ll\frac{1}{GM}.
\ee
For $\lambda<\Lambda$ we use the radial channel, while the region $\lambda>\Lambda$ must be described by the inner completion of the lens. As in Eq.~\eqref{eq:Jtot}, the channel function is the sum of the two contributions,
\be \label{eq:v:e:Jtot}
	\mathcal J^{\rm newt,tot}_p(k)=\mathcal J^{{\rm newt},(\Lambda)}_p(k)+\mathcal J^{\rm newt,UV}_p(k),
\ee
and, since $\Lambda$ is arbitrary, the dependence of $\mathcal J^{\rm newt,UV}_p$ on $\Lambda$ must cancel that of $\mathcal J^{{\rm newt},(\Lambda)}_p$. We first compute the exterior contribution, and then the inner one.

In the exterior region, $\mathcal J^{{\rm newt},(\Lambda)}_p$ is given by Eq.~\eqref{eq:v:a:one-dimensional-integral} with $I^{\rm newt}_p$ replaced by
\be \label{eq:v:e:Jcutdef}
	I^{{\rm newt},(\Lambda)}_p(k)\equiv\int_0^\Lambda\dd\lambda\,f_p(\lambda),
\ee
where $f_p$ is the integrand of Eq.~\eqref{eq:Ip}. To isolate the divergence, we expand $f_p$ at large $\lambda$. Its three factors behave as follows: $\lambda^{p+\ii w-4}$ is a power; $[1-\lambda/(2\ii\omega)]^{-\ii w}$ is equal to $(-2\ii\omega)^{\ii w}\lambda^{-\ii w}(1-2\ii\omega/\lambda)^{-\ii w}$, which we expand in powers of $2\ii\omega/\lambda$; and the ${}_2F_1$ is a series in powers of $-k^2/\lambda^2$. Hence, for $\lambda>2\omega$ and $\lambda>k$,
\be \label{eq:v:e:tail-expansion}
	f_p(\lambda)=(-2\ii\omega)^{\ii w}\sum_{j,n\geq0}a_{jn}\,\lambda^{p-4-j-2n},
\ee
with
\be \label{eq:v:e:ajn}
	a_{jn}=\frac{(\ii w)_j}{j!}\,\frac{(1-\ii w/2)_n^2}{(n!)^2}\,(2\ii\omega)^j(-k^2)^n.
\ee
Since this expansion holds only for $\lambda>2\omega$ and $\lambda>k$, it cannot be integrated over the whole range of Eq.~\eqref{eq:v:e:Jcutdef}, and the integral has no closed form for finite $\Lambda$. We therefore write the integral from $0$ to $\Lambda$ as the integral from $0$ to $\infty$ minus the integral from $\Lambda$ to $\infty$. The first is $I^{\rm newt}_p$ of Eq.~\eqref{eq:Ip}: it does not depend on $\Lambda$, and is given by the closed form of Sec.~\ref{sec:v:b}, the sum of the three families of Eq.~\eqref{eq:v:b:three-families}, continued analytically in $p$. In the second, $\lambda>\Lambda$ is larger than $2\omega$ and $k$, so that the expansion~\eqref{eq:v:e:tail-expansion} converges over the whole range of integration and can be integrated term by term. As long as no term is proportional to $1/\lambda$, i.e.\ for $p\neq3+j+2n$, each term gives a power $\Lambda^{p-3-j-2n}/(p-3-j-2n)$, and
\begin{align}
	I^{{\rm newt},(\Lambda)}_p={}&I^{\rm I}_p+I^{\rm II}_p+I^{\rm III}_p \nonumber \\
	\label{eq:v:e:cutoff-identity} &+(-2\ii\omega)^{\ii w}\sum_{j,n\geq0}a_{jn}\,\frac{\Lambda^{p-3-j-2n}}{p-3-j-2n}.
\end{align}
This is the analogue of Eq.~\eqref{eq:Jcut-exact}. None of the three families depends on $\Lambda$. The families $I^{\rm I}_p$ and $I^{\rm II}_p$ are not analytic in $k^2$, and are determined by the exterior potential, while $I^{\rm III}_p$ is a power series in $k^2$. All the dependence on $\Lambda$ is in the double sum, whose terms are polynomials in $k^2$. The terms with $j+2n<p-3$ are the power divergences, while those with $j+2n>p-3$ vanish for $\Lambda\to\infty$.

The analogue of the logarithmic channels $p=2n+3$ of Sec.~\ref{sec:leading} is the case in which a term of Eq.~\eqref{eq:v:e:tail-expansion} is proportional to $1/\lambda$, i.e.\ $p=3+j+2n$. Its integral then gives a logarithm instead of a power, and Eq.~\eqref{eq:v:e:cutoff-identity} does not apply. For the first integer channels, the pairs $(j,n)$ that give a logarithm are $(0,0)$ for $p=3$, $(1,0)$ for $p=4$, and $(2,0)$ and $(0,1)$ for $p=5$. At $w=0$, all the coefficients with $j\geq1$ vanish, since the Pochhammer symbol $(\ii w)_j=\ii w(\ii w+1)\cdots(\ii w+j-1)$ contains the factor $\ii w$. Only the pairs $(0,n)$ survive, which give a logarithm for $p=3+2n$: as for the free channels, logarithms appear only for odd $p$, where the Fourier transform of $r^{-p}$, defined as a distribution, has its poles~\cite{Frank:1971xx}; similarly, in the Born series for scattering off a black hole the first ultraviolet divergence of the partial wave $\ell$ appears for a $1/r^{3+2\ell}$ term of the potential~\cite{Caron-Huot:2025tlq}. For $w\neq0$, instead, they appear at every integer $p\geq3$, and the Newtonian dressing introduces logarithms in channels that did not have them. These come from the terms with $j\geq1$, i.e.\ from the modification of the reference waves close to the lens. At first order in $w$, the two ${}_1F_1$ in Eq.~\eqref{eq:v:a:channel-kernel} give $1-w\omega r$ near the origin, up to terms proportional to $\kk\cdot\rr$. Multiplied by $1/r^p$, the correction $-w\omega r$ acts as an additional channel $1/r^{p-1}$. For $p=4$, this is a $p=3$ term, which has a logarithmic divergence with the coefficient $a_{10}=-2w\omega$. The Newtonian potential is also extrapolated to $r\to0$ in the reference wave, but this alone does not produce a divergence, because $1/r$ is integrable and the Coulomb waves are regular at the origin.

In the logarithmic case, the integral $I^{{\rm newt},(\Lambda)}_p$ is finite, while the terms of Eq.~\eqref{eq:v:e:cutoff-identity} are not separately finite. When $p$ approaches $3+j+2n$, the coefficient $1/(p-3-j-2n)$ of the corresponding term of the double sum diverges, and $I^{\rm III}_p$ has a pole, from the gamma functions in $P^{\rm III}_p$, which cancels it. $I^{\rm III}_p$ is the only family with poles there, while $I^{\rm I}_p$ and $I^{\rm II}_p$ are finite and can simply be evaluated at that value of $p$. To write the constant that accompanies the logarithm, we therefore write $I^{\rm III}_p$ as a sum, $I^{\rm III}_p=\sum_{n\geq0}c_n(p)$, with
\begin{align}
	c_n(p){}&= (-2\ii\omega)^{p+\ii w-3} \frac{ \lp 1-\frac{\ii w}{2}\rp _n^2\,}{(n!)^2}\, \times \nonumber \\
	\label{eq:v:e:cn} &\times\frac{\Gamma(p+\ii w-3-2n)\,\Gamma(3-p+2n)}{\Gamma(\ii w)} \lp \frac{k^2}{4\omega^2} \rp^n.
\end{align}
At an integer $p\geq3$, the terms $c_n$ with $2n\leq p-3$ are those with a pole, and each of them combines with the term of the double sum with the pair $(p-3-2n,n)$; the terms with $2n>p-3$ are finite. Taking the limit, we find
\begin{align}
	{}&I^{{\rm newt},(\Lambda)}_p=I^{\rm I}_p+I^{\rm II}_p+\sum_{\substack{n\geq0\\2n>p-3}}c_n(p) + \nonumber \\
	\label{eq:v:e:Jlog} &~~+(-2\ii\omega)^{\ii w}\!\!\!\!\!\!\!\sum_{\substack{j,n\geq0\\ j+2n\neq p-3}}\!\!\!\!\!\!\!a_{jn}\,\frac{\Lambda^{p-3-j-2n}}{p-3-j-2n} + \\
	&~~+(-2\ii\omega)^{\ii w}\!\!\!\!\!\!\!\sum_{\substack{j,n\geq0\\ j+2n=p-3}}\!\!\!\!\!\!\!a_{jn}\Big[\ln\frac{\ii \Lambda}{2\omega} -\psi(j+\ii w)-\gamma_{\rm E}+H_j\Big], \nonumber
\end{align}
where $\psi$ is the digamma function, and $\gamma_{\rm E}-H_j$ comes from the expansion of $\Gamma(3-p+2n)$ around its pole. The second line contains the power divergences, $j+2n<p-3$, and the terms that vanish for $\Lambda\to\infty$, $j+2n>p-3$, while the last line contains the pairs with $j+2n=p-3$, which give the logarithm. This is the analogue of Eq.~\eqref{eq:Jcut}. At $w=0$, Eqs.~\eqref{eq:v:e:cutoff-identity} and~\eqref{eq:v:e:Jlog} give the free channels with the cutoff on the inverse range instead of the radius, e.g.\ $4\pi\ln(\Lambda/k)$, $2\pi\Lambda-\pi^2k$ and $(2\pi/3)[\Lambda^2/2-k^2\ln(\Lambda/k)]$ for $p=3,4,5$, to be compared with Eq.~\eqref{eq:J345cut}.

We now turn to the inner region, $\lambda>\Lambda$, i.e.\ to radii much smaller than the wavelength. There the Newtonian Green functions can be expanded around their values at the origin, $G_{\rm newt}(\rr_{\rm O},\rr)G_{\rm newt}(\rr,\rr_{\rm S})\simeq G_{\rm newt}(\rr_{\rm O},0)G_{\rm newt}(0,\rr_{\rm S})\,(1-w\omega r+\dots)$. The inner region therefore acts as in the free case, Eqs.~\eqref{eq:inner-series} and~\eqref{eq:local-terms}, as a set of local terms, now multiplied by the Newtonian Green functions at the origin, as for a contact interaction between Coulomb waves~\cite{Kong:1999sf}, as the horizon term of Eq.~\eqref{eq:v:f:horizon-explicit}; the corrections in powers of $\omega r$ give local terms proportional to powers of $\omega$. The terms of Eq.~\eqref{eq:v:e:Jlog} that depend on $\Lambda$ are polynomials in $k^2$ times powers of $\omega$, as are the local terms, which cancel them as in Eq.~\eqref{eq:ghat}. For $p=3+j+2n$, the logarithm $\ln\Lambda$ is absorbed in the local coefficient, which introduces the scale $\bar\mu$ as in Eq.~\eqref{eq:MSfree}: we replace $\ln\Lambda\to\ln\bar\mu+\kappa_p$, with $\kappa_{2n+3}=H_n/2$, so that at $w=0$ we recover Eq.~\eqref{eq:Jodd}, and $\kappa_p=0$ for even $p$. As for the free channels, the derivative of the integral of Eq.~\eqref{eq:v:e:Jcutdef} with respect to $\ln\Lambda$ is $\Lambda f_p(\Lambda)=(-2\ii\omega)^{\ii w}\sum_{j,n}a_{jn}\Lambda^{p-3-j-2n}$, a sum of constants times powers of $\Lambda$, so that the local coefficients obey an equation of the same type as Eq.~\eqref{eq:rg-free}, and the logarithms are again single logarithms.

The first three renormalized channels are then given by Eq.~\eqref{eq:v:a:one-dimensional-integral} with
\begin{align}
	{}&I^{\rm newt}_3=I^{\rm I}_3+I^{\rm II}_3+\sum_{n\geq1}c_n(3) + \nonumber \\
	&~~~~+(-2\ii\omega)^{\ii w}\llp \ln\frac{\ii \bar\mu}{2\omega} -\psi(\ii w)-\gamma_{\rm E}\rrp, \nonumber \\
	{}& I^{\rm newt}_4= I^{\rm I}_4+I^{\rm II}_4+\sum_{n\geq1}c_n(4) + \nonumber \\
	\label{eq:v:e:first-subtracted-channels} &~~~~+(-2\ii\omega)^{\ii w}a_{10}\llp \ln\frac{\ii \bar\mu}{2\omega} - \psi(1+\ii w)-\gamma_{\rm E}+1\rrp, \\
	{}& I^{\rm newt}_5= I^{\rm I}_5+I^{\rm II}_5+\sum_{n\geq2}c_n(5) + \nonumber \\
	&~~~~+(-2\ii\omega)^{\ii w}a_{20}\llp \ln\frac{\ii \bar\mu}{2\omega}-\psi(2+\ii w)-\gamma_{\rm E}+2\rrp \nonumber \\
	&~~~~+(-2\ii\omega)^{\ii w}a_{01}\llp \ln\frac{\ii \bar\mu}{2\omega}-\psi(\ii w)-\gamma_{\rm E}+\frac12\rrp, \nonumber
\end{align}
where $a_{10}=-2w\omega$, $a_{20}=2w(w-\ii)\omega^2$ and $a_{01}=-(1-\ii w/2)^2k^2$ from Eq.~\eqref{eq:v:e:ajn}, and $c_n$ is given in Eq.~\eqref{eq:v:e:cn}. The sums of the terms $c_n$ without poles can be resummed. Writing $n=n_0+m$, with $n_0$ the first term of the sum, the ratio of two consecutive terms is a rational function of $m$ times $k^2/(4\omega^2)$, so that each sum is a generalized hypergeometric function, in the notation of Eq.~\eqref{eq:v:b:three-families}:
\begin{align}
	\sum_{n\geq1}c_n(3)={}&(-2\ii\omega)^{\ii w}\,\frac{(2-\ii w)\,k^2}{16\,\omega^2(1-\ii w)} \nonumber \\
	&\times{}_4F_3\!\left(\begin{matrix}1,1,\frac32,\frac{4-\ii w}2\\[2pt]2,2,\frac{3-\ii w}2\end{matrix};\frac{k^2}{4\omega^2}\right), \nonumber \\[0.6em]
	\sum_{n\geq1}c_n(4)={}&(-2\ii\omega)^{\ii w}\,\frac{\ii\,(2-\ii w)^2\,k^2}{8\,\omega(1-\ii w)} \nonumber \\
	\label{eq:v:e:cn-sums} &\times{}_5F_4\!\left(\begin{matrix}1,1,\frac12,\frac{4-\ii w}2,\frac{4-\ii w}2\\[2pt]2,2,\frac{2-\ii w}2,\frac{3-\ii w}2\end{matrix};\frac{k^2}{4\omega^2}\right), \\[0.6em]
	\sum_{n\geq2}c_n(5)={}&-(-2\ii\omega)^{\ii w}\,\frac{(2-\ii w)(4-\ii w)^2\,k^4}{256\,\omega^2(1-\ii w)} \nonumber \\
	&\times{}_5F_4\!\left(\begin{matrix}1,1,\frac32,\frac{6-\ii w}2,\frac{6-\ii w}2\\[2pt]3,3,\frac{3-\ii w}2,\frac{4-\ii w}2\end{matrix};\frac{k^2}{4\omega^2}\right). \nonumber
\end{align}
At $w=0$, Eqs.~\eqref{eq:v:e:first-subtracted-channels} give back $\mathcal J_3$, $\mathcal J_4$ and $\mathcal J_5$ of Eq.~\eqref{eq:J15}. For $w\neq0$, the scale in the logarithm is the frequency rather than $k$: the logarithm comes from the family $I^{\rm III}_p$, which is analytic in $k^2$, while the dependence on $k$ is carried by $I^{\rm I}_p$ and $I^{\rm II}_p$,  where the Newtonian dressing turns the free logarithm $\ln k$ into the phase $k^{\ii w}$. For $w\to0$ the poles in $1/w$ of $I^{\rm I}_p$ and of $\psi(\ii w)$ cancel, and $\ln(\bar\mu/\omega)$ combines with $\ln(\omega/k)$ into the free logarithm $\ln(\bar\mu/k)$. In the $p=4$ channel, the Newtonian dressing introduces a new logarithm, proportional to $a_{10}=-2w\omega$.

As for the free wave, the subtracted terms are local terms at the centre of the lens. Their finite values, and the scale $\bar\mu$, are not fixed by the exterior calculation: they must be determined by matching to the inner structure of the lens, i.e.\ to the interior of a resolved body or to the ingoing condition at a black-hole horizon, as we do in Section~\ref{sec:v:f}; see \cite{Goldberger:2005cd,Ivanov:2024sds} for the corresponding matching in the worldline effective theory, and \cite{Unruh:1976fm,Sasaki:2003xr} for the matching of the low-frequency expansion to the horizon in black-hole perturbation theory.

\subsection{Small angle and the Newtonian dressing of radial channels}
\label{sec:v:c}

To connect the general Fraunhofer kernels with the small scattering angles relevant to lensing, we extract their leading small-angle behaviour and examine how the Newtonian interaction modifies the free-wave response. This makes the resummation of the previously identified logarithm explicit and allows us to interpret the correction through its amplitude and phase relative to the scattered point-lens wave. So let us focus on the small-angle limit, $\theta\ll1$, i.e.\ $k/\omega\ll1$, which is the relevant regime for lensing and, as discussed in Sec.~\ref{sec:v:b}, is compatible with the Fraunhofer approximation for $\sqrt\varepsilon\ll k/\omega\ll1$. In this limit the Newtonian dressing of each radial channel becomes explicit. We first extract the leading kernel, then recover the mixed Born logarithm by expanding it in $w$. Finally, we translate the kernel into a response of the amplification factor and compare it with the scattered contribution already present in the Newtonian reference signal. We continue to assume $1<p<3$ and $k\neq0$.

At small angle, the estimates of Sec.~\ref{sec:v:b} give, for the three families of Eq.~\eqref{eq:v:b:three-families},
\begin{align}
	\label{eq:v:c:families-small-angle} I_p^{\rm I}&=P_p^{\rm I}\left[1+\mathcal O\left(\frac{k^2}{\omega^2}\right)\right]
\end{align}
while
\be
	\frac{I_p^{\rm II}}{I_p^{\rm I}} \propto \frac{k}{\omega} \,, \qquad \frac{I_p^{\rm III}}{I_p^{\rm I}} \propto \lp \frac{k}{\omega}\rp^{3-p}\,.
\ee
Hence, for $1<p<3$ and $k/\omega\ll1$, we approximate $I^{\rm newt}_p\simeq P_p^{\rm I}$, and Eq.~\eqref{eq:v:a:one-dimensional-integral} gives
\begin{align}
	\label{eq:v:c:leading-kernel} \mathcal J^{\rm newt}_{p, {\rm Fraun}}\simeq 2\pi \, k^{p-3} (k^2r_{\rm S}r_{\rm O})^{\ii w/2} \frac{\Gamma\!\left(\frac{p-1}{2}\right)^2} {\Gamma(p-1)} \frac{\Gamma\!\left(\frac{3-p-\ii w}{2}\right)} {\Gamma\!\left(\frac{p-1+\ii w}{2}\right)}.
\end{align}
Here $k^2r_{\rm S}r_{\rm O}$ is dimensionless, and $\mathcal J^{\rm newt}_{p, {\rm Fraun}}$ has the dimension $k^{p-3}$ of the free channel function $\mathcal J_p$. The approximation sign refers to the small-angle limit. For $2<p<3$ the largest neglected correction is $(k/\omega)^{3-p}$, from $I_p^{\rm III}$, and it is not suppressed as $p\to3$. The physical reason is that the channel $1/r^p$ becomes more concentrated at small radii as $p$ grows: $I_p^{\rm I}$ comes from radii $r\sim1/k$, while $I_p^{\rm III}$ comes from radii $r\sim1/\omega$, and as $p\to3$ the region $r\sim1/\omega$ contributes as much as the region $r\sim1/k$, until both diverge logarithmically at $p=3$. In that case the all-angle result should be used. For $p>3$, instead, $(k/\omega)^{3-p}$ is large, and at small angle $I_p^{\rm III}$ dominates over $I_p^{\rm I}$. Since $P_p^{\rm III}$ does not depend on $k$, $I_p^{\rm III}$ is a power series in $k^2$, i.e.\ a local term, which comes from radii $r\sim1/\omega$ and contains the short-distance poles of Sec.~\ref{sec:v:e}. The leading kernel of Eq.~\eqref{eq:v:c:leading-kernel} therefore holds only for the convergent channels, $1<p<3$.

\begin{figure*}[t]
\centering
\IfFileExists{figure_secV_2.png}{\includegraphics[width=\textwidth]{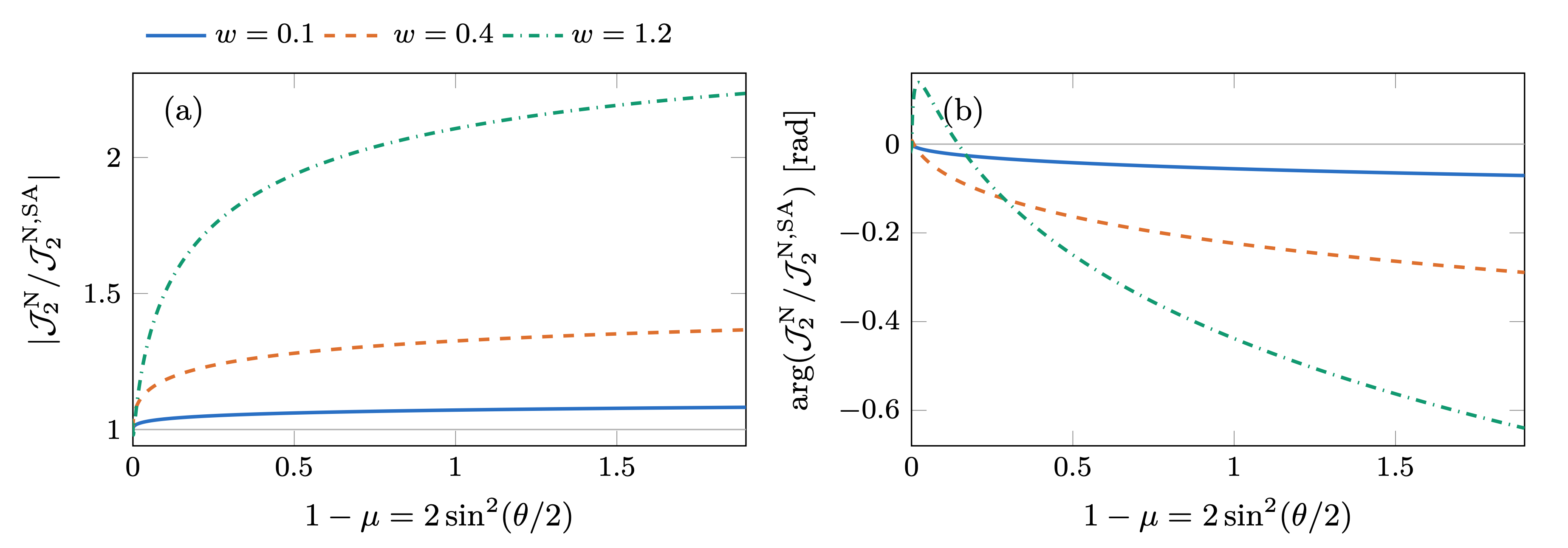}}{\fbox{\parbox{0.9\linewidth}{\centering Missing figure: \texttt{BH\_figures/figure\_secV\_2.png}}}}
\caption{Angular accuracy of the leading small-angle $p=2$ kernel. The ratio is the all-angle Fraunhofer kernel from Eqs.~\eqref{eq:v:a:one-dimensional-integral} and \eqref{eq:v:b:p-two}, divided by its small-angle limit, Eq.~\eqref{eq:v:c:leading-kernel} at $p=2$, i.e.\ $I^{\rm newt}_2/P^{\rm I}_2$. Panels show modulus (a) and principal phase (b) for $w=0.1$ (solid), $0.4$ (dashed), and $1.2$ (dash-dotted). Endpoint normalizations cancel. The ratio tends to unity at zero angle and is identically unity when the reference interaction is switched off. This tests the angular reduction within the same Fraunhofer model, not an exact finite-distance or full Schwarzschild calculation.}
\label{fig:v:angular}
\end{figure*}

Figure~\ref{fig:v:angular} shows the accuracy of this approximation for $p=2$. It divides the all-angle kernel, given by Eq.~\eqref{eq:v:a:one-dimensional-integral} with Eq.~\eqref{eq:v:b:p-two}, by Eq.~\eqref{eq:v:c:leading-kernel} at $p=2$, so that the plotted ratio is $I^{\rm newt}_2/P^{\rm I}_2$. The deviation comes mainly from $I_2^{\rm II}$ and $I_2^{\rm III}$, which are each suppressed by one power of $k/\omega$ and partially cancel each other; at $w=0$ they cancel exactly, as shown in Sec.~\ref{sec:v:b}, and the ratio is equal to one. The corrections in the hypergeometric function of $I_2^{\rm I}$ are smaller, of order $k^2/\omega^2$. For $w=0.1$ the deviation stays below ten per cent at all angles, while for $w\gtrsim1$ it is of order one at large angles. Since $I_2^{\rm III}$ comes from radii $r\sim1/\omega$, the large-angle result is reliable for a black hole only for $w\ll1$.

The Newtonian dressing is particularly transparent for $p=2$. Dividing Eq.~\eqref{eq:v:c:leading-kernel} by the free channel function $\mathcal J_2$ of Eq.~\eqref{eq:J15} gives
\begin{align}
	& \frac{\mathcal J^{\rm newt}_{2, {\rm Fraun}}}{\mathcal J_2}\simeq (k^2r_{\rm S}r_{\rm O})^{\ii w/2} \frac{\Gamma((1-\ii w)/2)}{\Gamma((1+\ii w)/2)} \nonumber \\
	\label{eq:v:c:p-two-dressing} &=\exp\left\{\ii\left[\frac{w}{2}\ln(k^2r_{\rm S}r_{\rm O}) -2\arg\Gamma\Big(\frac{1+\ii w}{2}\Big)\right]\right\},
\end{align}
In the second step we used that, for real $w$, the two gamma functions are complex conjugates, so that the Newtonian dressing only changes the phase of the leading small-angle kernel, not its modulus. The first term of the phase depends on $r_{\rm S}$ and $r_{\rm O}$: it comes from the long-range potential between the source, the lens and the observer, and resums to all orders in $w$ the mixed Born logarithm. The second term does not depend on the distances: it comes from the region close to the lens, as a Coulomb phase shift~\cite{Gordon:1928,Kong:1999sf,DiVecchia:2023frv}.

To connect with the free-wave Born calculation, we expand Eq.~\eqref{eq:v:c:p-two-dressing} in $w$ at fixed $k$, $r_{\rm S}$ and $r_{\rm O}$. With $\psi(z)=\dd\ln\Gamma(z)/\dd z$ denoting the digamma function, $\arg\Gamma((1+\ii w)/2)=w\psi(1/2)/2+\mathcal O(w^3)$, and the identity $\psi(1/2)=-\gamma_{\rm E}-2\ln2$ gives
\begin{align}
	\frac{\mathcal J^{\rm newt,Fraun}_2}{\mathcal J_2}={}&1+\frac{\ii w}{2} \left[\ln(k^2r_{\rm S}r_{\rm O})+2\gamma_{\rm E}+4\ln2\right] \nonumber \\
	\label{eq:v:c:mixed-born} &+\mathcal O(w^2).
\end{align}
Its linear term reproduces the mixed second Born contribution of Eq.~\eqref{eq:F21} with one Newtonian and one $p=2$ insertion, in either order, and also fixes its finite part, $C_2(\theta)=2\gamma_{\rm E}+4\ln2$ at small angle. Keeping Eq.~\eqref{eq:v:c:p-two-dressing} unexpanded retains all further Newtonian insertions on both sides of the single $p=2$ interaction. The large logarithm is thereby included in a phase rather than approximated by a short Taylor polynomial.

\begin{figure}[tb]
\centering
\includegraphics[width=\columnwidth]{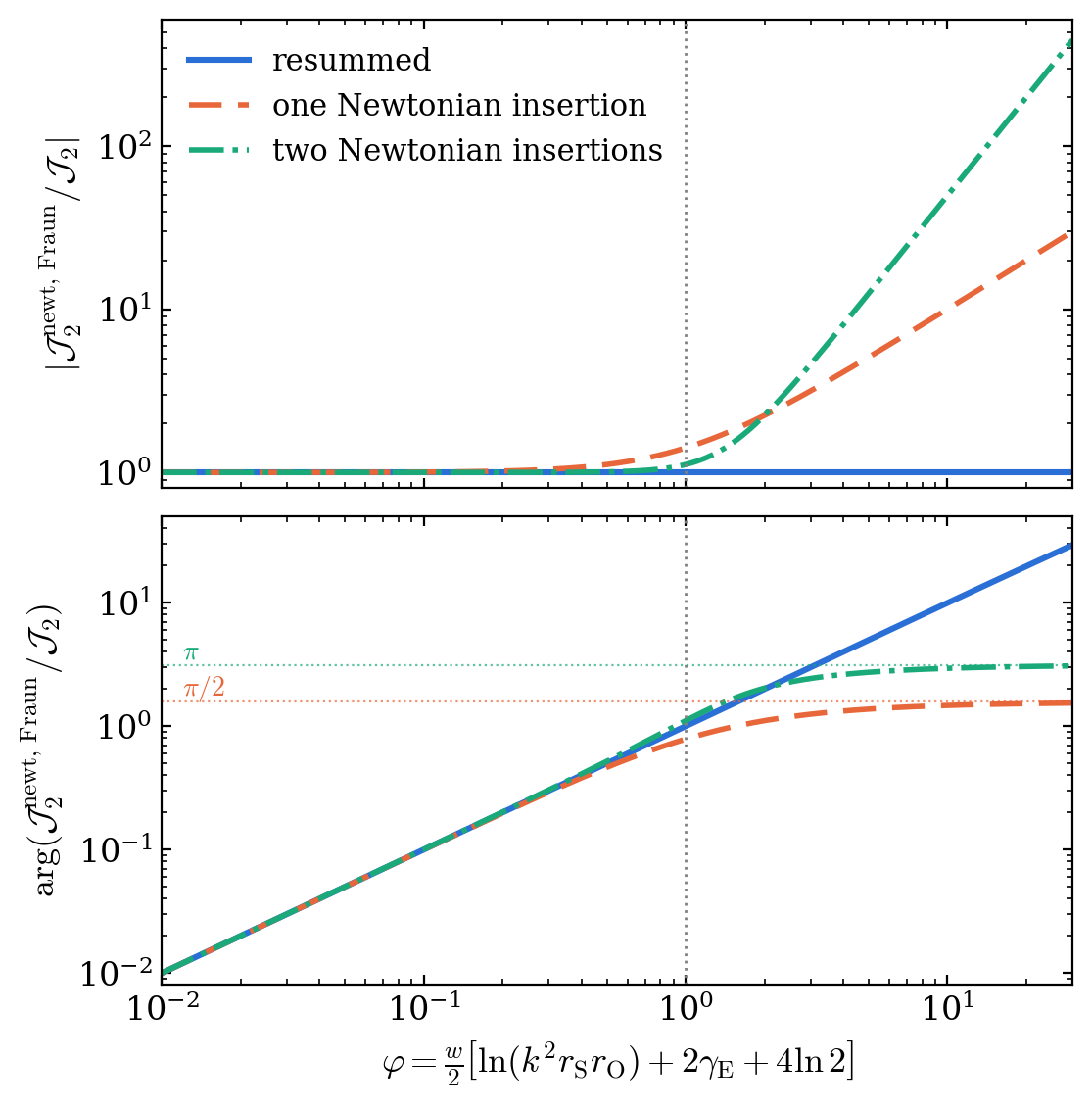}
\caption{Ratio $\mathcal J^{\rm newt,Fraun}_2/\mathcal J_2$ of Eq.~\eqref{eq:v:c:p-two-dressing} (solid), compared with its expansion in $w$ truncated at first order (dashed, one Newtonian insertion) and at second order (dash-dotted, two Newtonian insertions), as a function of the Newtonian phase $\varphi=w[\ln(k^2r_{\rm S}r_{\rm O})+2\gamma_{\rm E}+4\ln2]/2$. Top: modulus. Bottom: phase. For $\ln(k^2r_{\rm S}r_{\rm O})\gg1$ the curves depend on $w$ and $k^2r_{\rm S}r_{\rm O}$ only through $\varphi$. The dotted vertical line marks $\varphi=1$.}
\label{fig:v:resummation}
\end{figure}

Figure~\ref{fig:v:resummation} shows why the resummation matters. It compares the resummed ratio of Eq.~\eqref{eq:v:c:p-two-dressing} with its expansion in $w$ truncated at first and at second order. The first-order truncation is Eq.~\eqref{eq:v:c:mixed-born}: besides the $p=2$ insertion, it contains one Newtonian insertion, i.e.\ it corresponds to the free-wave Born terms $F^{(2)}_{{\rm free},21}+F^{(2)}_{{\rm free},12}$. The second-order truncation also contains the terms with two Newtonian insertions, i.e.\ the third-order free-wave Born terms with one $p=2$ and two $p=1$ insertions. In terms of the phase $\varphi=w[\ln(k^2r_{\rm S}r_{\rm O})+2\gamma_{\rm E}+4\ln2]/2$, the resummed ratio is $\eu^{\ii\varphi}$, up to corrections of order $w^3$, and the two truncations are $1+\ii\varphi$ and $1+\ii\varphi-\varphi^2/2$. When $\varphi$ becomes of order one, the truncations depart from the resummed result: their modulus grows, while the resummed ratio keeps unit modulus, and their phase saturates at $\pi/2$ and $\pi$, while the resummed phase grows linearly. Because of the large logarithm this happens already at small $w$: for $\ln(k^2r_{\rm S}r_{\rm O})=46$, at $w\simeq0.04$.

The figure shows a ratio of channel functions, which is not an observable. The observable is the total amplification factor, Eq.~\eqref{eq:FNseries}. Although the Newtonian dressing changes only the phase of the $p=2$ contribution, it does change the intensity, because the intensity depends on the interference between $F^{(0)}_{\rm newt}$ and $F^{(1)}_{\rm newt}$. At first order in $\delta U_{\rm newt}$,
\be \label{eq:v:c:intensity-response}
	\delta|F|^2 =2\operatorname{Re}\!\left(F^{(0)*}_{\rm newt}F^{(1)}_{\rm newt}\right) +\mathcal O(\delta U_{\rm newt}^2).
\ee

\subsubsection{Radial channels in dimensionless quantities}
As for the free wave in Sec.~\ref{sec:leading}, we write the first Born term of Eq.~\eqref{eq:v:a:fraunhofer-response} in the dimensionless variables of Section~\ref{sec:DimensionlessVariables}, for the convergent channels, $1<p<3$. With $c_p=\gamma_p(GM)^p$, the functions $T_p$ of Eq.~\eqref{eq:Tpdimless}, the endpoint factor of Eq.~\eqref{eq:Psmallangle} and the normalization of Eq.~\eqref{eq:XYphase}, dividing the leading kernel~\eqref{eq:v:c:leading-kernel} by the free channel function $\mathcal J_p$ of Eq.~\eqref{eq:jp}, we obtain
\begin{align}
	\eu^{-\ii\frac{w}{2}\ln\frac{X}{Y}}F^{(1)}_{\rm newt}\simeq{}&-\frac{\eu^{\ii s/2}}{4y^2}\left(\frac s2\right)^{\ii w} \nonumber \\
	\label{eq:v:c:dimensionless-response} &\times\sum_p\gamma_pT_p\,\frac{\Gamma\left(\frac{p-1}{2}\right)\Gamma\left(\frac{3-p-\ii w}{2}\right)} {\Gamma\left(\frac{3-p}{2}\right)\Gamma\left(\frac{p-1+\ii w}{2}\right)},
\end{align}
which is the free result of Eq.~\eqref{eq:FleadingDimless} multiplied, channel by channel, by a ratio of gamma functions that contains the Newtonian dressing without expanding in $w$. For $p\geq3$, the family $I_p^{\rm III}$ is not suppressed, as discussed above, and the subtracted functions of Sec.~\ref{sec:v:e} must be used instead.

\subsubsection{Response relative to the scattered point-lens wave}
Equation~\eqref{eq:v:c:dimensionless-response} carries the phase $\eu^{\ii s/2}$ of the scattered wave $B$ of Eq.~\eqref{eq:v:c:scattered-branch}. In the Fraunhofer limit the first Born term therefore modifies only the scattered wave, i.e.\ the secondary image when the images are resolved, while the response of the direct wave requires the wavefront curvature, Sec.~\ref{sec:v:d}. We therefore measure the effect relative to $B$, rather than relative to $F^{(0)}_{\rm newt}$, which at large $s$ is dominated by the direct wave. Dividing Eq.~\eqref{eq:v:c:dimensionless-response} by $B$, we find, for $1<p<3$,
\begin{align}
	\frac{\eu^{-\ii\frac{w}{2}\ln\frac{X}{Y}}F^{(1)}_{\rm newt}}{B}\simeq{}& -\frac14\sum_p\gamma_pT_p \frac{\Gamma(1+\ii w/2)}{\Gamma(1-\ii w/2)} \nonumber \\
	\label{eq:v:c:branch-response} &\times \frac{ \Gamma\!\left(\frac{p-1}{2}\right) \Gamma\!\left(\frac{3-p-\ii w}{2}\right)}{ \Gamma\!\left(\frac{3-p}{2}\right) \Gamma\!\left(\frac{p-1+\ii w}{2}\right)}.
\end{align}
At first order in $\delta U_{\rm newt}$, $B$ is multiplied by one plus the right-hand side, whose real and imaginary parts are the fractional changes of the amplitude and of the phase of the scattered wave. At $w=0$, $B$ coincides with the free Newtonian term, $\gamma_1T_1=-4$ in Eq.~\eqref{eq:FleadingDimless}, and Eq.~\eqref{eq:v:c:branch-response} reduces to the comparison $\sum_p\gamma_pT_p/(\gamma_1T_1)$ of Sec.~\ref{sec:leading}. For the Schwarzschild $p=2$ channel, $\gamma_2=-15/2$ and $T_2=\pi x/2$, the ratio of gamma functions has unit modulus, so that the response has modulus $15\pi x/16$. This ratio is equal to one at $w=0$ and tends to $\ii$ for large $w$: the $p=2$ channel changes mainly the amplitude of the scattered wave at small $w$, and its phase at large $w$.

\subsection{Fresnel propagation and the response of both wave branches}
\label{sec:v:d}

The Fraunhofer calculation describes the scattered response, but neglecting wavefront curvature can omit a contribution to the direct wave and produces an unphysical divergence at alignment. We therefore retain curvature in a small-angle Fresnel treatment, using the eikonal form of the Coulomb factors to obtain an analytic response that remains finite at alignment and reveals when both wave branches contribute at comparable strength. To do so, we now keep also the quadratic term of Eq.~\eqref{eq:distanceExpansion}, which describes the curvature of the wavefronts. With this term, different transverse interaction points have different travel times even at zero momentum transfer, and the first Born term acquires a contribution arriving with the direct wave, which is absent in the Fraunhofer result. To keep the calculation analytical, we combine the wavefront curvature with two further approximations: the small-angle limit for the geometry, and the eikonal form of the Coulomb factors, which keeps only their incident part, carrying the gravitational delay. The two constructions therefore make different approximations, and they overlap in the small-angle Fraunhofer limit, as we show below.

We start from the Green function of Eq.~\eqref{eq:Cwave}, which is the product of a free Green function and of a Coulomb factor given by a ${}_1F_1$ function, and we make one approximation on each of the two factors.

We keep the coordinates centred on the lens, and use the small-angle decomposition $\rr=\ell\,\bm e+\boldsymbol\zeta$ of Sec.~\ref{sec:eikonal}, for which $\kk\cdot\rr=\kk\cdot\boldsymbol\zeta$.

First, in the phase of the free Green functions, we use Eq.~\eqref{eq:distanceExpansion} including its quadratic term. In the small-angle geometry described above,  the quadratic Fresnel contribution is given by $r^2-(\bn_{\rm S,O}\cdot\rr)^2\simeq r^2-\ell^2=\zeta^2$ and it describes the curvature of the wavefronts. The linear terms of Eq.~\eqref{eq:distanceExpansion} are kept exactly and give, as in Eq.~\eqref{eq:planewaves}, the plane-wave factors $\eu^{\ii\kk_i\cdot\rr}$ and $\eu^{-\ii\kk_f\cdot\rr}$. In the amplitudes we keep $1/d_{\rm S,O}(\rr)\simeq1/r_{\rm S,O}$.

Second, we approximate the Coulomb factors of Eq.~\eqref{eq:Cwave} and of its observer counterpart as
\begin{align}
	&\Gamma\left(1-\frac{\ii w}{2}\right)(-2\ii\omega r_{\rm S})^{\frac{\ii w}{2}}\,{}_1F_1\left(\frac{\ii w}{2};1;\ii\omega(r+\bn_{\rm S}\cdot\rr)\right) \nonumber \\
	&\qquad\simeq\left(\frac{2r_{\rm S}}{r-\ell}\right)^{\ii w/2}, \nonumber \\
	\label{eq:v:d:eikonal-coulomb} &\Gamma\left(1-\frac{\ii w}{2}\right)(-2\ii\omega r_{\rm O})^{\frac{\ii w}{2}}\,{}_1F_1\left(\frac{\ii w}{2};1;\ii\omega(r+\bn_{\rm O}\cdot\rr)\right) \\
	&\qquad\simeq\left(\frac{2r_{\rm O}}{r+\ell}\right)^{\ii w/2}. \nonumber
\end{align}
These follow from the large-argument expansion of Eq.~\eqref{eq:pointlensKummerLarge}, with $s/2$ replaced by $\omega(r+\bn_{\rm S,O}\cdot\rr)\simeq\omega(r\mp\ell)$, keeping only its first term, which describes the incident wave, i.e.\ the eikonal form of the Coulomb wave, and dropping the second term, proportional to $\eu^{\ii z}$, which describes a wave scattered by the point mass on that leg. The expansion requires $\omega(r\mp\ell)\gg1+w^2$. For the source leg, $r-\ell=\zeta^2/(r+\ell)$ is of order $\zeta$ for $|\ell|\lesssim\zeta$, and of order $2|\ell|$ for $\ell<0$, i.e.\ between the source and the lens. Behind the lens, for $\ell\gg\zeta$, it is instead $\zeta^2/(2\ell)$, and the condition fails within $\zeta^2\lesssim(1+w^2)\ell/\omega$ of the line through the lens parallel to $\bm e$. This is the region where the rays deflected by the point mass, which are bent towards this line by the angle $4GM/\zeta$, cross it: there the incident and the scattered waves overlap and cannot be separated, and the ${}_1F_1$ cannot be replaced by its incident part. The same holds for the observer leg, with $\ell\to-\ell$. Since the integral over $\ell$ of $1/r^p$ is dominated by $|\ell|\lesssim\zeta$, the condition in the region that contributes to the integral is $\omega\zeta\gg1+w^2$: for $\zeta\sim1/k$ it becomes $(1+w^2)k/\omega\ll1$, i.e.\ the small-angle limit at fixed $w$, and for $\zeta\sim b$ it is satisfied more easily, since $\omega b=kr_{\rm eff}\gg1$.

Combining the two approximations, the product of the Green functions that enters Eq.~\eqref{eq:F1newt} is
\begin{align}
	&G_{\rm newt}(\rr_{\rm O},\rr) \,G_{\rm newt}(\rr,\rr_{\rm S}) \simeq \nonumber \\
	\label{eq:v:d:leg-product} &~~~~\simeq\frac{\eu^{\ii\omega(r_{\rm S}+r_{\rm O})}}{16\pi^2r_{\rm S}r_{\rm O}}\,\eu^{-\ii\kk\cdot\boldsymbol\zeta} \, \eu^{\frac{\ii \omega \zeta^2}{2 r_{\rm eff}}} \, \left(\frac{4r_{\rm S}r_{\rm O}}{\zeta^2}\right)^{\ii w/2},
\end{align}
where the plane-wave factors combine into $\eu^{\ii(\kk_i-\kk_f)\cdot\rr}=\eu^{-\ii\kk\cdot\boldsymbol\zeta}$, the curvature phases add, since $1/r_{\rm S}+1/r_{\rm O}=1/r_{\rm eff}$, and the Coulomb factors combine, since $(r-\ell)(r+\ell)=\zeta^2$. The last two factors are the extra time delay of the wave passing through the point $\rr$, with respect to the unlensed wave. For real $w$ they are pure phases,
\begin{align}
	& \eu^{\frac{\ii \omega \zeta^2}{2 r_{\rm eff}}} \left(\frac{4r_{\rm S}r_{\rm O}}{\zeta^2}\right)^{\ii w/2} \nonumber \\
	&\qquad=\exp\left\{\ii\omega\left[\frac{\zeta^2}{2r_{\rm eff}}+2GM\ln\frac{4r_{\rm S}r_{\rm O}}{\zeta^2}\right]\right\}.
\end{align}
The first term in the square brackets is the geometric delay due to the curvature of the wavefronts, and the second is the gravitational (Shapiro) delay accumulated along the two legs, which depends on the transverse position as $-4GM\ln\zeta$, i.e.\ as the lensing potential of a point mass. Equation~\eqref{eq:v:d:leg-product} therefore contains the same time delay as the standard diffraction integral of a point lens. Indeed, inserting it in Eq.~\eqref{eq:F1newt}, and using $\mathcal P\simeq\eu^{\ii s/2}/r_{\rm eff}$, $\eu^{\ii s/2}\eu^{-\ii\kk\cdot\boldsymbol\zeta}\eu^{\ii\omega\zeta^2/(2r_{\rm eff})}=\eu^{\ii\omega|\boldsymbol\zeta-\bm b|^2/(2r_{\rm eff})}$ and $c_p\int\dd\ell\,r^{-p}/4=\int\dd\ell\,\delta U_{\rm newt}$, we obtain exactly Eq.~\eqref{eq:v:d:F1-diffraction}, up to the constant phase $(4r_{\rm S}r_{\rm O})^{\ii w/2}$. The Fresnel construction of this subsection is therefore the eikonal picture of Sec.~\ref{sec:overview}, and in geometric optics its results can be read in terms of the two images of the point lens.

The conditions on the other approximations made on the propagation legs are discussed below Eq.~\eqref{eq:v:a:observer-leg}.

Inserting Eq.~\eqref{eq:v:d:leg-product} and the channels of Eq.~\eqref{eq:UNsplit} in Eq.~\eqref{eq:F1newt}, the prefactors combine into the endpoint factor of Eq.~\eqref{eq:endpointFactor}, as in Sec.~\ref{sec:v:a}, and we obtain $F^{(1)}_{\rm newt}$ in the form of Eq.~\eqref{eq:v:a:fraunhofer-response}, but with $\mathcal J^{\rm newt}_{p, {\rm Fraun}}$ replaced by
\be \label{eq:v:d:fresnel-kernel}
	\mathcal J^{\rm newt}_{p, {\rm Fres}} \simeq (4r_{\rm S}r_{\rm O})^{\ii w/2} \int\dd^3r\, \frac{\eu^{-\ii\kk\cdot\boldsymbol\zeta}}{r^p\zeta^{\ii w}}\,\eu^{\frac{\ii \omega \zeta^2}{2 r_{\rm eff}}} .
\ee
The potential $\delta U_{\rm newt}$ is still inserted only once, and the three propagation factors of Eq.~\eqref{eq:v:d:fresnel-kernel} describe, in order, the momentum transfer, the gravitational delay and the curvature of the wavefronts, as in Eq.~\eqref{eq:v:d:leg-product}.

In terms of the Fresnel scale of Eq.~\eqref{eq:rF}, the curvature phase is $\zeta^2/(2r_{\rm F}^2)$, and Eqs.~\eqref{eq:bdef} and~\eqref{eq:sepsx} show that it is of order $1/(2s)$ in the region $\zeta\sim1/k$, and of order $s/2$ in the region $\zeta\sim b$. For $s\gg1$ it is negligible in the region $\zeta\sim1/k$, which gives the Fraunhofer result, but it is large near the unlensed line of sight, $\zeta\sim b$. Dropping it everywhere can therefore remove the contribution of a different region of the same integral. This is why the Fraunhofer condition at $\zeta\sim1/k$ does not guarantee that the Fraunhofer result is complete for a power law that extends to large radii.

The integral of Eq.~\eqref{eq:v:d:fresnel-kernel} can be done in closed form. It converges for $1<p<3$; for $p\geq3$ it diverges at small $\zeta$, and the closed form below is its analytic continuation in $p$, i.e.\ its finite part, as in Sec.~\ref{sec:v:e}. In that case, however, the region $\zeta\lesssim1/\omega$, where the eikonal form of the Coulomb factors does not hold, contributes, and the local terms must be fixed as discussed in Sec.~\ref{sec:v:e}. Writing $\dd^3r=\dd^2\zeta\,\dd\ell$, only the potential depends on $\ell$, and
\begin{align}
	\mathcal J^{\rm newt}_{p, {\rm Fres}}\simeq{}&(4r_{\rm S}r_{\rm O})^{\ii w/2}\int\dd^2\zeta\,\eu^{-\ii\kk\cdot\boldsymbol\zeta}\,\zeta^{-\ii w}\,\eu^{\frac{\ii\omega\zeta^2}{2r_{\rm eff}}} \nonumber \\
	\label{eq:v:d:fresnel-kernel-split} &\times\int_{-\infty}^{\infty}\frac{\dd\ell}{(\zeta^2+\ell^2)^{p/2}}.
\end{align}
The integral over $\ell$ is
\be \label{eq:v:d:projection}
	\int_{-\infty}^{\infty} \frac{\dd\ell}{(\zeta^2+\ell^2)^{p/2}} =\frac{\sqrt\pi\,\Gamma\!\left(\frac{p-1}{2}\right)} {\Gamma(p/2)}\zeta^{1-p},
\ee
which projects the channel onto the plane transverse to the line of sight. The angular integral over the direction of $\boldsymbol\zeta$ gives $2\pi J_0(k\zeta)$, with $J_0$ the Bessel function, and the remaining radial integral follows from
\begin{align}
	\label{eq:v:d:bessel-integral} \int_0^\infty\dd\zeta\,\zeta^{c-1}\eu^{-\beta\zeta^2}J_0(k\zeta)=\frac{\Gamma(c/2)}{2\beta^{c/2}}\,{}_1F_1\!\left(\frac c2;1;-\frac{k^2}{4\beta}\right),
\end{align}
valid for ${\rm Re}\,c>0$ and ${\rm Re}\,\beta>0$, with $c=3-p-\ii w$ and $\beta\to-\ii\omega/(2r_{\rm eff})$, the limit being taken from ${\rm Re}\,\beta>0$. Since $-k^2/(4\beta)=-\ii s/2$, we obtain
\begin{align}
	&\mathcal J^{\rm newt}_{p, {\rm Fres}}\simeq (4r_{\rm S}r_{\rm O})^{\ii w/2}\,\frac{\pi^{3/2}\Gamma\left(\frac{p-1}{2}\right)}{\Gamma\left(\frac p2\right)}\,\Gamma\left(\frac{3-p-\ii w}{2}\right) \nonumber \\
	\label{eq:v:d:evaluated-kernel} &~~~~\times\left(-\frac{\ii\omega}{2r_{\rm eff}}\right)^{-\frac{3-p-\ii w}{2}} {}_1F_1\!\left(\frac{3-p-\ii w}{2};1;-\frac{\ii s}{2}\right).
\end{align}
The kernel $\mathcal J^{\rm newt}_{p, {\rm Fres}}$ has dimension $({\rm length})^{3-p}$, as the free channel functions. For $1<p<3$ this kernel is finite at $k=0$, since ${\rm Re}(3-p-\ii w)>0$ and the ${}_1F_1$ is equal to one at $s=0$: the curvature phase regulates the divergence at $k=0$ of the Fraunhofer kernel, discussed in Sec.~\ref{sec:v:a}.

\subsubsection{Comparison with the Fraunhofer result}

The Fraunhofer result of Sec.~\ref{sec:v:a} and Eq.~\eqref{eq:v:d:evaluated-kernel} are obtained with different approximations, and they must agree where both apply. The Fraunhofer kernel neglects the curvature of the wavefronts but keeps the full Coulomb factors, and it is therefore compared at small angle, i.e.\ through its leading kernel of Eq.~\eqref{eq:v:c:leading-kernel}, which, at small angle, we denote by $\mathcal J^{\rm newt}_{p, {\rm Fraun}}$ as well. Equation~\eqref{eq:v:d:evaluated-kernel} uses the eikonal form of the Coulomb factors but keeps the curvature, and it is therefore compared when the curvature phase $\zeta^2/(2r_{\rm F}^2)$ is negligible in the region $\zeta\sim1/k$, i.e.\ for $s\gg1$, when the impact parameter is much larger than the Fresnel scale.

To compare the two results, we divide Eq.~\eqref{eq:v:d:evaluated-kernel} by the leading small-angle Fraunhofer kernel of Eq.~\eqref{eq:v:c:leading-kernel}. Using the duplication formula of the gamma function and Kummer's transformation of the ${}_1F_1$~\cite{DLMF}, the ratio becomes a function of $s$, $w$ and $p$ only, which contains the whole effect of the wavefront curvature,
\begin{align}
	\frac{\mathcal J^{\rm newt}_{p, {\rm Fres}}}{\mathcal J^{\rm newt}_{p, {\rm Fraun}}} = {}& \Gamma\!\left(\frac{p-1+\ii w}{2}\right) \eu^{-\frac{\ii s}{2}} \left(\frac{\ii s}{2}\right)^{\frac{3-p-\ii w}{2}} \times \nonumber \\
	\label{eq:v:d:curvature-ratio} &\times {}_1F_1\!\left(\frac{p-1+\ii w}{2};1;\frac{\ii s}{2}\right).
\end{align}
For $s\to0$ the ratio vanishes as $s^{(3-p)/2}\propto k^{3-p}$, which compensates the divergence $k^{p-3}$ of the Fraunhofer kernel, so that $\mathcal J^{\rm newt}_{p, {\rm Fres}}$ is finite at alignment. For $s\gg1$, the large-argument expansion of the ${}_1F_1$, Eq.~\eqref{eq:pointlensKummerLarge} with the first parameter $(p-1+\ii w)/2$, gives
\begin{align}
	\label{eq:v:d:two-asymptotic-branches} \frac{\mathcal J^{\rm newt}_{p, {\rm Fres}}}{\mathcal J^{\rm newt}_{p, {\rm Fraun}}}\simeq{}&1+\ii\left(\frac s2\right)^{2-p-\ii w}\eu^{-\ii s/2}\,\frac{\Gamma\left(\frac{p-1+\ii w}{2}\right)}{\Gamma\left(\frac{3-p-\ii w}{2}\right)},
\end{align}
with corrections of relative order $1/s$ in each term. The first term, equal to one, comes from the region $\zeta\sim1/k$, where the curvature phase is of order $1/(2s)\ll1$: there the two constructions apply, and they agree, for any $w$. The second term comes instead from the region $\zeta\sim b$ around the unlensed line of sight, where the curvature phase is of order $s/2\gg1$, so that the Fraunhofer approximation does not hold there for any $s$. It carries the phase $\eu^{-\ii s/2}$, which cancels the phase $\eu^{\ii s/2}$ of the endpoint factor $\mathcal P$: as discussed after Eq.~\eqref{eq:pointlensLargeS}, it therefore arrives together with the direct wave, i.e.\ it is the response of the direct wave, which is absent in the Fraunhofer result. Its size relative to the first term is $s^{2-p}$: it is negligible for $2<p<3$, comparable for $p=2$, and dominant for $1<p<2$. Hence, the condition $s\gg1$ guarantees that the Fraunhofer result correctly describes the contribution of the region $\zeta\sim1/k$, but not that this is the complete result: for $p\leq2$ the contribution of the region $\zeta\sim b$ is as large or larger. Notice that the effect of the wavefront curvature is controlled by $s$, i.e.\ by the impact parameter in units of the Fresnel scale, and not by $\varepsilon$, Eq.~\eqref{eq:sepsx}, which would control the finite-distance corrections that we neglected, such as the cubic terms of Eq.~\eqref{eq:distanceExpansion}.

A faint image need not give a negligible correction to the total wave. In geometric optics and far from alignment, the primary image has amplitude of order unity and passes at $\zeta_+\simeq R_Ey$, whereas the secondary has amplitude of order $y^{-2}$ and passes at $|\zeta_-|\simeq R_E/y$. Since the additional phase of a radial channel scales as $\zeta^{1-p}$, the ratio of the primary to secondary response scales as $y^{2(2-p)}$, at leading order in the deformation. For $p=2$, the larger phase shift of the secondary image compensates its smaller reference amplitude, giving equal leading response magnitudes. Eq.~\eqref{eq:v:d:two-asymptotic-branches} extends this comparison to wave propagation without requiring separately resolved images.

\subsubsection{The $p=2$ channel}
We now evaluate the Fresnel result for the $p=2$ channel, which is the leading channel of a Schwarzschild black hole, Sec.~\ref{sec:v:f}. Since the wavefront curvature is retained, the result holds for any offset $y$, and it contains the response of both images of the point lens. We write it in terms of the phase $\delta\Phi(R_E)=-\pi\gamma_2w\alpha_E/32$ of Eq.~\eqref{eq:v:d:image-phases}, i.e.\ the additional phase of a ray passing at the Einstein radius. Inserting Eq.~\eqref{eq:v:d:evaluated-kernel} with $p=2$ in Eq.~\eqref{eq:v:a:fraunhofer-response}, in the dimensionless variables of Sec.~\ref{sec:leading} and with Eq.~\eqref{eq:XYsmallangle}, the correction to the amplification factor is
\begin{align}
	\eu^{-\ii\frac w2\ln\frac XY} {}& F^{(1)}_{\rm newt}\simeq\delta\Phi(R_E)\,\sqrt{\frac w2}\,\eu^{\frac{\pi w}{4}+\frac{\ii\pi}{4}}\,\Gamma\!\left(\frac{1-\ii w}{2}\right) \nonumber \\
	\label{eq:v:d:p-two-response} &\times\left(\frac{wy^2}{2}\right)^{\ii w/2}{}_1F_1\!\left(\frac{1+\ii w}{2};1;\frac{\ii wy^2}{2}\right).
\end{align}
The distances of source and observer enter only through $\delta\Phi(R_E)$, i.e.\ through $\alpha_E=4GM/R_E$, in agreement with Eq.~\eqref{eq:v:d:delta-phi-p}. To understand this result, we evaluate it in the two configurations of Fig.~\ref{fig:v:images}: at alignment, where the two images merge into the Einstein ring at $\zeta=R_E$, and far from alignment, where one image passes far from the lens and the other close to it.

At alignment, $y\to0$, the hypergeometric function in Eq.~\eqref{eq:v:d:p-two-response} tends to one. We divide by the point-lens amplification factor of Eq.~\eqref{eq:FhatNewt}, which at $y\to0$ is $\eu^{\pi w/4}\Gamma(1-\ii w/2)(wy^2/2)^{\ii w/2}$. The factor $\eu^{-\ii(w/2)\ln(X/Y)}$, which is ill-defined at exact alignment since $Y\to0$, multiplies both and cancels in the ratio, as does $(wy^2/2)^{\ii w/2}$, and we find
\begin{align}
	\label{eq:v:d:aligned-ratio} \lim_{y\to0}\frac{F^{(1)}_{\rm newt}}{F^{(0)}_{\rm newt}}\simeq{}&\delta\Phi(R_E)\,\sqrt{\frac w2}\,\eu^{\ii\pi/4}\,\frac{\Gamma\left(\frac{1-\ii w}{2}\right)}{\Gamma\left(1-\frac{\ii w}{2}\right)}
\end{align}
which, in geometric optics $w\gg1$, gives
\be
	\lim_{y\to0}\frac{F^{(1)}_{\rm newt}}{F^{(0)}_{\rm newt}} \simeq \ii\,\delta\Phi(R_E).
\ee
In geometric optics, the Einstein ring is therefore multiplied by $1+\ii\,\delta\Phi(R_E)$, as in the image picture. Since this is a phase of the whole signal, its absolute value depends on the arrival-time convention, and away from alignment the relevant observable is the relative delay between the images. Moreover, a linear treatment requires $|\delta\Phi(R_E)|\ll1$. At low frequency, instead, the ratio of gamma functions tends to $\sqrt\pi$, and the correction is $\sqrt{\pi w/2}\,\eu^{\ii\pi/4}\,\delta\Phi(R_E)$, smaller by $\sqrt w$, because the wave averages $\delta\Phi$ over the Fresnel disk, as discussed in Sec.~\ref{sec:overview}.

Far from alignment, $wy^2\gg1$, Eq.~\eqref{eq:pointlensKummerLarge} applied to Eq.~\eqref{eq:v:d:p-two-response} gives
\be \label{eq:v:d:far-offset-response}
	\eu^{-\ii\frac w2\ln\frac XY}F^{(1)}_{\rm newt}\simeq\ii\,\delta\Phi(\zeta_+)+\eu^{\ii\chi(w)}\,\delta\Phi(\zeta_-)\,B,
\ee
with $\delta\Phi(\zeta_\pm)$ of Eq.~\eqref{eq:v:d:image-phases}, $B$ of Eq.~\eqref{eq:v:c:scattered-branch}, and corrections of relative order $1/(wy^2)$; at large $w$ this also requires $y^2\gg w$. The phase
\be
	\chi(w)\equiv2\left[\arg\Gamma\Big(1+\frac{\ii w}{2}\Big)-\arg\Gamma\Big(\frac{1+\ii w}{2}\Big)\right]
\ee
comes from the ratio of gamma functions of Eq.~\eqref{eq:v:c:branch-response} for $p=2$, which has unit modulus, and it grows from $\chi\simeq2w\ln2$ for $w\ll1$ to $\pi/2$ for $w\gg1$. Since the reference signal is the sum of the direct wave and of the scattered wave $B$, Eq.~\eqref{eq:pointlensLargeS}, each term of Eq.~\eqref{eq:v:d:far-offset-response} is the change of one of them. The first term is the response of the direct wave, i.e.\ of the primary image, which requires the curvature phase: the direct wave is multiplied by $1+\ii\,\delta\Phi(\zeta_+)$, i.e.\ its phase is shifted by $\delta\Phi$ at its position, at any frequency. The second term is the response of the scattered wave, i.e.\ of the secondary image, Eq.~\eqref{eq:v:c:branch-response} for $p=2$, which was already obtained in the Fraunhofer limit: $B$ is multiplied by $1+\eu^{\ii\chi}\delta\Phi(\zeta_-)$. For $w\gg1$, $\eu^{\ii\chi}\to\ii$, and each image is multiplied by $1+\ii\,\delta\Phi$ at its position, as in Eq.~\eqref{eq:v:d:F1-saddle}. For $w\ll1$, instead, $\eu^{\ii\chi}\to1$: the relative change of the scattered wave still has modulus $|\delta\Phi(\zeta_-)|$, but it is a change of its amplitude rather than of its phase. The secondary image passes close to the lens, and its relative change, $\delta\Phi(\zeta_-)\propto y$, is large, while the primary image passes far from the lens, and its relative change, $\delta\Phi(\zeta_+)\propto1/y$, is small. However, the secondary image is faint, $|B|=1/y^2$, so that for $p=2$ the two terms of Eq.~\eqref{eq:v:d:far-offset-response} have the same magnitude, $|\delta\Phi(R_E)|/y$.

\subsection{Summary}
\label{sec:beyondborn:summary}

In this section we have built the Born series around the point lens, Eq.~\eqref{eq:FNseries}, in which the Newtonian interactions are included to all orders in the reference Green function, and only the channels with $p\geq2$ are treated as perturbations, Eq.~\eqref{eq:UNsplit}. The exact Hostler--Pratt Green function, Eq.~\eqref{eq:HP}, gives at large distances the familiar point-lens amplification, Eq.~\eqref{eq:FhatNewt}, which is the zeroth order of the expansion.

The first-order term, Eq.~\eqref{eq:F1newt}, describes how an additional radial channel changes the point-lens wave. In geometric optics, the deformation changes the phase along each image path in proportion to the additional time delay it produces, Eq.~\eqref{eq:v:d:F1-saddle}. At lower frequencies the wave samples a transverse region of Fresnel size, so that the correction depends on how the potential varies across this region. This explains, for instance, why the aligned inverse-square correction is weaker at low frequency than an estimate based on the Einstein ring would suggest.

In the Fraunhofer limit, writing each power law as a superposition of screened potentials allows us to do the spatial integral once, using a result of Nordsieck, and the radial channel enters only through the weight of a single integral over the inverse screening length, Eqs.~\eqref{eq:v:a:one-dimensional-integral} and~\eqref{eq:Ip}. For $1<p<3$ this gives closed forms with the full angular dependence, Sec.~\ref{sec:v:b}. Two divergences appear. A forward divergence at zero momentum transfer comes from the large-radius tail of the channel and is an artifact of the Fraunhofer limit. For $p\geq3$, the short-distance divergence of Sec.~\ref{sec:leading} is renormalized in the same way, Sec.~\ref{sec:v:e}, with local terms that are fixed by the inner structure of the lens; the Newtonian dressing introduces logarithmic divergences also in even channels, such as $p=4$.

In the small-angle limit, Sec.~\ref{sec:v:c}, the dressing of the inverse-square channel leaves the modulus of the free kernel unchanged and multiplies it by a phase, Eq.~\eqref{eq:v:c:p-two-dressing}. Expanding this phase recovers the mixed Born logarithm of Eq.~\eqref{eq:F21}, together with its finite part. The measured amplification depends on the phase of the correction relative to the point-lens wave, so that the dressing can change the observed intensity through interference.

Retaining the curvature of the wavefronts, Sec.~\ref{sec:v:d}, gives a finite response at alignment for the convergent channels, and captures, besides the scattered wave, the response of the direct wave near the unlensed line of sight. Far from alignment, the correction to the direct wave dominates for power laws shallower than inverse square, and is suppressed for steeper ones. For the inverse-square channel the two corrections have equal magnitudes, Eq.~\eqref{eq:v:d:far-offset-response}: the secondary image passes closer to the lens and receives a larger phase shift, which compensates for its smaller amplitude. The inverse-square Fresnel response, Eq.~\eqref{eq:v:d:p-two-response}, including its finite aligned limit, Eq.~\eqref{eq:v:d:aligned-ratio}, is the leading optical benchmark for the Schwarzschild lens of Sec.~\ref{sec:v:f}.

\section{Application to a Schwarzschild black hole lens}
\label{sec:v:f}

The dressed kernels describe how the exterior potential modifies the point-lens wave, but they cannot determine the inner boundary information required by the singular radial channels. A Schwarzschild black hole provides a concrete test of how to complete this description, since the missing information is fixed by the condition that waves enter the horizon. Working with the scalar field at low frequency, we compute the correction quadratic in $GM$, including both the inverse-square optical potential and the steeper derivative term generated by the field redefinition. Matching the exterior solution to an ingoing horizon solution fixes the local contribution and recovers the scalar absorption cross section \cite{Das:1996we,Starobinskii:1973vzb,Page:1976df,Sanchez:1977si}. We then combine these terms at finite source and observer distances to establish the accuracy of the leading Fresnel prediction and show why the horizon response must be evaluated together with optical corrections of the same order.

\subsection{The potential at order $(GM)^2$}
\label{sec:v:f:potential}

A calculation complete at a given order in $GM$ must include every potential term at that order, even when the terms have different radial falloffs. We therefore identify the quadratic Schwarzschild contributions, distinguishing the inverse-square optical term from the inverse-fourth-power derivative term, whose short-distance sensitivity requires horizon matching. For the Schwarzschild metric the coefficients $c_p^{\rm Sch}$ of the radial expansion~\eqref{eq:cp_expansion} are given in Eq.~\eqref{eq:cpSchw}. For definiteness, we keep only the terms of order $(GM)^2$:
\be \label{eq:v:f:optical-potential}
	\delta U_2=\frac{c_2^{\rm Sch}}{4r^2}=-\frac{15(GM)^2}{8r^2},\qquad \delta U_{\rm der}=-\frac{(GM)^2}{16\omega^2r^4}.
\ee
The first is the $p=2$ channel. The second is the $\omega^{-2}$ part of $c_4^{\rm Sch}$, which comes from the derivative terms of Eq.~\eqref{eq:Umaster}; it is of order $(GM)^2$ although it multiplies $r^{-4}$.

We can then write
\be \label{eq:v:f:Fsplit}
	F=F^{(0)}_{\rm newt}+F^{(1)}_{{\rm newt},2}+F^{(1)}_{{\rm newt},{\rm der}}+\mathcal O\big((GM)^3\big),
\ee
where $F^{(1)}_{{\rm newt},2}$ and $F^{(1)}_{{\rm newt},{\rm der}}$ are the single insertions of $\delta U_2$ and $\delta U_{\rm der}$ in Eq.~\eqref{eq:F1newt}, and we compute them separately. The $p=2$ channel is not divergent, and its response was already computed in Sec.~\ref{sec:v:d}; that result is valid for every value of $w$ (see Sec.~\ref{sec:v:f:optical}). The derivative term, instead, has $p=4$: its Born integral diverges at $r\to0$. It  requires the renormalization procedure discussed in Sec.~\ref{sec:v:e}, which leaves local terms to be fixed by the inner structure of the lens.

At this order we can also replace the Newtonian Green functions in the integral by free ones, since $G_{\rm newt}=G_{\rm free}+\mathcal O(GM)$ and $\delta U_{\rm der}$ is already of order $(GM)^2$. The difference is of order $(GM)^3$, and it contains the logarithm of the dressed $p=4$ channel, Eq.~\eqref{eq:v:e:first-subtracted-channels}, whose coefficient $a_{10}=-2w\omega$ carries one more power of $GM$. At order $(GM)^2$ the derivative term therefore has only a power divergence, and it is renormalized as the free $p=4$ channel of Sec.~\ref{sec:leading}, with local terms to be fixed by the inner structure of the lens. For a black hole the inner structure is the horizon, and we fix these terms by an explicit matching calculation which, for the reasons given at the beginning of this section, we perform at $w\ll1$ (Sec.~\ref{sec:v:f:matching2}).

\subsection{The $p=2$ term}
\label{sec:v:f:optical}

The inverse-square channel provides the leading optical correction and a benchmark against which the derivative and horizon contributions can be assessed. In the eikonal limit it gives the deflection angle $15\pi(GM)^2/(4b^2)$ at second order in the mass~\cite{Epstein:1980dw,Keeton:2005jd}, which also follows from the one-loop scattering amplitude~\cite{Bjerrum-Bohr:2014zsa,Correia:2024jgr}. We use its Fresnel response to determine how it changes the amplitude and phase of the point-lens wave, and explain how diffraction controls its size at low frequency.

Choosing $p=2$ the channel is not divergent, so $F^{(1)}_{{\rm newt},2}$ is given by Eq.~\eqref{eq:v:d:p-two-response} with $\gamma_2=-15/2$, for any offset $y$ and any $w$. Its ratio to $F^{(0)}_{\rm newt}$ is easy to read: since $F=F^{(0)}_{\rm newt}\big(1+F^{(1)}_{{\rm newt},2}/F^{(0)}_{\rm newt}\big)$, at first order the real part of the ratio is the fractional change of the amplitude of $F$, and its imaginary part is the change of its phase, in radians.

At low frequency, $w\ll1$ at fixed $s=wy^2$, dividing Eq.~\eqref{eq:v:d:p-two-response} by Eq.~\eqref{eq:FhatNewt} and using ${}_1F_1(\frac12;1;\frac{\ii s}2)=e^{\ii s/4}J_0(s/4)$, with $J_0$ the Bessel function, we get
\be \label{eq:v:f:E2-limits}
	\frac{F^{(1)}_{{\rm newt},2}}{F^{(0)}_{\rm newt}}\simeq\frac{15\pi^{3/2}}{128}(1+\ii)\,\alpha_Ew^{3/2}\,e^{\ii s/4}J_0\!\left(\frac s4\right),
\ee
up to relative corrections of order $w(1+s)$. At alignment, $s=0$, the real and imaginary parts are equal: the amplitude increases by the fraction $15\pi^{3/2}\alpha_Ew^{3/2}/128$, and the phase by the same number of radians. The size of the effect follows from the Fresnel scale: since $r_{\rm F}=R_E/\sqrt w>R_E$, the wave does not resolve the Einstein ring, but averages the phase $\psi_2$ of Eq.~\eqref{eq:v:d:delta-phi-p} over a region of size $r_{\rm F}$, where $\psi_2(r_{\rm F})\sim w\,GM/r_{\rm F}\sim\alpha_Ew^{3/2}$.

\subsection{Low-frequency horizon matching}
\label{sec:v:f:matching2}

The derivative term $\delta U_{\rm der}$ is a $p=4$ channel, and its insertion diverges at $r\to0$, as in Secs.~\ref{sec:leading} and~\ref{sec:v:e}. We renormalize it in the same way: we split the integral at a radius, describe the inner region by a local term, and fix the local term by matching to the solution close to the horizon. This follows the low-frequency expansion of black-hole perturbations, where the expansion in $GM\omega$ gives flat-space equations without a horizon, and the horizon condition is imposed by matching to a solution valid close to it~\cite{Unruh:1976fm,Sasaki:2003xr}. The matching also gives the absorptive part of the local term, which we check against the known horizon-area cross section.

The expansion of the potential in radial channels assumes $r\gg GM$. We therefore split the integral of Eq.~\eqref{eq:F1newt} defining $F^{(1)}_{{\rm newt},{\rm der}}$ at a radius $r_{\rm m}$, which plays the role of $r_0$ in Eq.~\eqref{eq:r0range}, with
\be \label{eq:v:f:rm}
	GM\ll r_{\rm m}\ll\frac{1}{\omega} ,
\ee
which exists since $w\ll1$. For $r>r_{\rm m}$ we use $\delta U_{\rm der}$ of Eq.~\eqref{eq:v:f:optical-potential}, while for $r<r_{\rm m}$ we do not know the potential. In this inner region the Green functions are constant. Indeed, the source and the observer are far away, $r_{\rm S},r_{\rm O}\gg r_{\rm m}$, and Eq.~\eqref{eq:distanceExpansion} gives $|\rr_{\rm O}-\rr|=r_{\rm O}-\bn_{\rm O}\cdot\rr+\dots$; for $r<r_{\rm m}$ the amplitude $1/|\rr_{\rm O}-\rr|$ therefore changes by a fraction of order $r_{\rm m}/r_{\rm O}$, and the phase $\omega|\rr_{\rm O}-\rr|$ by at most $\omega r_{\rm m}$. Both are small, so $G_{\rm newt}(\rr_{\rm O},\rr)=G_{\rm newt}(\rr_{\rm O},0)\,[1+\mathcal O(\omega r_{\rm m})]$, and similarly for the source. The inner region then contributes
\be \label{eq:v:f:ball}
	\frac{G_{\rm newt}(\rr_{\rm O},0)\,G_{\rm newt}(0,\rr_{\rm S})}{G_{\rm free}(\rr_{\rm O},\rr_{\rm S})}\,g_{4,0}(r_{\rm m}),
\ee
with
\be \label{eq:v:f:g40}
	g_{4,0}(r_{\rm m})=\int_{r<r_{\rm m}}\dd^3r\;4\omega^2\,\delta U_{\rm newt}(r).
\ee
This is the contribution of the first local term of Eq.~\eqref{eq:local-terms}, $g_{4,0}(r_{\rm m})\,\delta^{(3)}(\rr)$, i.e.\ the zeroth moment of the inner potential in Eq.~\eqref{eq:inner-series}. The other local terms of Eq.~\eqref{eq:local-terms}, $(\nabla^2)^n\delta^{(3)}(\rr)$ with $n\geq1$, correspond to the higher moments $\int\dd r\,r^{2+2n}\,\delta U_{\rm newt}$; they act on the derivatives of the Green functions at the origin, and are suppressed by $(\omega r_{\rm m})^{2n}$. Similarly, the part of $\delta U_{\rm newt}$ other than the derivative term contributes to $g_{4,0}$ terms smaller by $(\omega r_{\rm m})^2$ than those we keep. For a black hole $g_{4,0}$ cannot be computed as the integral~\eqref{eq:v:f:g40}, since the region $r<r_{\rm m}$ contains the horizon, and it is fixed below by matching.

Splitting the integral at $r_{\rm m}$ is therefore equivalent to replacing $\delta U_{\rm der}$ in Eq.~\eqref{eq:F1newt} by the regularized potential
\be \label{eq:v:f:Vmodel}
	4\omega^2\,\delta U_{\rm der}^{\rm reg}=-\frac{(GM)^2}{4r^4}\,\Theta(r-r_{\rm m})+g_{4,0}(r_{\rm m})\,\delta^{(3)}(\rr).
\ee
This has the form of Eq.~\eqref{eq:FPgeneral} at $p=4$ and finite $r_0=r_{\rm m}$, with $\omega^2c_4=-(GM)^2/4$, except that the coefficient of the delta function is not fixed to the counterterm $-4\pi\omega^2c_4/r_{\rm m}=\pi(GM)^2/r_{\rm m}$, but is left free.

As in Sec.~\ref{sec:leading}, the dependence of $g_{4,0}$ on $r_{\rm m}$ must cancel that of the exterior integral. Using free Green functions, as discussed in Sec.~\ref{sec:v:f:potential}, and the Fraunhofer limit, Eqs.~\eqref{eq:F1-split} and~\eqref{eq:JUV} with $p=4$ give
\be \label{eq:v:f:Fder-split}
	F^{(1)}_{{\rm newt},{\rm der}}\simeq-\frac{\mathcal P}{4\pi}\Big[\omega^2c_4\,\mathcal J_4^{(r_{\rm m})}(k)+g_{4,0}(r_{\rm m})\Big],
\ee
where, by Eq.~\eqref{eq:J345cut},
\be \label{eq:v:f:J4ext}
	\omega^2c_4\,\mathcal J_4^{(r_{\rm m})}(k)=-\frac{\pi(GM)^2}{r_{\rm m}}+\frac{\pi^2(GM)^2}{4}\,k .
\ee
The second term is $\omega^2c_4\mathcal J_4$, with $\mathcal J_4=-\pi^2k$ of Eq.~\eqref{eq:J15}, and is determined by the exterior potential. The first term diverges for $r_{\rm m}\to0$, and only the combination $g_{4,0}(r_{\rm m})-\pi(GM)^2/r_{\rm m}$ enters the amplification factor. The same holds beyond the Fraunhofer limit, since the term proportional to $1/r_{\rm m}$ comes from radii $r\sim r_{\rm m}$, where the Green functions are constant.

To find $g_{4,0}$, we compare two solutions of the wave equation for the spherically symmetric part of the wave: the one with the potential of Eq.~\eqref{eq:v:f:Vmodel}, valid for $r\gg GM$, and the solution of the full equation close to the horizon, which contains the information on the inner structure of the black hole. At low frequency both solutions hold in an overlap interval around $r_{\rm m}$, and requiring that they coincide there fixes $g_{4,0}$.

\subsubsection{Solution from the expansion at $r\gg GM$}

We solve the wave equation with the potential $\delta U_{\rm der}^{\rm reg}$ for $r\ll1/\omega$. There the incident wave is constant, and we normalize it to one. Since $\delta U_{\rm der}^{\rm reg}$ is spherically symmetric, we look at the spherically symmetric part of the wave, $\chi(r)$; this is sufficient, because the local term only acts on the value of the wave at the origin. At first order we write $\chi=1+\chi_1$. Inserting this in Eq.~\eqref{eq:helmholtz} with the potential~\eqref{eq:v:f:Vmodel}, and neglecting $\omega^2\chi_1$ with respect to $\nabla^2\chi_1$ -- which is allowed because $\chi_1$ varies over distances of order $r\ll1/\omega$ -- we get the Poisson equation
\be \label{eq:v:f:poisson}
	\nabla^2\chi_1=4\omega^2\,\delta U_{\rm der}^{\rm reg}.
\ee
(The other terms of the potential, $\delta U_2$ and the Newtonian one, are proportional to $\omega^2$; they only produce corrections that grow with $r$, of relative size $(\omega r)^2$ and $w\,\omega r$, which do not contain the $1/r$ term that we compare below.) For $r>r_{\rm m}$, a particular solution is $-(GM)^2/(8r^2)$, since $\nabla^2r^{-2}=2r^{-4}$, and the solutions of the homogeneous equation are a constant, which vanishes because of the normalization, and $1/r$. The coefficient of $1/r$ follows from Gauss' theorem: integrating Eq.~\eqref{eq:v:f:poisson} over a ball of radius $r>r_{\rm m}$, the left-hand side gives the flux $4\pi r^2\chi_1'(r)$, and the right-hand side the total source inside the ball,
\be
	4\pi r^2\chi_1'(r)=g_{4,0}(r_{\rm m})-\frac{\pi(GM)^2}{r_{\rm m}}+\frac{\pi(GM)^2}{r}.
\ee
The last term is the flux of $-(GM)^2/(8r^2)$, while a term $B/r$ gives the constant flux $-4\pi B$. Hence
\be \label{eq:v:f:chimodel}
	\chi(r)=1-\frac{(GM)^2}{8r^2}-\frac{1}{4\pi r}\left[g_{4,0}(r_{\rm m})-\frac{\pi(GM)^2}{r_{\rm m}}\right].
\ee
The coefficient of $1/r$ is fixed by the source concentrated at small radii: the local term, minus the part $\pi(GM)^2/r_{\rm m}$ of the $r^{-4}$ source that was removed from the region $r<r_{\rm m}$. It is the same combination that enters the amplification factor, Eqs.~\eqref{eq:v:f:Fder-split} and~\eqref{eq:v:f:J4ext}.

\subsubsection{Solution close to the horizon}

On the other hand, the wave near the black hole can be found from the full equation of motion, without expanding the potential. It is convenient to use the areal radius $R$, related to the isotropic radius by
\be \label{eq:v:f:radius-map2}
	R=r\left(1+\frac{GM}{2r}\right)^2,\qquad f(R)=1-\frac{2GM}{R},
\ee
in which the Schwarzschild metric reads $\dd s^2=-f\,\dd t^2+\dd R^2/f+R^2\dd\Omega^2$. Once again we look at the spherically symmetric part of the physical field, $\phi=e^{-\ii\omega t}\phi(R)$. The equation $\Box\phi=0$ of Sec.~\ref{sec:waveeq} then becomes~\cite{Unruh:1976fm}
\be \label{eq:v:f:radial-equation2}
	\frac{\dd}{\dd R}\left(R^2f\frac{\dd\phi}{\dd R}\right)+\frac{\omega^2R^2}{f}\,\phi=0.
\ee

We first find the solution close to the horizon, $R=2GM$, where $f=0$. There the physical condition is that nothing can come out of the black hole: the wave can only enter the horizon. To impose it, we introduce the tortoise coordinate $R_*$, defined by
\be
	\frac{\dd R_*}{\dd R}=\frac1f,\qquad R_*=R+2GM\ln\left(\frac{R}{2GM}-1\right).
\ee
It is the radial coordinate in which radial light rays move as in flat space, $R_*=\pm t+{\rm const}$, and it goes to $-\infty$ at the horizon. Close to the horizon, $R\simeq2GM$ and $R_*\simeq2GM\ln f$. Replacing $\dd/\dd R=f^{-1}\dd/\dd R_*$ in Eq.~\eqref{eq:v:f:radial-equation2} and setting $R\simeq2GM$, we get
\be
	\frac{\dd^2\phi}{\dd R_*^2}+\omega^2\phi=0,
\ee
whose solutions are $e^{\pm\ii\omega R_*}$. With the time dependence $e^{-\ii\omega t}$, the solution $e^{-\ii\omega(t+R_*)}$ moves towards decreasing $R_*$, i.e.\ into the black hole, and it is the only one allowed. Using $R_*\simeq2GM\ln f$, and expanding for small frequency,
\be \label{eq:v:f:phi-exact}
	\phi=e^{-\ii\omega R_*}=f^{-\ii w/2}\simeq1-\frac{\ii w}{2}\ln f .
\ee
This expression is not only valid close to the horizon. Both $1$ and $\ln f$ solve Eq.~\eqref{eq:v:f:radial-equation2} when its $\omega^2$ term is neglected, as can be checked by direct substitution, so Eq.~\eqref{eq:v:f:phi-exact} is the solution with the correct horizon condition in the whole region $R\ll1/\omega$.

We now write Eq.~\eqref{eq:v:f:phi-exact} for $r\gg GM$, in the variables of Eq.~\eqref{eq:v:f:chimodel}. Using Eq.~\eqref{eq:v:f:radius-map2},
\be
	f=\left(\frac{1-GM/2r}{1+GM/2r}\right)^2,\qquad \ln f=-\frac{2GM}{r}+\mathcal O\left(\frac{(GM)^3}{r^3}\right).
\ee
The field used in the rest of the paper is the rescaled one, $\chi=(AD)^{1/4}\phi$, Eq.~\eqref{eq:helmholtz}, and for the Schwarzschild metric in isotropic coordinates
\be
	(AD)^{1/4}=\sqrt{1-\frac{(GM)^2}{4r^2}}\simeq1-\frac{(GM)^2}{8r^2}.
\ee
Hence, to order $(GM)^2$,
\be \label{eq:v:f:chitrue}
	\chi(r)=1-\frac{(GM)^2}{8r^2}+\frac{4\ii(GM)^2\omega}{r}.
\ee

\subsubsection{Matching}

The two solutions just found are valid in different regions. Equation~\eqref{eq:v:f:chimodel} uses the expansion of the potential in $GM/r$ and the low-frequency approximation, so it holds for $r_{\rm m}<r\ll1/\omega$. Equation~\eqref{eq:v:f:phi-exact} uses only the low-frequency approximation and holds for all $R\ll1/\omega$, down to the horizon; its form~\eqref{eq:v:f:chitrue} holds for $GM\ll r\ll1/\omega$. The two regions overlap around the radius $r_{\rm m}$ of Eq.~\eqref{eq:v:f:rm}, and there both expressions describe the same wave, so they must coincide term by term. The constant terms agree because both waves are normalized to one. The terms $-(GM)^2/(8r^2)$ also agree, and this is a check of the calculation: in Eq.~\eqref{eq:v:f:chitrue} this term comes from the rescaling factor $(AD)^{1/4}$, while in Eq.~\eqref{eq:v:f:chimodel} it comes from $\delta U_{\rm der}$, which was itself generated by the same rescaling in Eq.~\eqref{eq:Umaster}. The only nontrivial condition is the equality of the $1/r$ terms, which gives
\be \label{eq:v:f:C}
	g_{4,0}(r_{\rm m})=\frac{\pi(GM)^2}{r_{\rm m}}-16\pi\ii(GM)^2\omega .
\ee
The first term depends on $r_{\rm m}$ and is the counterterm of Eq.~\eqref{eq:ghat} at $p=4$, $n=0$: it cancels the divergent term of the exterior integral, Eq.~\eqref{eq:v:f:J4ext}. The remaining part,
\be \label{eq:v:f:ghat}
	\hat g_{4,0}\equiv g_{4,0}(r_{\rm m})-\frac{\pi(GM)^2}{r_{\rm m}}=-16\pi\ii(GM)^2\omega ,
\ee
does not depend on the cutoff, as required by Eq.~\eqref{eq:ghat}, and is the finite local term that the exterior potential cannot determine. The derivative term then contributes to the amplification factor as in Eq.~\eqref{eq:F1-renormalized},
\be
	F^{(1)}_{{\rm newt},{\rm der}}\simeq-(\mathcal P/4\pi)\,[\omega^2c_4\mathcal J_4(k)+\hat g_{4,0}] \,.
\ee

The finite part $\hat g_{4,0}$ comes entirely from the horizon. Indeed, the $1/r$ term of the exact solution~\eqref{eq:v:f:chitrue} comes from the coefficient of $\ln f$ in Eq.~\eqref{eq:v:f:phi-exact}, i.e.\ from the horizon condition. For an object without horizon, such as a star, the static solution must instead be regular at the centre, which excludes the $\ln f$ solution: the exact solution would have no $1/r$ term, and the matching would give $\hat g_{4,0}=0$ at this order, i.e.\ only the counterterm. For the black hole, instead, $\hat g_{4,0}$ is imaginary, and it describes the part of the wave that enters the horizon and does not come back, i.e.\ absorption. Neither the point-lens reference nor $F^{(1)}_{{\rm newt},2}$ contain this effect.

The normalization of $\hat g_{4,0}$ can be checked against the known absorption cross section of a black hole. From Eq.~\eqref{eq:helmholtz}, the current $\mathbf j={\rm Im}(\chi^*\nabla\chi)$ satisfies
\be
	\nabla\cdot\mathbf j={\rm Im}\left(\chi^*\nabla^2\chi\right)={\rm Im}\left(4\omega^2U_{\rm eff}\right)|\chi|^2 .
\ee
For a real potential the current is conserved, $\nabla\cdot\mathbf j=0$. Here the only imaginary part of the potential is the local term, ${\rm Im}(4\omega^2U_{\rm eff})={\rm Im}\,\hat g_{4,0}\,\delta^{(3)}(\rr)$, so the current is conserved everywhere except at the origin, where part of the wave disappears into the black hole. Integrating over all space, the flux absorbed per unit time is
\be
	-\int\dd^3r\,\nabla\cdot\mathbf j=-{\rm Im}\,\hat g_{4,0}\,|\chi(0)|^2 .
\ee
For an incident plane wave $\chi=e^{\ii\omega z}$, $|\chi(0)|=1$ and the incident current is $\mathbf j=\omega\,\hat{\mathbf z}$. The absorption cross section, i.e.\ the absorbed flux divided by the incident flux per unit area, is then
\be \label{eq:v:f:absorption-check2}
	\sigma_{\rm abs}=-\frac{{\rm Im}\,\hat g_{4,0}}{\omega}=16\pi(GM)^2=4\pi(2GM)^2,
\ee
the area of the horizon, which is the known low-frequency result for a massless scalar field~\cite{Das:1996we}. The sign of ${\rm Im}\,\hat g_{4,0}$ is fixed by the condition that the wave enters the horizon; an outgoing condition would reverse it and give a negative cross section.

The horizon term is specific to the scalar field: it is the absorption of the monopole, $\ell=0$~\cite{Page:1976df,Sanchez:1977si}. Indeed, in Eq.~\eqref{eq:v:f:ball} the Green functions are evaluated at the origin, so that the local term $g_{4,0}\,\delta^{(3)}(\rr)$ couples only to the value of the wave at the position of the lens, to which only the monopole contributes. Similarly, a contact interaction between Coulomb waves acts only in the S-wave, and a complex coupling describes absorption into other channels~\cite{Kong:1999sf}. Keeping the next orders in the expansion of the Green functions around the origin gives the local terms $(\nabla^2)^n\delta^{(3)}(\rr)$ of Eq.~\eqref{eq:local-terms}, which act on the derivatives of the wave at the origin and describe the absorption of the higher multipoles. For a black hole their coefficients are fixed by matching at the horizon, and they are suppressed by $(GM\omega)^{2n}$ with respect to $g_{4,0}$, i.e.\ they enter at higher orders in $GM$. Note that a spin-2 wave has no monopole or dipole, and the low-frequency absorption cross section of its lowest multipole, $\ell=2$, is suppressed by $(GM\omega)^4$ with respect to the horizon area~\cite{Page:1976df}. The inverse-square optical term, which comes from the propagation speed $D/A$ in the lens metric, is instead the same for any massless field.

\subsection{Complete quadratic amplification at finite distances}
\label{sec:v:f:complete}
With the inner response fixed by matching, we can now determine how it combines with exterior propagation to change the amplification at the observer. We assemble the inverse-square, finite derivative, and horizon contributions at quadratic order in $GM$, retaining the source–lens–observer geometry so that all three can be compared at consistent accuracy. Using the value of $g_{4,0}(r_m)$ in Eq.~\eqref{eq:v:f:C}, the regularized potential~\eqref{eq:v:f:Vmodel} can be written as
\be \label{eq:v:f:matched-insertion}
	\delta U_{\rm der}^{\rm reg}=\delta U_{\rm der}^{\rm finite}+\delta U_{\rm hor},
\ee
where we have defined
\begin{align}
	4\omega^2\,\delta U_{\rm der}^{\rm finite}&=-\frac{(GM)^2}{4}\left[\frac{\Theta(r-r_{\rm m})}{r^4}-\frac{4\pi}{r_{\rm m}}\,\delta^{(3)}(\rr)\right] \nonumber \\
	\label{eq:v:f:Uder-match} &\equiv-\frac{(GM)^2}{4}\,{\rm FP}\frac1{r^4}, \\
	\label{eq:v:f:Uhor} 4\omega^2\,\delta U_{\rm hor}&=-16\pi\ii(GM)^2\omega\,\delta^{(3)}(\rr).
\end{align}
The first is the $r^{-4}$ potential outside $r_{\rm m}$ together with the real part of the point term. We denote the bracket by ${\rm FP}\,r^{-4}$, the finite part of the $r^{-4}$ contribution, for the following reason. When it is integrated against a function that is smooth at the origin, such as the product of Green functions in Eq.~\eqref{eq:F1newt}, the integral of $\Theta(r-r_{\rm m})/r^4$ produces a term $4\pi/r_{\rm m}$ times the value of the function at the origin, which diverges for $r_{\rm m}\to0$; the delta function cancels it exactly. The result is therefore finite, and independent of $r_{\rm m}$ up to corrections of order $\omega r_{\rm m}$, which are negligible by Eq.~\eqref{eq:v:f:rm}. This is the same subtraction used for the free $p=4$ channel in Sec.~\ref{sec:leading}: the Fourier transform of ${\rm FP}\,r^{-4}$ is the cutoff transform $\mathcal J_4^{(r_0)}$ of Sec.~\ref{sec:leading} with $r_0=r_{\rm m}$, minus its cutoff-dependent term $4\pi/r_{\rm m}$, i.e.\ $-\pi^2k$, Eq.~\eqref{eq:J15}. The matching thus shows that, in this subtraction scheme, the only local term left at order $(GM)^2$ is $\delta U_{\rm hor}$. It condenses the eliminated horizon region in a low-frequency effective equation. It  is not a literal point contribution to the Schwarzschild metric. At higher orders in $GM$ the terms proportional to $w$ in the Newtonian dressing of the $p=4$ channel, Sec.~\ref{sec:v:e}, the $p=3$ channel, and the corresponding matching terms must also be included.

Inserting the two potentials in Eq.~\eqref{eq:F1newt}, the derivative term splits as
\be
	F^{(1)}_{{\rm newt},{\rm der}}=F^{(1)\,{\rm finite}}_{{\rm newt},{\rm der}}+F^{(1)}_{{\rm newt},{\rm hor}} \,,
\ee
with
\begin{align}
	F^{(1)\,{\rm finite}}_{{\rm newt},{\rm der}}&=\frac{4\omega^2}{G_{\rm free}(\rr_{\rm O},\rr_{\rm S})}\int\dd^3r\;G_{\rm newt}(\rr_{\rm O},\rr) \nonumber \\
	\label{eq:v:f:Fder-match} &\quad\times\delta U_{\rm der}^{\rm finite}(r)\,G_{\rm newt}(\rr,\rr_{\rm S}), \\
	F^{(1)}_{{\rm newt},{\rm hor}}&=\frac{4\omega^2}{G_{\rm free}(\rr_{\rm O},\rr_{\rm S})}\int\dd^3r\;G_{\rm newt}(\rr_{\rm O},\rr) \nonumber \\
	\label{eq:v:f:Fhor} &\quad\times\delta U_{\rm hor}(\rr)\,G_{\rm newt}(\rr,\rr_{\rm S}),
\end{align}
which, inserted in Eq.~\eqref{eq:v:f:Fsplit}, give the amplification factor at order $(GM)^2$.

At order $(GM)^2$ the Newtonian Green functions can be replaced by free ones in all three terms, since $G_{\rm newt}=G_{\rm free}+\mathcal O(GM)$ and the potentials are already of order $(GM)^2$. Then $F^{(1)}_{{\rm newt},2}$ coincides with the free-wave channel term $F^{(1)}_{{\rm free},2}$ of Eq.~\eqref{eq:Fref1p}, $F^{(1)\,{\rm finite}}_{{\rm newt},{\rm der}}$ is given by Eq.~\eqref{eq:v:f:Fder-match} with free Green functions, and, because of the delta function, the horizon term only requires the Green functions at the origin,
\begin{align}
	F^{(1)}_{{\rm newt},{\rm hor}}&=-16\pi\ii(GM)^2\omega\,\frac{G_{\rm newt}(\rr_{\rm O},0)\,G_{\rm newt}(0,\rr_{\rm S})}{G_{\rm free}(\rr_{\rm O},\rr_{\rm S})} \nonumber \\
	\label{eq:v:f:horizon-explicit} &\simeq4\ii(GM)^2\omega\,\frac{d_{\rm SO}}{r_{\rm O}r_{\rm S}}\,e^{\ii\omega(r_{\rm O}+r_{\rm S}-d_{\rm SO})},
\end{align}
where in the second line we used $G_{\rm newt}\simeq G_{\rm free}$ and Eq.~\eqref{eq:G0}. With free Green functions the three terms retain the full source--lens--observer geometry, with no Fraunhofer or Fresnel approximation.

The free-Green-function expressions use the normalization of Eq.~\eqref{eq:Fdef}, while Eqs.~\eqref{eq:FhatNewt} and~\eqref{eq:v:d:p-two-response} are multiplied by $e^{-\ii(w/2)\ln(X/Y)}$. A common phase applied to both the reference and the response leaves their ratio and intensities unchanged, and since this phase is $1+\mathcal O(GM)$ at fixed frequency and geometry, applying it to the quadratic response changes only cubic and higher orders.

\subsection{The horizon contribution at finite distance}
\label{sec:v:f:aligned}

The horizon acts on the wave only through a local term at the centre of the lens, whose effect at the observer decreases with the distances faster than the leading optical response: to see it, the finite distances of source and observer must be kept in all the terms.

The horizon term of Eq.~\eqref{eq:v:f:horizon-explicit} is proportional to $d_{\rm SO}/(r_{\rm S}r_{\rm O})$, which at small angle is $1/r_{\rm eff}$, so that it is of order $(GM)^2\omega/r_{\rm eff}=\varepsilon w^2/16$, with $\varepsilon=1/(\omega r_{\rm eff})$ of Eq.~\eqref{eq:sepsx}. The leading $p=2$ term, Eq.~\eqref{eq:v:f:E2-limits}, is instead of order $\alpha_Ew^{3/2}=\sqrt\varepsilon\,w^2$. The horizon term is therefore suppressed by $\sqrt\varepsilon$ with respect to the leading $p=2$ term, and it is of the same order as the corrections to the $p=2$ term due to the finite distances of source and observer, which are neglected in the Fresnel result of Sec.~\ref{sec:v:d}. Its effect on the field at the observer can therefore be assessed only if all the terms are computed at least to first order in $\sqrt\varepsilon$ beyond the leading one. The scattering of waves by a Schwarzschild black hole with the observer at finite distance, and its comparison with the diffraction integral, was also recently studied with black-hole perturbation theory in~\cite{Li:2025gou}; see also~\cite{Nambu:2015aea}.

For general positions of source and observer the integrals cannot be done in closed form. We therefore consider the aligned and symmetric configuration, with the lens midway between source and observer on their common axis, $\rr_{\rm O}=-\rr_{\rm S}$ and $r_{\rm O}=r_{\rm S}$. Then $y=0$, $r_{\rm eff}=r_{\rm S}/2$, $d_{\rm SO}=2r_{\rm S}$, and $\varepsilon=2/(\omega r_{\rm S})$.

Using the free Green function~\eqref{eq:G0} in Eqs.~\eqref{eq:Fref1p}, \eqref{eq:v:f:Fder-match} and~\eqref{eq:v:f:horizon-explicit}, with the potentials of Eq.~\eqref{eq:v:f:optical-potential}, the three terms of $F^{(1)}_{\rm newt}$ are
\begin{align}
	\label{eq:v:f:aligned-p2} F^{(1)}_{{\rm newt},2}&=\frac{15(GM)^2\omega^2r_{\rm S}}{4\pi}\int\dd^3r\,\frac{e^{\ii\omega(d_{\rm S}+d_{\rm O}-2r_{\rm S})}}{d_{\rm S}\,d_{\rm O}\,r^2}, \\
	\label{eq:v:f:aligned-der} F^{(1)\,{\rm finite}}_{{\rm newt},{\rm der}}&=\frac{(GM)^2r_{\rm S}}{8\pi}\int\dd^3r\,\frac{e^{\ii\omega(d_{\rm S}+d_{\rm O}-2r_{\rm S})}}{d_{\rm S}\,d_{\rm O}}\,{\rm FP}\frac{1}{r^4}, \\
	\label{eq:v:f:aligned-hor} F^{(1)}_{{\rm newt},{\rm hor}}&=\frac{8\ii(GM)^2\omega}{r_{\rm S}},
\end{align}
where ${\rm FP}$ is the finite part of Eq.~\eqref{eq:FPgeneral}. The integrals extend over the whole space, including the regions close to the source and to the observer. They can be reduced to one-dimensional integrals, which can be evaluated numerically for any $\varepsilon$ and expanded for $\varepsilon\ll1$, as shown in Appendix~\ref{app:aligned}. The expansion at fixed $w$ gives
\begin{align}
	\frac{F^{(1)}_{{\rm newt},2}}{w^2}={}&\frac{15\pi^{3/2}}{128}(1+\ii)\,\varepsilon^{1/2}-\frac{15\ii}{64}\,\varepsilon \nonumber \\
	&+\frac{15\pi^{3/2}}{4096}(1-\ii)\,\varepsilon^{3/2}-\frac{5}{128}\,\varepsilon^2+\cdots, \nonumber \\
	\label{eq:v:f:optical-asymptotic} \frac{F^{(1)\,{\rm finite}}_{{\rm newt},{\rm der}}}{w^2}={}&-\frac{\pi^{3/2}}{512}(1-\ii)\,\varepsilon^{3/2}+\cdots, \\
	\frac{F^{(1)}_{{\rm newt},{\rm hor}}}{w^2}={}&\frac{\ii}{4}\,\varepsilon . \nonumber
\end{align}
The leading term of $F^{(1)}_{{\rm newt},2}$ is the Fresnel result, Eq.~\eqref{eq:v:f:E2-limits} at $s=0$. At the next order, $\varepsilon w^2$, the finite-distance correction of the $p=2$ term, $-15\ii\varepsilon w^2/64$, and the horizon term, $+16\ii\varepsilon w^2/64$, have opposite signs, and their sum is only $\ii\varepsilon w^2/64$: fifteen sixteenths of the horizon term are cancelled by the finite-distance $p=2$ term. The two terms have a different origin. The inverse-square channel comes from the exterior metric, which is the same for any spherically symmetric lens of mass $M$, while the horizon term is fixed by the inner boundary condition and is absent for a non-absorbing lens. The difference between the amplification by a black hole and by a non-absorbing lens of the same mass is therefore the horizon term alone, $+i\varepsilon w^2/4$, and the cancellation does not reduce it. What the cancellation reduces is the total correction to the Fresnel result at this order: adding the horizon term without the finite-distance correction of the $p=2$ term would overestimate it by a factor of sixteen. The cancellation concerns the field at the observer, and not the flux absorbed by the horizon, Eq.~\eqref{eq:v:f:absorption-check2}, which remains positive. The derivative term starts at order $\varepsilon^{3/2}w^2$. Collecting the three terms,
\begin{align}
	\frac{F^{(1)}_{\rm newt}}{w^2}={}&\frac{15\pi^{3/2}}{128}(1+\ii)\,\varepsilon^{1/2}+\frac{\ii}{64}\,\varepsilon \nonumber \\
	\label{eq:v:f:complete-alpha} &+\frac{7\pi^{3/2}}{4096}(1-\ii)\,\varepsilon^{3/2}-\frac{5}{128}\,\varepsilon^2+\cdots,
\end{align}
so that the Fresnel result, i.e. the first term in the equation above, gives the complete quadratic response up to a relative correction of order $\sqrt\varepsilon$.

To compare the full amplification factor $F=F^{(0)}_{\rm newt}+F^{(1)}_{\rm newt}$ with the standard point-lens template, we must retain these finite distance corrections in $F^{(0)}_{\rm newt}$ as well. To this end, we improve Eq.~\eqref{eq:pointlensRawAmplification} as in Eq.~\eqref{eq:F0-finite-distance} and we evaluate the latter in the same configuration: at perfect alignment and equal distances, i.e. at $Y=0$ and $\omega X= 4 r_s = 8/\varepsilon$. In particular, we obtain,
\begin{align}
	\label{eq:F0-aligned} F^{(0)}_{\rm newt} &= e^{\pi w/4}\,\Gamma \lp 1-\frac{\ii w}{2} \rp \lp \frac{8}{\varepsilon} \rp^{\frac{\ii w}{2}} \llp 1 - \frac{w (2-\ii w )}{32} \varepsilon \rrp
\end{align}
up to ${\cal O}(\varepsilon^2)$ terms. The term linear in $w$ of Eq.~\eqref{eq:F0-aligned}, including $-w\varepsilon/16$, can be checked independently of the Hostler--Pratt solution, from the first Born term of the Newtonian channel with free Green functions, as shown in Appendix~\ref{app:aligned}. Expanding also in $w \ll 1$, we obtain the small frequency Newtonian solution
\begin{align}
	\label{eq:F0-aligned-smallfreq} F^{(0)}_{\rm newt} \simeq 1 + w \llp \frac{\pi}{4} - \frac{\varepsilon}{16} + \frac{\ii}{2} \lp \gamma_E + \ln\frac{8}{\varepsilon} \rp \rrp + {\cal O} \lp w^2, w\varepsilon^2\rp \,.
\end{align}

Combining Eqs.~\eqref{eq:v:f:complete-alpha} and~\eqref{eq:F0-aligned-smallfreq}, we estimate the fractional variation of the amplification factor when including the first-order-Born contributions
\begin{align}
	\frac{F}{F^{(0)}_{\rm newt}}={}1+w^2&\bigg[\frac{15\pi^{3/2}}{128}(1+\ii)\,\varepsilon^{1/2}+\frac{\ii}{64}\,\varepsilon \nonumber \\
	\label{eq:F-total-aligned} &+\frac{7\pi^{3/2}}{4096}(1-\ii)\,\varepsilon^{3/2}-\frac{5}{128}\,\varepsilon^2\bigg] \\
	&+ \mathcal O\big(w^3,\,w^2\varepsilon^{5/2}\big)\,, \nonumber
\end{align}
where all the corrections to $F^{(0)}_{\rm newt}=1+\mathcal O(w)$, including the finite-distance ones of Eq.~\eqref{eq:F0-aligned-smallfreq}, multiply $F^{(1)}_{\rm newt}=\mathcal O(w^2)$ and therefore enter at order $w^3$. The relative change of the wave due to the structure of the black hole is therefore $F^{(1)}_{\rm newt}$ alone: at order $\varepsilon w^2$ its coefficient $\ii/64$ is the sum of $+16\ii/64$ from the horizon and $-15\ii/64$ from the finite-distance inverse-square response.

At exact alignment the terms of order $\varepsilon$ discussed above are not separately observable. Since $w\varepsilon=4GM/r_{\rm eff}$ does not depend on frequency, the term $-w\varepsilon/16$ rescales the amplitude, and is degenerate with the distance of the source. Since $w^2\varepsilon=16(GM)^2\omega/r_{\rm eff}$ is linear in frequency, each term $i\,c\,w^2\varepsilon$ is the first-order expansion of $e^{i\omega\,\delta t}$ with $\delta t=16\,c\,(GM)^2/r_{\rm eff}$, i.e.\ a shift of the arrival time of the whole signal (as the infrared phase of the tail effect is removed by a redefinition of time~\cite{Galley:2015kus}), as for the phase removed below Eq.~\eqref{eq:pointlensSeparatedPhase}. This holds for the horizon term, for the finite-distance $p=2$ term and for the point-lens term separately, so that at $y=0$ the cancellation is one among arrival-time shifts. Away from alignment the horizon term is distinguishable: by Eq.~\eqref{eq:v:f:horizon-explicit} it carries the phase $\eu^{\ii\omega(r_{\rm O}+r_{\rm S}-d_{\rm SO})}$ of the wave scattered by the lens, and it modulates the amplification as a function of frequency. 

Finally, $F$ is the amplification of the rescaled field $\chi$ of Eq.~\eqref{eq:helmholtz}. The physical field is $\phi=(AD)^{-1/4}\chi$, and for the Schwarzschild metric $(AD)^{-1/4}\simeq1+(GM)^2/(8r^2)$. At the observer this is a frequency-independent factor $1+w^2\varepsilon^2(1-a)^2/128$, with an analogous factor fixed by the normalization of the source. It is degenerate with the overall amplitude of the signal, but it is of the same order as the $\varepsilon^2w^2$ term of Eq.~\eqref{eq:v:f:complete-alpha}, which should therefore not be read as a prediction for the physical field without it.

\begin{figure*}[t]
\centering
\includegraphics[width=\textwidth]{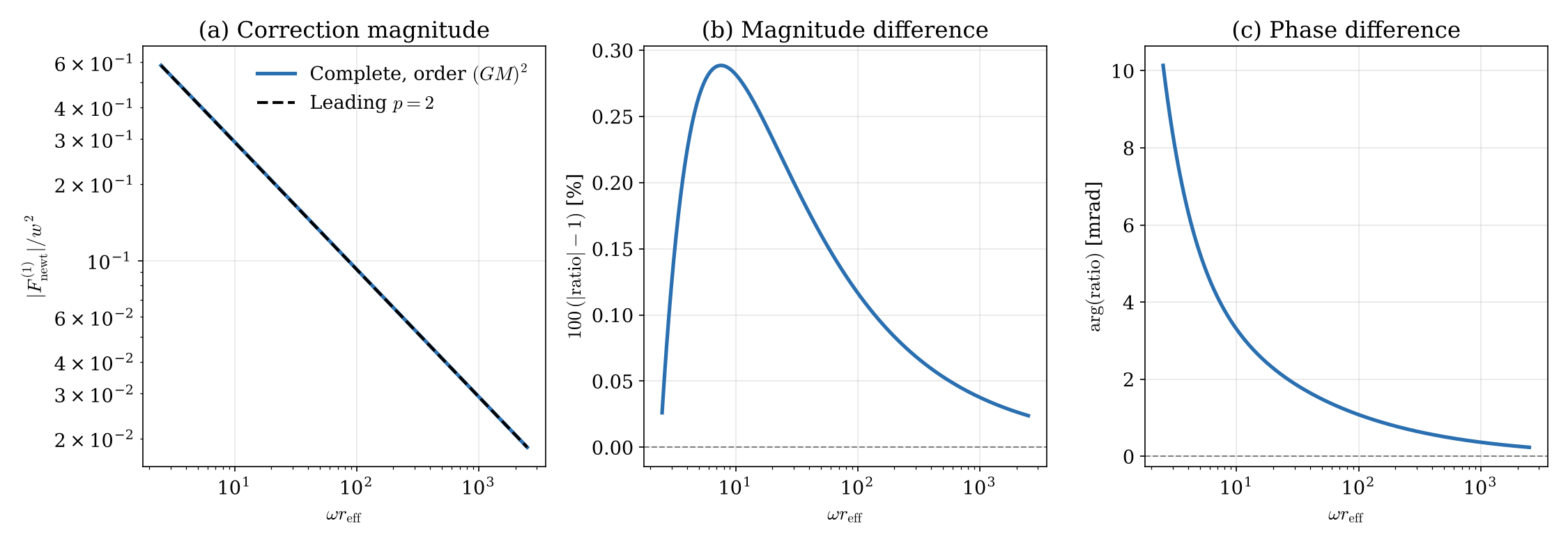}
\caption{Complete quadratic correction $F^{(1)}_{\rm newt}$, Eqs.~\eqref{eq:v:f:aligned-p2}--\eqref{eq:v:f:aligned-hor} evaluated with the one-dimensional integrals of Eq.~\eqref{eq:v:f:exact-aligned}, and the leading $p=2$ term, the first term of Eq.~\eqref{eq:v:f:complete-alpha}, for aligned source and observer at equal distances, as functions of $\omega r_{\rm eff}$. (a) Magnitudes divided by $w^2$: the complete result (solid) and the leading $p=2$ term (dashed) nearly overlap. (b) Percentage difference of the magnitudes. (c) Phase difference, in milliradians. The complete result includes the finite-distance $p=2$ terms, the matched derivative contribution and the horizon response. The curves compare coefficients quadratic in $GM$ at low frequency, not the full amplification factor or the observed strain.}
\label{fig:v:f:comparison}
\end{figure*}

Figure~\ref{fig:v:f:comparison} compares $F^{(1)}_{\rm newt}$ with the first term of Eq.~\eqref{eq:v:f:complete-alpha} at finite $\varepsilon$, from the numerical evaluation of the one-dimensional integrals of Eq.~\eqref{eq:v:f:exact-aligned}, which does not rely on the expansion in $\varepsilon$. Over $2.5\le\omega r_{\rm eff}\le2500$ the magnitudes of the complete correction and of the leading $p=2$ term differ by at most $0.29\%$, near $\omega r_{\rm eff}\simeq7.6$, and their phases by at most $10$ milliradians, at $\omega r_{\rm eff}=2.5$. These numbers compare the coefficients of $(GM)^2$, not the full amplification factors.

\subsection{Eikonal limit}
\label{sec:v:f:geometric}

The need for inner matching raises the question of why the usual eikonal diffraction integral can nevertheless provide an accurate leading prediction. We compare the relevant propagation scales to identify when inner-region effects are subleading and clarify the distinction between resumming the exterior phase and determining the physical horizon response. The divergences of the singular channels, and therefore the matching, come from truncating the Born series at a fixed order, and they are absent when the entire series can be resummed. This is possible in the eikonal limit, $\omega b\gg1$, where, as shown in Sec.~\ref{sec:eikonal}, the Born series is the expansion of the exponential of the diffraction integral, Eq.~\eqref{eq:overview-eikonal}, in powers of the lensing phase. For the singular channels the phase diverges at small $\zeta$, and the transverse integral of each Born term diverges at $\zeta\to0$, so that it requires local terms, as discussed below Eq.~\eqref{eq:breakdownCounting}. The resummed integrand $\eu^{\ii\psi}$, instead, has modulus one, so that the diffraction integral has no divergence, and no local terms are needed. This does not require geometric optics: the diffraction integral holds for any $w$, as long as $\omega\zeta\gg1$ in the regions that dominate it, such as $\zeta\sim r_{\rm F}$ and $\zeta\sim b$ in Sec.~\ref{sec:v:d}, for which $(\omega r_{\rm F})^2=1/\varepsilon$ and $\omega b=kr_{\rm eff}$.

In this Section, instead, we work at a fixed order in $GM$, i.e.\ with a single insertion, and the singular channels must be renormalized, with local terms fixed by matching. The local terms describe the region within a wavelength of the lens, where the eikonal description does not hold: $\omega\zeta\gg1$ fails, and the lens cannot be treated as thin, i.e.\ its finite size and the departures from the paraxial approximation matter. This region gives contributions suppressed by powers of $\varepsilon$, of the same order as the finite-distance corrections, which are also absent from the diffraction integral. The horizon term is suppressed by $\sqrt\varepsilon$ with respect to the leading $p=2$ term, Eq.~\eqref{eq:v:f:optical-asymptotic}, and the derivative term by $\varepsilon$, in agreement with the ratio of the phases of the two channels, Eq.~\eqref{eq:v:d:delta-phi-p}, at the Fresnel scale,
\be \label{eq:v:f:derivative-ray}
	\frac{\psi_{\rm der}(\zeta)}{\psi_2(\zeta)}=\frac{1}{60(\omega\zeta)^2}\,,\qquad \frac{\psi_{\rm der}(r_{\rm F})}{\psi_2(r_{\rm F})}=\frac{\varepsilon}{60}\,.
\ee
Our matched result cannot be extended to $w\gtrsim1$, because there the higher orders in $GM$ are not suppressed, as discussed at the beginning of this Section, and the low-frequency solution close to the horizon, Eq.~\eqref{eq:v:f:phi-exact}, does not hold.

\subsection{Size of the corrections}
\label{sec:size}
We estimate the size of the corrections in the one configuration for which all the contributions at order $(GM)^2$ have been computed: aligned source and observer at low frequency, $w\ll1$. At leading order in $\varepsilon$, Eq.~\eqref{eq:F-total-aligned} gives
\be \label{eq:size-low}
	\frac{F}{F^{(0)}_{\rm newt}}-1\simeq 0.65\,(1+\ii)\,\alpha_E\,w^{3/2},
\ee
with $\alpha_E=\sqrt{4GM/r_{\rm eff}}=\sqrt{w\varepsilon}$. The horizon and derivative terms enter at higher orders in $\sqrt\varepsilon$. The real part of Eq.~\eqref{eq:size-low} is the fractional change of the amplitude and its imaginary part is the change of the phase, in radians: the two are equal. The frequency is related to the lens mass by $w\simeq1.24\times10^{-4}\,(M/M_\odot)\,(f/{\rm Hz})$.

For a lens at cosmological distance the effect is negligible: with $r_{\rm eff}=0.5$~Gpc and $M=10^2$--$10^6\,M_\odot$ one has $\alpha_E\simeq 10^{-10} - 10^{-8}$, and the correction is below $10^{-7}$. It becomes appreciable only when the source, or the observer, is within $10^2$--$10^4$ gravitational radii of the lens, so that $r_{\rm eff}\simeq r_S$ and $\alpha_E\simeq2\sqrt{GM/r_S}$, as shown in Table~\ref{tab:size}.

\begin{table}[t]
\caption{\label{tab:size}Size of the correction at
alignment for $w=0.1$, from Eq.~\eqref{eq:size-low}. The last column
is the fractional change of the amplitude,
$\delta|F|/|F|\simeq0.65\,\alpha_E\,w^{3/2}$; the phase changes by the same number of radians.}
\begin{ruledtabular}
\begin{tabular}{cccc}
$r_{\rm eff}/GM$ & $\alpha_E$ & $\varepsilon$ & $\delta|F|/|F|$\\
\hline
$10^6$ & $2.0\times10^{-3}$ & $4.0\times10^{-5}$ & $4\times10^{-5}$\\
$10^4$ & $2.0\times10^{-2}$ & $4.0\times10^{-3}$ & $4\times10^{-4}$\\
$10^3$ & $6.3\times10^{-2}$ & $4.0\times10^{-2}$ & $1.3\times10^{-3}$\\
$10^2$ & $2.0\times10^{-1}$ & $4.0\times10^{-1}$ & $4\times10^{-3}$\\
\end{tabular}
\end{ruledtabular}
\end{table}

As a rough criterion, a correction $\delta F$ can be distinguished from the point-lens template when $|\delta F|\,\rho\gtrsim1$, with $\rho$ the signal-to-noise ratio, up to degeneracies with the lens and source parameters. Corrections at the $10^{-3}$ level and above therefore require $r_{\rm eff}\lesssim10^3\,GM$, as for a binary merging in the vicinity of a massive black hole, e.g.\ in hierarchical triples (see e.g. \cite{Pijnenburg:2024btj,Yu:2021dqx,Cardoso:2026ugm}), with $w\ll1$ in the band of the detector: $M\sim10^5$--$10^6\,M_\odot$ at $f\sim10^{-3}$~Hz, or $M\lesssim10^2\,M_\odot$ at $f\sim10$~Hz. In such compact configurations $\varepsilon$ is not very small, and the finite-distance terms of Eqs.~\eqref{eq:F-total-aligned} and~\eqref{eq:F0-aligned} must be retained.

\section{Conclusions}
\label{sec:conclusions}

We have developed a way to calculate how a radial deformation changes a waveform that has already been lensed by a Newtonian point mass. The calculation starts from the three-dimensional scalar wave equation and uses the exact Newtonian Green function for propagation on both sides of the additional interaction. The resulting distorted-wave Born expansion separates three pieces of the physics that enter the measured field. The reference wave carries the long-range phase and focusing of the point mass. The dressed radial kernels describe how the exterior deformation changes that wave. For a compact lens, matching supplies the inner response selected by the boundary condition.

The free-wave calculation in Sec.~\ref{sec:born} explains the choice of reference. The Newtonian potential is weak over most of the path, but its long tail gives an accumulated phase that grows logarithmically with the source and observer distances. In a free-wave Born expansion, even a single Newtonian insertion can strongly enhance the response to a faster-decaying channel. This is explicit in the mixed Newtonian and inverse-square term, which can exceed the single inverse-cube contribution at the same mass order. Using the exact Coulomb propagator incorporates this phase into the reference solution, along the lines suggested for relativistic corrections by Carrillo-Gonz\'alez, Heavens and Whiteway \cite{Carrillo-Gonzalez:2026}. Once this is done, the remaining accuracy question is how many insertions of the additional potential are required. The radial falloff, the mass order, and the number of insertions each supply different information about the size of a correction.

Section~\ref{sec:beyondborn} makes this organization concrete by calculating the linear response to individual radial powers. In the Fraunhofer approximation, the propagation integral for a screened potential can be evaluated once, leaving a single inverse-range integral for each power. For the convergent channels, we obtain closed expressions with the angular dependence and Newtonian coupling unexpanded. The inverse-square channel gives a particularly clear example of what the resummation does. At leading order in small angle, the dressed-to-free kernel ratio has unit modulus and carries the phase whose expansion produces the mixed Born logarithm. This phase changes how the correction interferes with the reference wave, so even a pure phase of the kernel can affect the observed intensity.

The comparison with the Fresnel calculation identifies a limitation of using the scattering contribution alone. Wavefront curvature can be small over the lens-centered region that determines the scattered response and essential around the unlensed line of sight, where it allows a separate contribution to add coherently to the direct wave. Retaining curvature also gives a finite aligned response for the convergent channels. Far from alignment, at fixed Newtonian coupling in the shared regime of the two approximations, the direct response dominates for convergent profiles shallower than inverse square and is relatively suppressed for steeper ones. The two corrections have equal leading magnitudes for the inverse-square channel. In the image limit this equality reflects a balance between the larger phase shift of the secondary image and its smaller amplitude. The reliability of the Fraunhofer approximation for the scattered wave therefore tells us only part of what we need to know about the coherent amplification.

The short-distance singularities of steeper radial powers have a different origin from the forward divergence regulated by wavefront curvature. They arise when an exterior power law is continued to the center. Subtraction isolates the exterior kernel, and the physical local contribution is fixed by the inner solution. The Schwarzschild application in Sec.~\ref{sec:v:f} illustrates this distinction. Its inverse-fourth-power derivative term already enters at quadratic order in the mass and requires a short-distance subtraction. Matching the low-frequency exterior monopole to a purely ingoing solution at the horizon fixes the local term and removes the dependence on the matching radius at the retained order. The matched scalar amplification combines the inverse-square optical response, the finite derivative contribution, and the horizon term. The absorptive part reproduces the known low-frequency cross section equal to the horizon area.

The aligned, equal-distance example shows how the horizon term appears in the field measured by an observer. The leading Fresnel inverse-square response dominates the quadratic correction. The first finite-distance optical correction and the leading horizon term enter at the same subleading distance order. The optical correction cancels fifteen sixteenths of the horizon contribution in the forward amplitude, while the flux into the horizon remains positive. This explains why the leading Fresnel result can be close to the matched quadratic correction even when both subleading terms are individually appreciable at their distance order. The comparison isolates the quadratic mass coefficient in this geometry. A prediction for the full waveform at finite frequency also has a separate error from the omitted mass orders.

The calculation gives a starting point for studying lenses with a specified exterior profile and different inner boundary conditions. An extended environment adds its own radial scales to the propagation problem, while a different compact-object boundary requires a new matching calculation~\cite{Nambu:2019sqn}. Higher-order scalar scattering and independent partial-wave calculations provide ways to test those extensions and the transition away from the low-frequency limit. For gravitational waves the leading inverse-square response is unchanged, while the horizon term computed here, the absorption of the scalar monopole, has no counterpart at order $(GM)^2$. A calculation with tensor polarizations requires the spin-2 wave equations and their own inner matching data \cite{Tambalo:2022wlm,Caron-Huot:2025tlq,Li:2025lvl,Dolan:2007ut,Dolan:2008kf,Bautista:2021wfy,Cusin:2019rmt,Andersson:2020gsj,Dalang:2021qhu,Kubota:2024zkv, Li:2025gou,Li:2026dzr}. Results covering these points will be presented elsewhere.

\acknowledgments
We thank Cédric Deffayet for insightful conversations. GT is partially supported by the STFC grant [UKRI4243], by the Royal Society grant IES R3 43186, and by the Leverhulme Trust grant RF-2026-166 9. Standard AI models  assisted us in brainstorming, improving the presentation,  and developing some of our technical   results, which we then  checked numerically and analytically.
\appendix

\section{Integral representations behind the reduction}
\label{app:angular}
This appendix records enough intermediate structure to check the reduction without treating a special-function identity as a new physical approximation.

Temporarily take $0<\operatorname{Re}\kappa<1$. Euler's representation is
\[
	{}_1F_1(\kappa;1;Z)= \frac1{\Gamma(\kappa)\Gamma(1-\kappa)} \int_0^1\dd u\,u^{\kappa-1}(1-u)^{-\kappa}\eu^{Zu}.
\]
Insert one copy for each Coulomb wave in $K(\lambda)$. The spatial integral becomes an elementary Yukawa transform with complex inverse range $A=\lambda-\ii\omega(u+v)$ and wavevector $\bm Q=\kk-\omega(u\bn_{\rm S}+v\bn_{\rm O})$:
\[
	\int\dd^3\rr\,\frac{\eu^{-Ar-\ii\bm Q\cdot\rr}}r =\frac{4\pi}{A^2+\bm Q^2}.
\]
For the elastic kinematics of this section its denominator is
\[
	D(u,v)=\lambda^2+k^2 -(k^2+2\ii\lambda\omega)(u+v) -(4\omega^2-k^2)uv .
\]
Consequently
\begin{align}
	K(\lambda)={}&\frac{4\pi} {[\Gamma(\kappa)\Gamma(1-\kappa)]^2} \nonumber \\
	\label{eq:EulerNordsieckHere} &\times\int_0^1\!\!\int_0^1 \frac{u^{\kappa-1}v^{\kappa-1} (1-u)^{-\kappa}(1-v)^{-\kappa}} {D(u,v)}\,\dd u\,\dd v .
\end{align}
The beta-integral reduction of this Nordsieck representation, followed by a Pfaff transformation of the Gauss function, gives Eq.~\eqref{eq:v:a:nordsieck} with $2\kappa=\ii w$. Continuation to imaginary $\kappa$ retains the outgoing prescription. This is a mathematical continuation, not an expansion in $w$. The free transform and direct three-dimensional integration give independent normalization checks.

\section{Mellin--Barnes evaluation of the power-law kernel}
\label{app:MB}
Temporarily continue $\alpha=\ii w$ to $0<\operatorname{Re}\alpha<3-p$ with $1<p<3$. Use
\[
	(1+\ii\lambda/2\omega)^{-\alpha} =\frac1{2\pi\ii\,\Gamma(\alpha)} \int_{\cal C}\dd t\,\Gamma(-t)\Gamma(\alpha+t) \pa{\frac{\ii\lambda}{2\omega}}^t .
\]
Choose a vertical contour with real part $c$ satisfying $\max(-\operatorname{Re}\alpha,1-p)<c<0$. The Mellin integral of the Gauss function is
\begin{align*}
	\int_0^\infty\dd u\,u^{v-1}{}_2F_1(a,a;1;-u) &=\frac{\Gamma(v)\Gamma(a-v)^2} {\Gamma(a)^2\Gamma(1-v)}, \\
	0<\operatorname{Re}v&<\operatorname{Re}a .
\end{align*}
With $u=k^2/\lambda^2$, one obtains
\begin{align}
	I_p={}&\frac{k^{p-3+\alpha}} {2\Gamma(1-\alpha/2)^2\Gamma(\alpha)} \frac1{2\pi\ii}\int_{\cal C}\dd t \nonumber \\
	\label{eq:MBmasterHere} &\times\Gamma(-t)\Gamma(\alpha+t)\pa{\frac{\ii k}{2\omega}}^t \\
	&\times \frac{\Gamma(\frac{3-p-\alpha-t}{2}) \Gamma(\frac{p-1+t}{2})^2} {\Gamma(\frac{p-1+\alpha+t}{2})}. \nonumber
\end{align}
For $k/(2\omega)<1$, closing to the right gives the even and odd poles of $\Gamma(-t)$ and the poles $t=3-p-\alpha+2n$. These produce the three families in Eq.~\eqref{eq:v:b:three-families}. Continue their sum to $\alpha=\ii w$. Colliding poles and limiting values require the sum, not an isolated family. This construction also explains why analytic continuation beyond $p=3$ must be accompanied by the short-distance subtractions in the main text.

\section{Finite-distance integrals at alignment}
\label{app:aligned}

In this appendix we evaluate the integrals~\eqref{eq:v:f:aligned-p2} and~\eqref{eq:v:f:aligned-der} of Sec.~\ref{sec:v:f:aligned}, and derive the expansion~\eqref{eq:v:f:optical-asymptotic}. We place the lens at the origin, the source at $z=-r_{\rm S}$ and the observer at $z=+r_{\rm S}$ on the $z$ axis, and use cylindrical coordinates $(\rho,z,\theta)$ around it, with $\dd^3r=\rho\,\dd\rho\,\dd z\,\dd\theta$. The distances of the integration point from the source, from the observer and from the lens are
\be
	d_{\rm S}=\sqrt{\rho^2+(z+r_{\rm S})^2},\qquad d_{\rm O}=\sqrt{\rho^2+(z-r_{\rm S})^2},
\ee
and the integrals in Eqs.~\eqref{eq:v:f:aligned-p2} and~\eqref{eq:v:f:aligned-der} to compute are
\begin{align}
	&\int\dd\theta\,\dd z\,\rho\,\dd\rho\;\frac{e^{\ii\omega(d_{\rm S}+d_{\rm O}-2r_{\rm S})}}{d_{\rm S}\,d_{\rm O}}\,\frac{1}{\rho^2+z^2}, \nonumber \\
	\label{eq:app:cyl} &\int\dd\theta\,\dd z\,\rho\,\dd\rho\;\frac{e^{\ii\omega(d_{\rm S}+d_{\rm O}-2r_{\rm S})}}{d_{\rm S}\,d_{\rm O}}\,{\rm FP}\frac{1}{(\rho^2+z^2)^2}.
\end{align}
The phase depends only on the total path length $d_{\rm S}+d_{\rm O}$. We therefore replace $\rho$ and $z$ by two variables $\tau\ge0$ and $-1\le\eta\le1$, defined by
\be
	d_{\rm S}=r_{\rm S}(1+\tau+\eta),\qquad d_{\rm O}=r_{\rm S}(1+\tau-\eta),
\ee
so that $\tau=(d_{\rm S}+d_{\rm O}-2r_{\rm S})/(2r_{\rm S})$ is the excess of the path length over the direct one, in units of $2r_{\rm S}$, and $\eta$ the difference of the two legs. Solving for the cylindrical coordinates gives
\be
	z=r_{\rm S}(1+\tau)\,\eta,\qquad \rho^2=r_{\rm S}^2\,\tau(2+\tau)(1-\eta^2).
\ee
In these coordinates, the integrals above become
\begin{align}
	&\int_0^{2\pi}\dd\theta\int_0^\infty\dd\tau\,e^{4\ii\tau/\varepsilon}\int_{-1}^{1}\frac{\dd\eta}{\tau(2+\tau)+\eta^2}, \nonumber \\
	&\int_0^{2\pi}\dd\theta\;{\rm FP}\!\int_0^\infty\dd\tau\,e^{4\ii\tau/\varepsilon}\int_{-1}^{1}\frac{\dd\eta}{\big[\tau(2+\tau)+\eta^2\big]^2} \,,
\end{align}
with $\varepsilon = 2 /(\omega r_{\rm S})$, the finite-distance parameter in the aligned, equidistant configuration. The integral over $\theta$ gives $2\pi$, and those over $\eta$ can be computed analytically, so that one is left with
\be \label{eq:v:f:exact-aligned}
	4 \pi \int_0^\infty\dd\tau\,e^{4\ii\tau/\varepsilon}\,\mathcal A_2(\tau),\quad 2 \pi \;{\rm FP}\!\int_0^\infty\dd\tau\,e^{4\ii\tau/\varepsilon}\,\mathcal A_4(\tau),
\ee
where
\begin{align}
	\mathcal A_2(\tau)&=\frac{\arctan\!\big(1/\sqrt{\tau(2+\tau)}\big)}{\sqrt{\tau(2+\tau)}}, \nonumber \\
	\label{eq:v:f:prolate-kernels} \mathcal A_4(\tau)&=\frac{1}{\tau(2+\tau)(1+\tau)^2}+\frac{\arctan\!\big(1/\sqrt{\tau(2+\tau)}\big)}{\big[\tau(2+\tau)\big]^{3/2}}.
\end{align}
The outgoing prescription is implemented by a convergence factor $e^{-\epsilon\tau}$ with $\epsilon\to0^+$. The second integral also requires a subtraction at $\tau=0$, because $\mathcal A_4\simeq\pi\tau^{-3/2}/(4\sqrt2)$; the matched finite part is
\be
	{\rm FP}\!\int_0^\infty\dd\tau\,e^{4\ii\tau/\varepsilon}\mathcal A_4 =\lim_{\delta\to0^+}\left[\int_\delta^\infty\dd\tau\,e^{4\ii\tau/\varepsilon}\mathcal A_4-\frac{\pi}{2\sqrt{2\delta}}\right],
\ee
which is the continuation of the subtraction in Eq.~\eqref{eq:v:f:Uder-match} to these variables and introduces no adjustable contact coefficient. For $\varepsilon\ll1$ the rapidly oscillating phase cancels the contributions of smooth regions at nonzero $\tau$, and the expansion is determined by the behavior at $\tau\to0$, where the path length approaches that of the direct path. To expand $\mathcal A_2$ and $\mathcal A_4$ we use
\begin{align}
	\frac{1}{\sqrt{\tau(2+\tau)}}&=\frac{1}{\sqrt{2\tau}}\Big(1-\frac\tau4+\dots\Big), \nonumber \\
	\arctan\frac1x&=\frac\pi2-x+\frac{x^3}{3}+\dots,
\end{align}
with $x=\sqrt{\tau(2+\tau)}$, so that $x^2=2\tau+\tau^2$. Hence
\begin{align}
	\mathcal A_2={}&\frac{\pi}{2\sqrt2}\tau^{-1/2}-1-\frac{\pi}{8\sqrt2}\tau^{1/2}+\frac23\tau+\mathcal O(\tau^{3/2}), \nonumber \\
	\label{eq:v:f:endpoint-series} \mathcal A_4={}&\frac{\pi}{4\sqrt2}\tau^{-3/2}-\frac{3\pi}{16\sqrt2}\tau^{-1/2}-\frac23+\mathcal O(\tau^{1/2}).
\end{align}
Each power is integrated with
\be \label{eq:v:f:Abel-integral}
	\int_0^\infty\dd\tau\,e^{4\ii\tau/\varepsilon}\tau^{a-1}=\Gamma(a)\,e^{\ii\pi a/2}\left(\frac\varepsilon4\right)^{a},
\ee
where analytic continuation to $a=-1/2$ implements the finite part. With this we find
\begin{align}
	{}&\int_0^\infty\dd\tau\,e^{4\ii\tau/\varepsilon}\mathcal A_2=\frac{\pi^{3/2}}{8}(1+\ii)\,\varepsilon^{1/2}-\frac{\ii}{4}\,\varepsilon \nonumber \\
	&\qquad +\frac{\pi^{3/2}}{256}(1-\ii)\,\varepsilon^{3/2}-\frac{1}{24}\,\varepsilon^2+\mathcal O(\varepsilon^{5/2}), \\
	{\rm FP}\!{}&\int_0^\infty\dd\tau\,e^{4\ii\tau/\varepsilon}\mathcal A_4=-\frac{\pi^{3/2}}{2}(1-\ii)\,\varepsilon^{-1/2} \nonumber \\
	\label{eq:app:A-integrals} &\qquad -\frac{3\pi^{3/2}}{64}(1+\ii)\,\varepsilon^{1/2}-\frac{\ii}{6}\,\varepsilon+\mathcal O(\varepsilon^{3/2}).
\end{align}
With the prefactors of Eqs.~\eqref{eq:v:f:aligned-p2} and~\eqref{eq:v:f:aligned-der}, these give Eq.~\eqref{eq:v:f:optical-asymptotic}. In particular, the term $-1$ in $\mathcal A_2$ gives the term $-\ii\varepsilon/4$, which is the finite-distance correction $-15\ii\varepsilon w^2/64$ of the $p=2$ term.

As a check of Eq.~\eqref{eq:F0-aligned}, the same variables give the first Born term of the Newtonian channel, $F^{(1)}_{{\rm free},1}$, i.e.\ the term linear in $w$ of $F^{(0)}_{\rm newt}$. Its integral is that of Eq.~\eqref{eq:app:cyl} with $1/r$ in place of $1/r^2$, and the integral over $\eta$ is
\be
	\int_{-1}^{1}\frac{\dd\eta}{\sqrt{\tau(2+\tau)+\eta^2}}=2\,{\rm arsinh}\frac{1}{\sqrt{\tau(2+\tau)}}=\ln\Big(1+\frac{2}{\tau}\Big).
\ee
Expanding $\ln(1+2/\tau)=\ln2-\ln\tau+\tau/2+\dots$, and using Eq.~\eqref{eq:v:f:Abel-integral}, with the derivative with respect to $a$ at $a=1$ for $\ln\tau$,
\be \label{eq:first-born-newton-aligned}
	\int_0^\infty\dd\tau\,e^{4\ii\tau/\varepsilon}\ln\Big(1+\frac{2}{\tau}\Big)=\frac{\ii\varepsilon}{4}\Big[\ln\frac{8}{\varepsilon}+\gamma_{\rm E}\Big]+\frac{\pi\varepsilon}{8}-\frac{\varepsilon^2}{32}+\mathcal O(\varepsilon^3).
\ee
With the prefactor of the Newtonian first Born term, $2w/\varepsilon$, the first two terms are the expansion of $F^{(0)}_{{\rm newt},\infty}$ to first order in $w$, and the third gives the finite-distance correction $-w\varepsilon/16$ of Eq.~\eqref{eq:F0-aligned}.

\bibliographystyle{apsrev4-2}
\bibliography{bibliography}

\end{document}